\documentclass[aps,prc,twocolumn,showpacs,preprintnumbers,
nofootinbib,float,superscriptaddress,longbibliography]{revtex4-2}
\usepackage[multiple]{footmisc}
\usepackage{graphicx}
\usepackage{amsmath}
\usepackage{amssymb}
\usepackage[dvipsnames]{xcolor}
\usepackage[colorlinks=true, pdfstartview=FitV, linkcolor=RedOrange, citecolor=ForestGreen, urlcolor=blue]{hyperref}

\usepackage{bm}

\begin{document}

\title{Transport-based initial conditions for heavy-ion collisions at finite densities}

\author{H. Roch}\email{Hendrik.Roch@wayne.edu} 
\affiliation{Department of Physics and Astronomy, Wayne State University, Detroit, MI 48201.}

\author{G. Pihan}
\thanks{Associate Member of JETSCAPE}
\email{gpihan@uh.edu}
\affiliation{Department of Physics, University of Houston, 3507 Cullen Blvd, Houston, TX 77204.}

\author{A. Monnai}
\thanks{Associate Member of JETSCAPE}
\affiliation{Department of General Education, Faculty of Engineering, Osaka Institute of Technology, Osaka 535-8585, Japan.}

\author{S. Ryu}
\thanks{Associate Member of JETSCAPE}
\affiliation{Department of Physics and Astronomy, Wayne State University, Detroit, MI 48201.}

\author{N. Senthilkumar}
\thanks{Associate Member of JETSCAPE}
\affiliation{Department of Physics and Astronomy, Wayne State University, Detroit, MI 48201.}

\author{J. Staudenmaier}
\affiliation{Institute for Theoretical Physics, Goethe University, 60438 Frankfurt am Main, Germany.}

\author{H. Elfner}
\affiliation{GSI Helmholtzzentrum f\"{u}r Schwerionenforschung, 64291 Darmstadt, Germany.}
\affiliation{Institute for Theoretical Physics, Goethe University, 60438 Frankfurt am Main, Germany.}
\affiliation{Frankfurt Institute for Advanced Studies, 60438 Frankfurt am Main, Germany.}

\author{B. Schenke}
\affiliation{Physics Department, Brookhaven National Laboratory, Upton, NY 11973.}

\author{J. H. Putschke}
\affiliation{Department of Physics and Astronomy, Wayne State University, Detroit, MI 48201.}

\author{C. Shen} \email{chunshen@wayne.edu}
\affiliation{Department of Physics and Astronomy, Wayne State University, Detroit, MI 48201.}

\author{S.~A.~Bass}
\affiliation{Department of Physics, Duke University, Durham, NC 27708.}

\author{M.~Chartier}
\affiliation{Oliver Lodge Laboratory, University of Liverpool, Liverpool L69 7ZU, United Kingdom.}

\author{Y.~Chen}
\affiliation{Department of Physics and Astronomy, Vanderbilt University, Nashville, TN 37235.}

\author{R.~Datta}
\affiliation{Department of Physics and Astronomy, Wayne State University, Detroit, MI 48201.}

\author{R.~Dolan}
\affiliation{Department of Physics and Astronomy, Wayne State University, Detroit, MI 48201.}

\author{L.~Du}
\affiliation{Department of Physics, University of California, Berkeley, CA 94270.}
\affiliation{Nuclear Science Division, Lawrence Berkeley National Laboratory, Berkeley, CA 94270.}
\affiliation{Department of Physics, McGill University, Montr\'{e}al, QC H3A\,2T8, Canada.}

\author{R.~Ehlers}
\affiliation{Department of Physics, University of California, Berkeley, CA 94270.}
\affiliation{Nuclear Science Division, Lawrence Berkeley National Laboratory, Berkeley, CA 94270.}


\author{R.~J.~Fries}
\affiliation{Cyclotron Institute, Texas A\&M University, College Station, TX 77843.}
\affiliation{Department of Physics and Astronomy, Texas A\&M University, College Station, TX 77843.}

\author{C.~Gale}
\affiliation{Department of Physics, McGill University, Montr\'{e}al, QC H3A\,2T8, Canada.}

\author{D.~A.~Hangal}
\affiliation{Lawrence Livermore National Laboratory, Livermore, CA 94550.}
\affiliation{Department of Physics and Astronomy, Wayne State University, Detroit, MI 48201.}


\author{B.~V.~Jacak}
\affiliation{Department of Physics, University of California, Berkeley, CA 94270.}
\affiliation{Nuclear Science Division, Lawrence Berkeley National Laboratory, Berkeley, CA 94270.}

\author{P.~M.~Jacobs}
\affiliation{Department of Physics, University of California, Berkeley, CA 94270.}
\affiliation{Nuclear Science Division, Lawrence Berkeley National Laboratory, Berkeley, CA 94270.}

\author{S.~Jeon}
\affiliation{Department of Physics, McGill University, Montr\'{e}al, QC H3A\,2T8, Canada.}

\author{Y.~Ji}
\affiliation{Department of Statistical Science, Duke University, Durham, NC 27708.}

\author{F.~Jonas}
\affiliation{Department of Physics, University of California, Berkeley, CA 94270.}
\affiliation{Nuclear Science Division, Lawrence Berkeley National Laboratory, Berkeley, CA 94270.}


\author{M.~Kordell~II}
\affiliation{Cyclotron Institute, Texas A\&M University, College Station, TX 77843.}
\affiliation{Department of Physics and Astronomy, Texas A\&M University, College Station, TX 77843.}

\author{A.~Kumar}
\affiliation{Department of Physics, University of Regina, Regina, SK S4S 0A2, Canada.}

\author{R.~Kunnawalkam-Elayavalli}
\affiliation{Department of Physics and Astronomy, Vanderbilt University, Nashville, TN 37235.}

\author{J.~Latessa}
\affiliation{Department of Computer Science, Wayne State University, Detroit, MI 48202.}

\author{Y.-J.~Lee}
\affiliation{Laboratory for Nuclear Science, Massachusetts Institute of Technology, Cambridge, MA 02139.}
\affiliation{Department of Physics, Massachusetts Institute of Technology, Cambridge, MA 02139.}


\author{M.~Luzum}
\affiliation{Instituto  de  F\`{i}sica,  Universidade  de  S\~{a}o  Paulo,  C.P.  66318,  05315-970  S\~{a}o  Paulo,  SP,  Brazil. }

\author{A.~Majumder}
\affiliation{Department of Physics and Astronomy, Wayne State University, Detroit, MI 48201.}

\author{S.~Mak}
\affiliation{Department of Statistical Science, Duke University, Durham, NC 27708.}

\author{A.~Mankolli}
\affiliation{Department of Physics and Astronomy, Vanderbilt University, Nashville, TN 37235.}

\author{C.~Martin}
\affiliation{Department of Physics and Astronomy, University of Tennessee, Knoxville, TN 37996.}

\author{H.~Mehryar}
\affiliation{Department of Computer Science, Wayne State University, Detroit, MI 48202.}

\author{T.~Mengel}
\affiliation{Department of Physics and Astronomy, University of Tennessee, Knoxville, TN 37996.}

\author{C.~Nattrass}
\affiliation{Department of Physics and Astronomy, University of Tennessee, Knoxville, TN 37996.}

\author{J.~Norman}
\affiliation{Oliver Lodge Laboratory, University of Liverpool, Liverpool L69 7ZU, United Kingdom.}

\author{M.~Ockleton}
\affiliation{Oliver Lodge Laboratory, University of Liverpool, Liverpool L69 7ZU, United Kingdom.}

\author{C.~Parker}
\affiliation{Cyclotron Institute, Texas A\&M University, College Station, TX 77843.}
\affiliation{Department of Physics and Astronomy, Texas A\&M University, College Station, TX 77843.}

\author{J.-F.~Paquet}
\affiliation{Department of Physics and Astronomy, Vanderbilt University, Nashville, TN 37235.}



\author{G.~Roland}
\affiliation{Laboratory for Nuclear Science, Massachusetts Institute of Technology, Cambridge, MA 02139.}
\affiliation{Department of Physics, Massachusetts Institute of Technology, Cambridge, MA 02139.}


\author{L.~Schwiebert}
\affiliation{Department of Computer Science, Wayne State University, Detroit, MI 48202.}

\author{A.~Sengupta}
\affiliation{Cyclotron Institute, Texas A\&M University, College Station, TX 77843.}
\affiliation{Department of Physics and Astronomy, Texas A\&M University, College Station, TX 77843.}


\author{M.~Singh}
\affiliation{Department of Physics and Astronomy, Vanderbilt University, Nashville, TN 37235.}

\author{C.~Sirimanna}
\affiliation{Department of Physics, Duke University, Durham, NC 27708.}


\author{R.~A.~Soltz}
\affiliation{Lawrence Livermore National Laboratory, Livermore, CA 94550.}
\affiliation{Department of Physics and Astronomy, Wayne State University, Detroit, MI 48201.}

\author{I.~Soudi}
\affiliation{University of Jyväskylä, Department of Physics, P.O. Box 35, FI-40014 University of Jyväskylä, Finland.}
\affiliation{Helsinki Institute of Physics, P.O. Box 64, FI-00014 University of Helsinki, Finland.}

\author{Y.~Tachibana}
\affiliation{Akita International University, Yuwa, Akita-city 010-1292, Japan.}

\author{J.~Velkovska}
\affiliation{Department of Physics and Astronomy, Vanderbilt University, Nashville, TN 37235.}

\author{G.~Vujanovic}
\affiliation{Department of Physics, University of Regina, Regina, SK S4S 0A2, Canada.}

\author{X.-N.~Wang}
\affiliation{Key Laboratory of Quark and Lepton Physics (MOE) and Institute of Particle Physics, Central China Normal University, Wuhan 430079, China.}
\affiliation{Department of Physics, University of California, Berkeley, CA 94270.}
\affiliation{Nuclear Science Division, Lawrence Berkeley National Laboratory, Berkeley, CA 94270.}

\author{X.~Wu}
\affiliation{Department of Physics, McGill University, Montr\'{e}al, QC H3A\,2T8, Canada.}

\author{J.~Zhang}
\affiliation{Department of Physics and Astronomy, Vanderbilt University, Nashville, TN 37235.}

\author{W.~Zhao}
\affiliation{Department of Physics, University of California, Berkeley, CA 94270.}
\affiliation{Nuclear Science Division, Lawrence Berkeley National Laboratory, Berkeley, CA 94270.}

\collaboration{The JETSCAPE Collaboration}

\begin{abstract}
We employ the SMASH transport model to provide event-by-event initial conditions for the energy-momentum tensor and conserved charge currents in hydrodynamic simulations of relativistic heavy-ion collisions. We study the fluctuations and dynamical evolution of three conserved charge currents (net baryon, net electric charges, and net strangeness) with a 4D lattice-QCD-based equation of state, NEOS-4D, in the hydrodynamic phase. Out-of-equilibrium corrections at the particlization are generalized to finite densities to ensure the conservation of energy, momentum, and the three types of charges. These theoretical developments are integrated within the X-SCAPE code as a unified framework for studying the nuclear matter properties in the Beam Energy Scan program.
\end{abstract}

\maketitle

\section{Introduction}
\label{sec:Intro}

Relativistic nuclear collision experiments provide a versatile laboratory for studying the emergent properties of strongly interacting matter under extreme conditions~\cite{Busza:2018rrf, Shen:2020mgh, Achenbach:2023pba, Arslandok:2023utm}. 
At high temperatures and low net baryon densities, such as those probed at the top energies of the Relativistic Heavy-Ion Collider (RHIC) and the Large Hadron Collider (LHC), the produced quark-gluon plasma (QGP) behaves as a type of strongly-coupled fluid, whose collective space-time evolution is successfully described by relativistic hydrodynamics initialized with models based on nuclear geometry and parton saturation physics~\cite{Heinz:2013th, Gale:2013da, Petersen:2014yqa, Elfner:2022iae, Schenke:2020mbo}. 
Precision comparisons between theories and experiments require a unified approach with robust uncertainty quantification, which the JETSCAPE Collaboration has been pursuing~\cite{Putschke:2019yrg, JETSCAPE:2024dgu}. 
Meanwhile, as the field advances toward a more complete mapping of the phase diagram of Quantum Chromodynamics (QCD) --- exploring the uncharted regions at finite temperatures and large net baryon densities --- new theoretical challenges emerge~\cite{Bzdak:2019pkr, Monnai:2021kgu, An:2021wof, Almaalol:2022xwv, Sorensen:2023zkk, Du:2024wjm}. 
These include the need to model systems far from boost invariance, with significant baryon stopping, interactions between the fluid and spectators, and substantial non-equilibrium effects in the early stages of the collision~\cite{Schlichting:2019abc, Shen:2020gef, Shen:2021nbe}.

In this intermediate to low beam energy regime, relevant to the RHIC Beam Energy Scan (BES) program~\cite{Mohanty:2011nm, STAR:2017sal} and upcoming experiments at the Facility for Antiproton and Ion Research (FAIR)~\cite{CBM:2016kpk}, the assumptions of eikonal collisions and significant Lorentz contraction in traditional initial condition models become increasingly unreliable~\cite{Karpenko:2015xea, Shen:2017bsr, Shen:2017fnn, Shen:2021nbe}. 
For example, as the saturation scales become non-perturbative, the parton saturation-based models lose their predictive power at low collision energies~\cite{Mantysaari:2025tcg, Mantysaari:2025cos}. 
The traditional Glauber-type models lack descriptions of the complex baryon transport and longitudinal fluctuations arising from the initial stopping of individual nucleon-nucleon collisions~\cite{Shen:2020jwv, Ryu:2021lnx, Du:2022yok}. 
To explore phenomenological impacts from these aspects, transport-based models offer a microscopic and dynamical approach for generating (3+1)D initial conditions at finite densities.

Microscopic transport models --- such as UrQMD~\cite{Bass:1998ca, Bleicher:1999xi}, JAM~\cite{Nara:1999dz}, SMASH~\cite{SMASH:2016zqf}, and AMPT~\cite{Lin:2004en} --- simulate, in a Monte-Carlo setup, the non-equilibrium evolution of hadronic or partonic degrees of freedom through the early phases of the collision. 
These models naturally incorporate key features relevant for relativistic nuclear collisions at finite net baryon density, including nucleon stopping, finite charge diffusion, non-trivial longitudinal structure, and the development of out-of-equilibrium pressure anisotropies. 
By evolving the system from the initial nuclear overlap through a cascade of interactions, these models generate energy-momentum tensors and conserved charge currents that can be used to initialize hydrodynamic evolution on an event-by-event basis~\cite{Petersen:2008dd, Pang:2012he, Pang:2015zrq, Karpenko:2015xea, Du:2018mpf, Akamatsu:2018olk, Schafer:2021csj}.

However, the transport model description is effective when the system's averaged mean free path is longer than the particles' typical thermal wavelength, which is necessary for well-defined quasi-particles in the system~\cite{Cassing:2009vt}. 
It is challenging to include the effects of mixed/changing degrees of freedom between hadrons and partons. 
The transport-to-hydro matching procedure introduces model complexities, such as determining the appropriate time to switch to hydrodynamics~\cite{Karpenko:2015xea, Kanakubo:2019ogh}, implementing consistent smearing or coarse-graining procedures~\cite{Galatyuk:2015pkq}, and handling off-equilibrium corrections to the energy-momentum tensor and conserved currents~\cite{Inghirami:2022afu}.
Nevertheless, this hybrid framework enables a realistic treatment of the event-by-event initial conditions for relativistic nuclear collisions with multiple types of conserved charges, namely net baryon $B$, electric charge $Q$, and strangeness $S$. 
It enables systematic phenomenological studies of the QCD equation of state and the QGP transport coefficients, including its viscosity and charge diffusion coefficients~\cite{Greif:2017byw, Fotakis:2019nbq, Danhoni:2022xmt, Danhoni:2024kgi, Plumberg:2024leb}, within a 4D QCD phase diagram~\cite{Monnai:2024pvy, Abuali:2025tbd, Shen:2023aeg, Wu:2025psu}.

In this work, we investigate the construction and characterization of transport-based initial conditions with the SMASH transport model for heavy-ion collisions at finite densities. 
We focus on the event-by-event fluctuations of multiple conserved charges ($B$, $Q$, $S$) in the initial condition, their following collective evolution in the hydrodynamic phase with a 4D lattice-QCD-based equation of state~\cite{Monnai:2024pvy, Monnai:2025nyg}, and finally, the mapping from fluid to hadrons and their subsequent propagation to the final state. 
Our goal is to establish a multi-stage framework that enhances the phenomenological studies of QCD matter in the high-baryon-density region of the QCD phase diagram.

\section{Model Description}
\label{sec:model}
In this work, we employ the microscopic non-equilibrium hadronic transport model SMASH~\cite{SMASH:2016zqf} (version 3.1) as the initial conditions for relativistic viscous hydrodynamics, followed by particlization using the Cooper-Frye procedure~\cite {Cooper:1974mv} when the energy densities of fluid cells drop below a switching energy density. 
The produced hadrons are then fed back to SMASH and evolved until their kinetic freeze-out and final strong decays are performed. 
This setup is similar to other hybrid approaches with hadronic transport initial conditions, like the SMASH-vHLLE-Hybrid~\cite{Schafer:2021csj, Elfner:2025ojd} or setups using UrQMD in the initial state~\cite{Petersen:2008dd, Karpenko:2015xea, Du:2018mpf, Rathod:2025gvj}.

We apply this setup to Au+Au collisions at $\sqrt{s_{\rm NN}}=200\;\mathrm{GeV}$ collision energy with an impact parameter of $b=0$ or sampled in a range $b=6-8\;\mathrm{fm}$.
We also show examples for the evolution of the hydrodynamic medium at lower energies,~i.e., $\sqrt{s_{\rm NN}}=19.6\;\mathrm{GeV}$ and $7.7\;\mathrm{GeV}$.

\subsection{Hadronic Transport}
\label{subsec:SMASH}
The hadronic transport model SMASH is designed to evolve a hadronic out-of-equilibrium system, as it can be found in the initial and final state of heavy-ion collisions, in phase space~\cite{SMASH:2016zqf,weil_2025_15837933}. In this work, we run SMASH with physical particles in the quantum molecular dynamics (QMD) mode without mean field and test particle method.
SMASH incorporates all hadronic degrees of freedom up to a mass of $2.35\;\mathrm{GeV}$ listed in the PDG table~\cite{ParticleDataGroup:2018ovx}.
The high-energy hadronic interactions in SMASH are treated with string formation and fragmentation using Pythia 8~\cite{Sjostrand:2006za,Sjostrand:2007gs}.
Interactions at lower energies are implemented using resonances with vacuum Breit-Wigner spectral functions.
These have mass-dependent widths implemented with the Manley-Saleski ansatz~\cite{Manley:1992yb}.
The collisions between hadrons use a geometric covariant collision criterion~\cite{Hirano:2012yy} including $1\rightarrow 2$, $2\rightarrow 1$, and $2\rightarrow 2$ processes. 

\subsubsection{Initial Conditions}
\label{subsubsec:SMASH_IC}
The colliding nuclei in SMASH are sampled from a Woods–Saxon distribution, shifted by the impact parameter, and propagated along the $z$-direction.
To prevent nuclei without potentials from disintegrating before the collision, the initial Fermi motion in the transverse plane is frozen. 
It is reactivated as soon as the first nucleon–nucleon scattering occurs.

For the initialization of relativistic hydrodynamics, particles are extracted from SMASH on a constant proper time hypersurface $\tau_0$, defined by the nuclear passing time~\cite{Karpenko:2015xea}:
\begin{align}
\tau_0=\frac{R_{\rm p}+R_{\rm t}}{\sqrt{\left(\tfrac{\sqrt{s_{\rm NN}}}{2m_{\rm N}}\right)^2-1}},
\end{align}
where $R_{\rm p/t}$ are the projectile and target radii, $\sqrt{s_{\rm NN}}$ the center-of-mass energy, and $m_{\rm N}$ the nucleon mass.
At high collision energies, for example $\sqrt{s_{\rm NN}}=200\;\mathrm{GeV}$, we impose a lower bound of $\tau_0=0.5\;\mathrm{fm}/c$ similar to Ref.~\cite{Schafer:2021csj}.\footnote{For comparison, the values $\sqrt{s_{\rm NN}}=19.6\,\mathrm{GeV}$ and $7.7\,\mathrm{GeV}$ correspond to initialization times of $\tau_0=1.23\,\mathrm{fm}/c$ and $3.2\,\mathrm{fm}/c$, respectively, based on the nucleon mass $m_\mathrm{N}=0.938\,\mathrm{GeV}$ and gold radius $R_{\rm Au}=6.37\,\mathrm{fm}$ as implemented in SMASH.}

At high energies, most of the hadrons are produced by string fragmentation~\cite{Mohs:2019iee}. 
SMASH assumes that strings are decayed into hadrons instantaneously through Pythia.
Therefore, this setup does not include overlapping and interacting strings~\cite{Bierlich:2024odg}.
Using SMASH as an initial condition effectively sets the hadron formation time to $\tau_0$.
The physical process of string fragmentation requires finite time, which could be larger than $\tau_0$.
Our approach assumes a fast chemical equilibration from a gluon-dominated phase. 
We use only the space-time information of the hadrons' energies and the quantum numbers of the conserved charges to initialize the hydrodynamic fields.

\subsubsection{Hadronic Rescatterings}
\label{subsubsec:Afterburner}
After the particlization from the hydrodynamic freeze-out surface, we feed all the produced hadrons back into SMASH for the dilute non-equilibrium evolution in the afterburner phase.
For this evolution, the hadrons are back-propagated to the time of the first produced hadron, and they appear in the transport evolution when they are formed.
Before reaching their formation time, the hadrons free-stream.
During the evolution, which is mainly rescatterings and resonance decays, the system dilutes until there are no more interactions.
At the end of the evolution time, SMASH performs all strong decays of the hadrons.

\subsection{Hydrodynamics}
\label{subsec:MUSIC}

We consider individual hadrons extracted from SMASH at a constant proper time as wave packs. They contribute as source terms for the (3+1)D relativistic viscous hydrodynamic evolution using MUSIC~\cite{Schenke:2010nt, Schenke:2010rr, Paquet:2015lta, Denicol:2018wdp}.
The hydrodynamics solves the following equations with source terms on the right-hand side:
\begin{align}
    \partial_\mu T^{\mu\nu} &= J^\nu,\label{eq:EM_cons} \\
    \partial_\mu J^\mu_\alpha &= \rho_\alpha,\label{eq:charge_cons}
\end{align}
where the energy-momentum tensor can be decomposed as
\begin{align}
    T^{\mu\nu} = eu^\mu u^\nu - (P+\Pi)\Delta^{\mu\nu} + \pi^{\mu\nu}.
\end{align}
Here $\Delta^{\mu\nu}=g^{\mu\nu}-u^\mu u^\nu$ denotes a projection operator, $u^\mu$ the four-velocity, and $g^{\mu\nu}=\mathrm{diag}(1,-1,-1,-1)$ is the space-time metric.
Dissipative effects are described by the bulk viscous pressure $\Pi$ and the shear stress tensor $\pi^{\mu\nu}$.
The index $\alpha$ in Eq.~\eqref{eq:charge_cons} indicates the three different types of conserved charges: baryon number $B$, electric charge $Q$, and strangeness $S$.
The hadron source terms on the right-hand side of Eqns.~\eqref{eq:EM_cons}-\eqref{eq:charge_cons} are smeared with the kernel $\mathcal{K}(x^\mu;x_{\rm h}^\mu)$:
\begin{align}
    J^\nu(x^\mu) &= \frac{1}{{\rm d} t} \mathcal{K}(x^\mu; x^\mu_{\rm h}) P_{\rm h}^\nu, \label{eq:K_Jmu} \\
    \rho_\alpha(x^\mu) &= \frac{1}{{\rm d} t} \mathcal{K}(x^\mu; x^\mu_{\rm h}) Q_\alpha.
    \label{eq:K_rho}
\end{align}
The four-position of the hadron is indicated by $x^\mu_{\rm h}$.
Since we are dealing with hadrons, there is no pre-equilibrium color field contribution to the hydrodynamic source terms.
In this work, we employ a covariant smearing kernel and compare its results with those from a regular Gaussian smearing kernel.
The covariant smearing kernel in Cartesian coordinates was employed in Refs.~\cite{Oliinychenko:2015lva, Akamatsu:2018olk}:
\begin{align}
    \mathcal{K}_{\rm cov}(\mathbf{r}; \mathbf{r}_{\rm h}) &= \frac{\gamma}{(2\pi\sigma^2)^{3/2}} \nonumber \\
    & \times \exp\left(-\frac{(\mathbf{r} - \mathbf{r}_{\rm h})^2 + ((\mathbf{r} - \mathbf{r}_{\rm h}) \cdot \mathbf{u})^2}{\sigma^2}\right),
    \label{eq:K_covariant}
\end{align}
where $\mathbf{u} = \gamma \mathbf{v}$ is the three spatial components of the hadron's velocity four vector.
Here, we generalize this kernel to Milne coordinates by considering a wave packet on a constant proper time surface. In this case, the smearing kernel remains in the same form by defining $\mathbf{r} = (x, y, \tau \eta_s)$ and $\mathbf{r}_{\rm h} = (x_{\rm h}, y_{\rm h}, \tau \eta_{s, \mathrm{h}})$, where we explicitly include a factor of $\tau$ in the $\eta_s$ components such that the metric used for the dot product is $g_{ij} = \mathrm{diag}(-1, -1, -1)$. The spatial components of the hadron's four-velocity are $\mathbf{u} = (u^x, u^y, \tau u^\eta) = (P^x_{\rm h}/m_{\rm h}, P^y_{\rm h}/m_{\rm h}, \tau P^\eta_{\rm h}/m_{\rm h})$, where $m_{\rm h}$ is the rest mass of the hadron, and $\gamma = \sqrt{1 + (u^x)^2 + (u^y)^2 + (\tau u^\eta)^2}$.

The second is a simple Gaussian-smearing kernel
\begin{align}
    \mathcal{K}_G&(x,y,\eta_s; x_{\rm h}, y_{\rm h}, \eta_{s, {\rm h}}) = \frac{1}{\pi \sigma_\perp^2} \sqrt{\frac{\gamma_\eta^2}{\pi \sigma_\eta^2}} \nonumber \\
    &\times\exp\left(-\frac{(x - x_{\rm h})^2 + (y - y_{\rm h})^2}{\sigma^2_{\perp}} - \gamma_\eta^2 \frac{(\eta_s - \eta_{s, {\rm h}})^2}{\sigma_\eta^2}\right)
    \label{eq:K_Gaussian}
\end{align}
with two separate widths $\sigma_\perp$ and $\sigma_\eta$ in the transverse plane and the longitudinal direction, respectively. Here, the longitudinal profile was Lorentz contracted with the hadron's longitudinal velocity, $\gamma_\eta = \mathrm{cosh}(y_{\rm h})$. Comparing these width parameters with the one in the covariant smearing kernel, we have the following relation $\sigma_\perp = \sigma$ and $\sigma_\eta = \sigma/\tau_{\rm h}$.

The covariant smearing kernel, compared to the Gaussian kernel, introduces a Lorentz contraction in the direction of movement of the smeared hadron, such that rapidly moving hadrons appear to have a more squeezed density distribution in the position space. During the hydrodynamic evolution, these larger spatial gradients could lead to a faster expansion of the fluid into the direction of the contraction.
The Gaussian kernel was used in a few works in the literature~\cite{Karpenko:2015xea, Schafer:2021csj, Du:2025fdp}.
In this work, we set $\sigma = \sigma_\perp = 0.5$\,fm and $\sigma_\eta = 0.5$.

\begin{figure}[h!]
    \includegraphics[width=\linewidth]{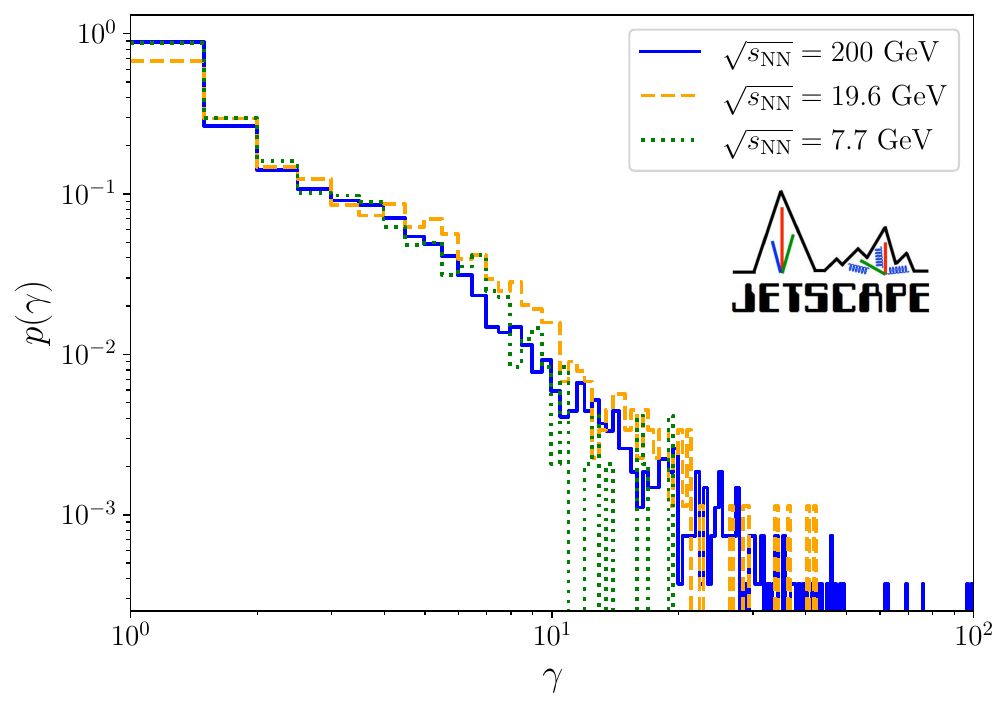}
    \caption{Probability distributions of the Lorentz factor $\gamma$ in the covariant smearing kernel Eq.~\eqref{eq:K_covariant} for a single Au+Au collision event at $b=0$, shown for three different center-of-mass energies $\sqrt{s_{\rm NN}}$.}
    \label{fig:gamma_hist}
\end{figure}

Compared to the Gaussian kernel, the covariant smearing kernel poses greater numerical challenges. 
The Lorentz contraction, particularly for fast-moving hadrons in the source terms, can amplify numerical inaccuracies arising from the finite spatial and rapidity spacing of the hydrodynamic grid. 
This effect can be especially significant for the evolution of conserved charges. Figure~\ref{fig:gamma_hist} shows the probability distribution of the Lorentz factor $\gamma$ in the covariant kernel~\eqref{eq:K_covariant} for three different $\sqrt{s_{\rm NN}}$ values, obtained from a single Au+Au collision event at $b=0$. 
In rare cases, $\gamma$ can reach values as large as $\sim 100$, resulting in a pronounced contraction of the source term. 
In practice, we find that the impacts from these fast-moving hadrons are small on the global charge conservation during hydrodynamic evolution.
In our simulations, we employ grid spacings of $\Delta\eta_s = 0.1$ and $\Delta x = \Delta y = 0.2\;\mathrm{fm}$, which yield a relative uncertainty per hydrodynamic step of $\mathcal{O}(10^{-4})$ for $N_B$ and $N_Q$, while the absolute uncertainty for $N_S$ is of the same order. 

The hydrodynamic equation of motion in Eqns.~\eqref{eq:EM_cons} and \eqref{eq:charge_cons} are solved together with an equation of state \textsc{neos-4D} based on lattice QCD and hadron resonance gas at finite densities~\cite{Monnai:2024pvy, Monnai:2025nyg}. 
This approach assumes that the system reaches chemical equilibrium at $\tau_0$.
We leave the exploration of dynamical chemical equilibration for future work. 
The complete 4-dimensional dependence for pressure on local energy density and three types of conserved charges, $P(e, n_B, n_Q, n_S)$, enables the independent propagation of net baryon, electric charge, and strangeness inside the QCD fluid. We consider viscous effects, namely shear and bulk viscosity, in fluid dynamic evolution. The shear stress tensor and bulk viscous pressure are evolved with the Denicol-Niemi-Molnar-Rischke (DNMR) theory~\cite{Denicol:2012cn, Denicol:2014vaa}.
The specific shear and bulk viscosity are parametrized as
\begin{align}
    \frac{\eta T}{e + \mathcal{P}} &= \eta_0 \left[1 + b \left(\frac{\mu_B}{\mu_{B, 0}}\right)^a \right] \,,\\
    \frac{\zeta T}{e + \mathcal{P}} &= \zeta_0 \exp \left[-\left(\frac{T - T_\mathrm{peak}}{T_{\mathrm{width}, \lessgtr}}\right)\right]\,,
\end{align}
where $T_\mathrm{peak} = 0.17\,\mathrm{GeV} - 0.15\;\mathrm{GeV}^{-1} \mu_B^2$ denotes the temperature at which the bulk viscosity peaks. The widths of the Gaussian profile are $T_{\mathrm{width},<} = 0.01$~GeV for $T < T_\mathrm{peak}$ and $T_{\mathrm{width},>} = 0.08$~GeV for $T > T_\mathrm{peak}$.
We use $\zeta_0 = 0.1$ for the peak value of bulk viscosity, and set the shear viscosity parameters to $\eta_0 = 0.08$, $b = 2$, $\mu_{B,0} = 0.6$~GeV, and $a = 0.7$~\cite{Pihan:2023dsb, Pihan:2024lxw}. We do not introduce explicit dependencies of the QGP's specific viscosity on $\mu_S$ and $\mu_Q$.
In this work, we do not consider the diffusion effects on multiple conserved charge currents and leave the full model development for future work.

The hydrodynamic evolution is performed until a switching energy density of $e_{\rm sw}=0.35\;\mathrm{GeV}/\mathrm{fm}^3$ is reached~\cite{Jahan:2024wpj}, followed by the construction of the freeze-out hypersurface $\Sigma$ using the Cornelius algorithm~\cite{Huovinen:2012is}.

\subsection{Particlization}
\label{subsec:iSS}

Before the further evolution of the hadrons in the afterburner, the cells at the freeze-out hypersurface have to be converted into hadronic degrees of freedom, where their spatial and momentum distributions are sampled.
This is done with a Cooper-Frye freeze-out procedure~\cite{Cooper:1974mv} for each hadron species $k$ via
\begin{equation}
    E_k\frac{\mathrm{d}N_k}{\mathrm{d}^3\mathbf{p}} = \int_{\Sigma}p^\mu\mathrm{d}^2\sigma_\mu \left[f^{\rm eq}_k(\mathbf{x},\mathbf{p})+\delta f_k(\mathbf{x},\mathbf{p})\right],
\end{equation}
where the out-of-equilibrium corrections $\delta f_k(\mathbf{x},\mathbf{p})=\delta f_k^{\rm shear}(\mathbf{x},\mathbf{p})+\delta f_k^{\rm bulk}(\mathbf{x},\mathbf{p})$ arise from the shear stress tensor and bulk viscous pressure. The $\mathrm{d}^3\sigma_\mu$ is the normal vector on the constant energy density hyper-surface $\Sigma$ and $f^{\rm eq}_k(\mathbf{x},\mathbf{p})$ denotes the thermal equilibrium distribution for the hadron species $k$.

We use the $(e,n_B,n_Q,n_S)$ in individual hydrodynamic cells and convert them to $(T,\mu_B,\mu_Q,\mu_S)$ using a hadron resonance gas (HRG) EOS including the hadronic content of the SMASH transport code.\footnote{The HRG does not include phase shift corrections from hadronic interactions, but the resonance states effectively take into account some of them~\cite{Venugopalan:1992hy}.}
This procedure ensures the conservation of energy-momentum and conserved charges during the particlization stage.
The official \textsc{neos-4D} matches to an HRG with a slightly different hadronic content, leading to a less than 5\% difference in the conversion.
Ref.~\cite{JETSCAPE:2020mzn} shows that the discrepancy is negligible in the final-state observables.

In this work, we derive the out-of-equilibrium $\delta f$ corrections with Grad's moment method and the Chapman-Enskog (CE) expansion with three types of conserved charges at finite densities~\cite{Fotakis:2022usk}. We start with the matching condition for a given fluid cell, the local energy and net charge densities are given by the equilibrium distributions as
\begin{align}
    e &= u_\mu u_\nu \sum_k g_k \int_\mathbf{p} p^\mu p^\nu f_k^\mathrm{eq}(T, \mu_k) \label{eq:feq_matching1} \\
    n_q &= u_\mu \sum_k q_k g_k \int_\mathbf{p} p^\mu f_k^\mathrm{eq}(T, \mu_k), \label{eq:feq_matching2} 
\end{align}
where $\int_\mathbf{p} = \int{\rm d}^3p/[(2\pi)^3 E_p]$, $g_k$ and $q_k$ are the degeneracy factor and quantum charge for particle species $k$, respectively.
The summation runs over all the hadronic species, including the cocktail resonance states. These conditions translate to the following requirement for $\delta f$:
\begin{align}
    0 &= u_\mu u_\nu \sum_k g_k \int_\mathbf{p} p^\mu p^\nu \delta f_k(T, \mu_k) \label{eq:df_matching1} \\
    0 &= u_\mu \sum_k q_k g_k \int_\mathbf{p} p^\mu \delta f_k(T, \mu_k). \label{eq:df_matching2}
\end{align}
And the viscous tensors of the fluid cell are mapped to the out-of-equilibrium corrections as
\begin{align}
    \pi^{\mu\nu} - \Pi \Delta^{\mu\nu} = \sum_k g_k \int_\mathbf{p} p^\mu p^\nu \delta f_k (T, \mu_k).
    \label{eq:df_matching3}
\end{align}
One can show that the matching conditions for charge diffusion currents are separated from the shear and bulk sectors in Eq.~\eqref{eq:df_matching3}.
We leave the derivation of the $\delta f$ for multi-charge diffusion to future work. 

\subsubsection{Grad's moment method}
\label{sec:Grad_df}

To consider multiple flavors of conserved charge currents at the particlization, we generalized Grad's moment expansion of $\delta f$ as~\cite{Monnai:2009ad, Monnai:2010qp, McNelis:2019auj}
\begin{align}
    \delta f_k &= f_k^\mathrm{eq} (1 \pm f_k^\mathrm{eq}) \nonumber \\
    &\left[ p^{\alpha} p^{\beta} \epsilon_{\alpha\beta} + p^{\alpha} \left( B_k \epsilon_{(B),\alpha} + S_k \epsilon_{(S),\alpha} + Q_k \epsilon_{(Q),\alpha} \right) \right],
    \label{eq:df_Grad}
\end{align}
where $B_k$, $S_k$, and $Q_k$ are the quantum numbers of the hadron species $k$.
Compared to the standard Grad's 14-moment method for a single conserved charge current, we introduce independent coefficients $\epsilon_{(q_i),\alpha}$ for individual conserved charge currents.
The coefficients $\epsilon_{\alpha\beta}$, $\epsilon_{(B),\alpha}$, $\epsilon_{(S),\alpha}$, $\epsilon_{(Q),\alpha}$ have $10 + 4 N_q$ independent moments
(22 for $N_q = 3$), which depend on local temperature and chemical potentials. They will be solved from the matching conditions in Eqns.~\eqref{eq:df_matching1} to \eqref{eq:df_matching3}.

The shear sector can be separated from Eq.~\eqref{eq:df_matching3} with the projection tensor as follows,
\begin{align}
    \pi^{\mu\nu} & = \sum_k g_k \int_\mathbf{p} \left[ \Delta^{\mu\alpha} \Delta^{\nu\beta} p_{\alpha} p_{\beta} - \frac{1}{3} \Delta^{\mu\nu} \left( \Delta^{\alpha\beta} p_{\alpha} p_{\beta} \right) \right] \delta f_k \nonumber \\
    & = 2 \mathcal{J}_{42} \left( \frac{1}{2} \Delta^{\mu}_{\alpha} \Delta^{\nu}_{\beta} + \frac{1}{2} \Delta^{\mu}_{\beta} \Delta^{\nu}_{\alpha} - \frac{1}{3} \Delta^{\mu\nu} \Delta_{\alpha\beta} \right) \epsilon^{\alpha\beta} \nonumber \\
    & \equiv 2 \mathcal{J}_{42} \epsilon^{\langle \mu\nu \rangle},
\end{align}
where the projection tensor $\Delta^{\mu\nu} \equiv g^{\mu\nu} - u^\mu u^\nu$ with $g^{\mu\nu} = \mathrm{diag}(1, -1, -1, -1)$. We define the following thermodynamic integrals for convenience,
\begin{align}
    & \mathcal{I}_{mn, k}(T, \{\mu_{q_i}\}) \equiv \nonumber \\
    & \qquad \int_\mathbf{p} (p^\mu u_\mu)^{m-2n} (-p^{\mu} p^\nu \Delta_{\mu\nu})^n f_k^\mathrm{eq}(1 \pm f_k^\mathrm{eq}),
\end{align}
\begin{align}
    \mathcal{J}_{mn}(T, \{\mu_{q_i}\}) & \equiv \frac{1}{(2n + 1)!!} \sum_k g_k  \mathcal{I}_{mn, k}(T, \{\mu_{q_i}\}) \\
    \mathcal{J}^{q_1}_{mn}(T, \{\mu_{q_i}\}) & \equiv \frac{1}{(2n + 1)!!} \sum_k q_{1k} g_k \mathcal{I}_{mn, k}(T, \{\mu_{q_i}\}) \\
    \mathcal{J}^{q_1 q_2}_{mn}(T, \{\mu_{q_i}\}) & \equiv \frac{1}{(2n + 1)!!} \sum_k q_{1k} q_{2k} g_k \mathcal{I}_{mn, k}(T, \{\mu_{q_i}\})
\end{align}
where $q_1$ and $q_2$ are the quantum charges of hadrons ($B$, $Q$, and $S$).
The bulk sector involves five scalar moments, namely $\epsilon_{**} = u_{\alpha} u_{\beta} \epsilon^{\alpha\beta}$, $\epsilon_{(B)*} = u_{\alpha} \epsilon_{(B)}^{\alpha}$, $\epsilon_{(S)*} = u_{\alpha} \epsilon_{(S)}^{\alpha}$, $\epsilon_{(Q)*} = u_{\alpha} \epsilon_{(Q)}^{\alpha}$ and $\textrm{Tr}(\epsilon) = g_{\alpha\beta} \epsilon^{\alpha\beta}$~\cite{Monnai:2009ad, Monnai:2010qp}. They can be solved from the matching condition Eqns.~\eqref{eq:df_matching1} to \eqref{eq:df_matching3} in the form of a matrix equation
\begin{widetext}
\begin{align}
    \left(\begin{array}{c}
    0 \vspace{5pt}\\
    -1 \vspace{5pt}\\
    0 \vspace{5pt}\\
    0 \vspace{5pt}\\
    0
    \end{array}\right)
    = \left(\begin{array}{ccccc}
    \mathcal{J}_{40} & \mathcal{J}_{41} & \mathcal{J}^{B}_{30} & \mathcal{J}^{S}_{30} &\mathcal{J}^Q_{30} \vspace{5pt}\\
    \mathcal{J}_{41} & \displaystyle \frac{5}{3} \mathcal{J}_{42} & \mathcal{J}^B_{31} & \mathcal{J}^S_{31} & \mathcal{J}^Q_{31} \vspace{5pt}\\
    \mathcal{J}^B_{30} & \mathcal{J}^B_{31} & \mathcal{J}^{BB}_{20} & \mathcal{J}^{BS}_{20} & \mathcal{J}^{BQ}_{20} \vspace{5pt}\\
    \mathcal{J}^S_{30} & \mathcal{J}^S_{31} & \mathcal{J}^{BS}_{20} & \mathcal{J}^{SS}_{20} & \mathcal{J}^{SQ}_{20} \vspace{5pt}\\
    \mathcal{J}^Q_{30} & \mathcal{J}^Q_{31} & \mathcal{J}^{BQ}_{20} & \mathcal{J}^{SQ}_{20} & \mathcal{J}^{QQ}_{20}
    \end{array}\right)
    \left(\begin{array}{c}
    \epsilon_{**} \vspace{5pt}\\
    \textrm{Tr}(\epsilon) - \epsilon_{**} \vspace{5pt}\\
    \epsilon_{(B)*} \vspace{5pt}\\
    \epsilon_{(S)*} \vspace{5pt}\\
    \epsilon_{(Q)*}
    \end{array}\right).
\end{align}
\end{widetext}

Finally, the out-of-equilibrium corrections are,
\begin{align}
    &\delta f_i^\mathrm{Grad, shear} = f_i^\mathrm{eq}(1 \pm f_i^\mathrm{eq}) \frac{p^\mu p^\nu \pi_{\mu\nu}}{2\mathcal{J}_{42}} \label{eq:df_Gradshear}
\end{align}
\begin{align}
    &\delta f_i^\mathrm{Grad, bulk} = f_i^\mathrm{eq}(1 \pm f_i^\mathrm{eq})\Pi \nonumber \\
    & \hspace{0.5cm} \times \bigg[\frac{1}{3}(4 \epsilon_{**} - \textrm{Tr}(\epsilon)) (p \cdot u)^2 + \frac{1}{3}(\textrm{Tr}(\epsilon) - \epsilon_{**}) m_i^2 \nonumber \\
    & \hspace{1cm} + (B_i \epsilon_{(B)*} + S_i \epsilon_{(S)*} + Q_i \epsilon_{(Q)*})(p \cdot u) \bigg].  \label{eq:df_Gradbulk}
\end{align}
We evaluate the six coefficients $\mathcal{J}_{42}$, $\frac{1}{3}(4 \epsilon_{**} - \textrm{Tr}(\epsilon))$, $\frac{1}{3}(\textrm{Tr}(\epsilon) - \epsilon_{**})$, $\epsilon_{(B)*}$, $\epsilon_{(S)*}$, and $\epsilon_{(Q)*}$ with a hadron resonance gas model as functions of $T, \mu_B, \mu_S, \mu_Q$.

\subsubsection{Chapman-Enskog expansion}
\label{sec:CE_df}

Taking the relaxation time approximation, we can obtain the leading-order $\delta f$ from the Chapman-Enskog expansion,
\begin{align}
    \delta f_k &= \frac{\tau_R}{(p \cdot u)} p^\mu \partial_\mu f_k^\mathrm{eq} \\
    &= \frac{\tau_R}{(p \cdot u)} f_k^\mathrm{eq}(1 \pm f_k^\mathrm{eq}) \nonumber \\ & \quad \times p^\mu \partial_\mu \left(\frac{p \cdot u - B_k \mu_B - S_k \mu_S - Q_k \mu_Q}{T}\right),
\end{align}
where $B_k$, $S_k$, and $Q_k$ are the quantum numbers of the hadron species $k$.
By separating the derivatives into scalar, vector, and tensor parts and using ideal hydrodynamic equations of motion, we can identify the shear and bulk viscous corrections as follows~\cite{Czajka:2017wdo, Czajka:2020mho},
\begin{align}
    &\delta f_k^\mathrm{CE, shear} = f_k^\mathrm{eq}(1 \pm f_k^\mathrm{eq}) \frac{ \pi_{\mu\nu}}{2 \hat{\eta}} \frac{p^\mu p^\nu}{(p \cdot u) T} \\
    &\delta f_k^\mathrm{CE, bulk} = - f_k^\mathrm{eq}(1 \pm f_k^\mathrm{eq}) \left( \frac{\Pi}{\hat{\zeta}} \right) \frac{1}{(p \cdot u) T} \nonumber \\
    & \hspace{2cm} \times \left[ \frac{1}{3} m^2 - \left( \frac{1}{3} - \hat{c}^2 \right) (p \cdot u)^2 \right],
\end{align}
where the relaxation time $\tau_R$ is casted into the coefficients $\hat{\eta}$, $\hat{\zeta}$, and $\hat{c}$ which are functions of $T, \mu_B, \mu_S, \mu_Q$. They can be determined with the matching conditions in Eqns.~\eqref{eq:df_matching1} to \eqref{eq:df_matching3},
\begin{align}
    \hat{\eta} &= \frac{1}{15 T} \sum_i g_i \int_\mathbf{p} \frac{|\mathbf{p}|^4}{E_p} f_i^\mathrm{eq}(1 \pm f_i^\mathrm{eq}) \\
    \hat{\zeta} &= \frac{1}{3 T} \sum_i g_i m_i^2 \int_\mathbf{p} f_i^\mathrm{eq}(1 \pm f_i^\mathrm{eq}) \left[ \frac{ m_i^2 }{ 3 E_p } - \left( \frac{1}{3} - \hat{c}^2 \right) E_p \right] \\
     \hat{c}^2 &= \frac{\mathcal{J}_{31}}{\mathcal{J}_{30}}
\end{align}
These three coefficients are computed using a hadron resonance gas model in the 4-dimensional $(T, \mu_B, \mu_S, \mu_Q)$ space.

The 4-dimensional $\delta f$ coefficients from Grad's and CE methods have been computed and implemented in the particle sampler \texttt{iSS}~\cite{ryu_2025_17088248, Shen:2014vra}.

\section{X-SCAPE Implementation}
\label{sec:XSCAPE}
The X-SCAPE framework implements the workflow chain \texttt{SMASH+MUSIC+iSS+SMASH}, orchestrated by the Bulk Dynamics Manager (BDM) module in version~2.0.0 in the executable \texttt{SMASHInitialCondition}~\cite{XSCAPErepo}.
Within X-SCAPE, all modules can operate on a time-stepped basis, enabling workflows where each module advances according to its own internal clock.

\begin{figure}[h!]
    \includegraphics[width=\linewidth]{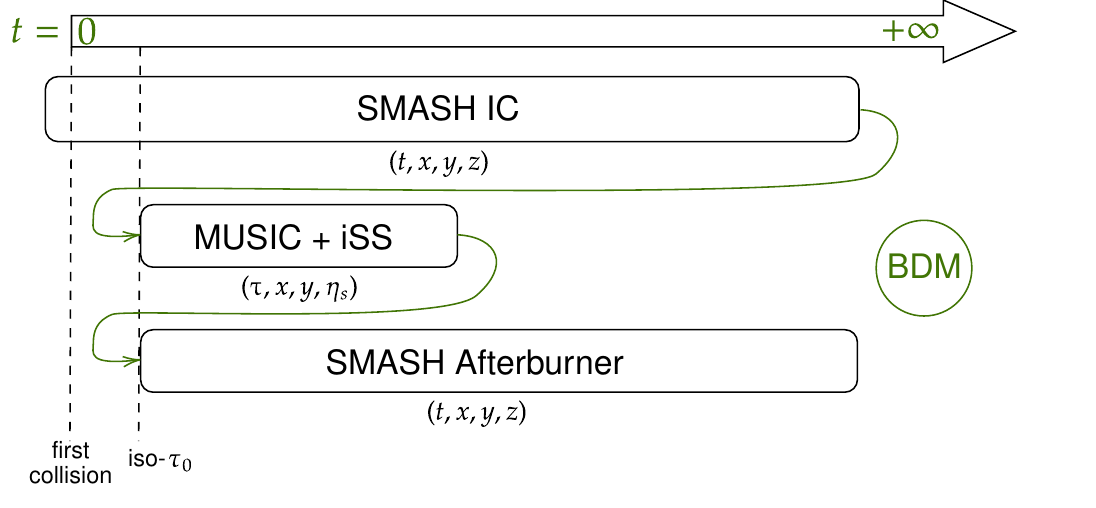}
    \caption{Schematic diagram of the simulation workflow for \texttt{SMASH+MUSIC+iSS+SMASH} model in the X-SCAPE framework and orchestrated by the BDM module. The simulation begins with SMASH pre-equilibrium dynamics in Minkowski coordinates, proceeds through hydrodynamic evolution and particlization in Milne coordinates, and concludes with hadronic rescattering in the SMASH afterburner.}
    \label{fig:FlowDiagramXSCAPE}
\end{figure}

The specific workflow corresponding to this study is illustrated in Fig.~\ref{fig:FlowDiagramXSCAPE}, where time flows from left to right.

The simulation begins by initializing the nuclei in SMASH according to a Woods–Saxon distribution, separated by the chosen impact parameter, which can be either fixed or randomly sampled within a range.
The time origin $t=0$ is then defined dynamically as the moment of the first nucleon–nucleon collision.
During the pre-equilibrium stage, SMASH propagates the produced hadrons in Minkowski coordinates, while the BDM module continuously extracts particles crossing the iso-$\tau_0=0.5\,\mathrm{fm}/c$ hypersurface.
Ref.~\cite{Schafer:2021csj} demonstrated that this removal, which leaves holes in the hadronic medium, does not affect final-state observables. 
To ensure accurate determination of hadron crossings on this surface, the pre-equilibrium hadronic transport evolution is performed with a very small timestep of $\Delta t=0.002\;\mathrm{fm}/c$.
Hadrons in the extracted hadron list are back-propagated using free-streaming, ensuring they are exactly on the iso-$\tau_0$ hypersurface when the source terms are created.
All participant hadrons are then given to a hadron liquefier module that smears the source terms according to the procedure described in Sec.~\ref{subsec:MUSIC}. At the same time, the spectator hadrons are stored for possible use in the afterburner phase.\footnote{In this work, we do not include the spectator hadrons in the afterburner phase.}

Since the hydrodynamic evolution is performed in Milne coordinates, it is not possible to concurrently feed hadrons from the transport stage into the hydrodynamic stage in real time.
Instead, after the SMASH initial stage is complete, the time is reset to $\tau_0$, and the subsequent evolution proceeds in proper time using timesteps of $0.1\;\mathrm{fm}/c$ in the BDM module.\footnote{Note that within one time step of the BDM module, the hydrodynamic evolution typically advances in multiple substeps. In our setup, a time step of $\Delta\tau = 0.02\;\mathrm{fm}/c$ is used to satisfy the Courant–Friedrichs–Lewy (CFL) condition~\cite{Courant1928}.}
At the first hydrodynamic timestep, the sources from the hadron liquefier module are deposited in MUSIC.

During the hydrodynamic evolution, the medium is continuously checked for cells crossing the freeze-out temperature hypersurface.
The hadrons produced at freeze-out are stored in the BDM module.
Once the entire system has frozen out, the time is reset again to the earliest hadron emission time in the BDM list.

Finally, the simulation resumes in Minkowski time during the late-stage hadronic evolution. 
The afterburner module (SMASH) propagates all hadrons emitted from the hydrodynamic phase, and optionally the initial-state spectators, until kinetic freeze-out.

Following the discussion in Sec.~\ref{subsec:MUSIC} on the numerical challenges of implementing the covariant smearing kernel -- particularly the accuracy loss caused by highly contracted source terms from fast-moving hadrons -- we introduce a mitigation strategy implemented in the BDM and hadron liquefier modules. 
The hadron liquefier includes a user-defined upper bound on the Lorentz factor $\gamma$, which can be set in the X-SCAPE configuration file. 
If a participant hadron designated as a source term exceeds this $\gamma$ threshold, it is rejected by the liquefier and returned to the BDM, where it is stored until the afterburner stage. 
At that point, it is propagated together with the hadrons emerging from the hydrodynamic evolution, ensuring energy–momentum conservation.

For the $\sqrt{s_{\rm NN}} = 200\;\mathrm{GeV}$ case with $\Delta\eta_s = 0.1$ and $\Delta x = \Delta y = 0.2\;\mathrm{fm}$, setting $\gamma_{\rm cutoff} = 10$ yields a substantial improvement: the numerical accuracy for global charge conservation increases by a factor of $\approx 2$.

We verify that the current X-SCAPE implementation produces consistent numerical results with those obtained by running standard SMASH and MUSIC code packages sequentially outside the X-SCAPE framework. 

\section{Results}
\label{sec:restults}

In this section, we investigate the hydrodynamic evolution with multiple conserved charge currents and their impact on the final-state observables using the SMASH+MUSIC+iSS+SMASH model.
Simulation outputs were analyzed using the Python analysis package SPARKX~\cite{Sass:2025opk,hendrik_roch_2025_15838371}.

\subsection{Hydrodynamic evolution with multiple conserved charge currents}
\label{sec:Dynamics}

\begin{figure*}[ht!]
    \includegraphics[width=0.49\linewidth]{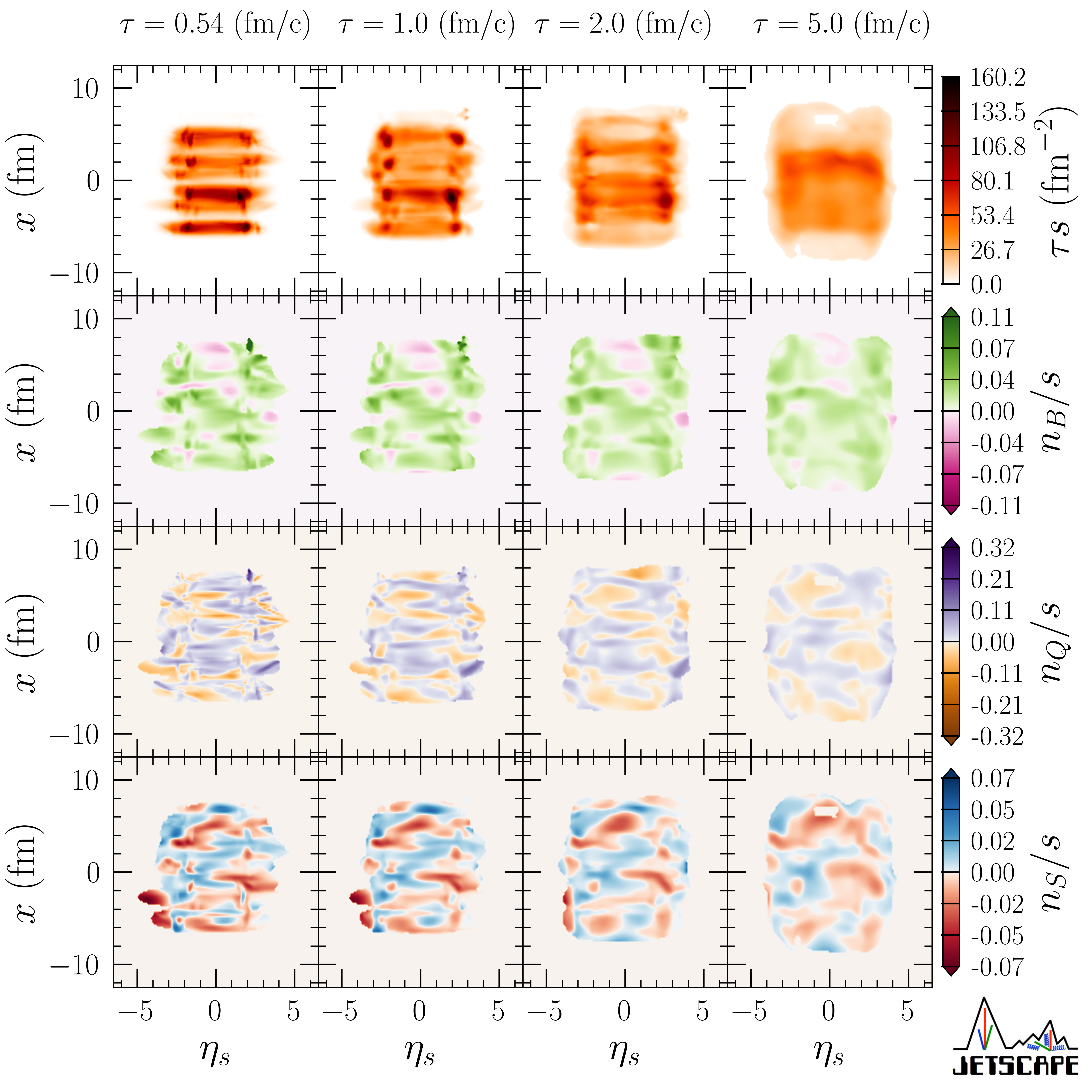}
    \includegraphics[width=0.49\linewidth]{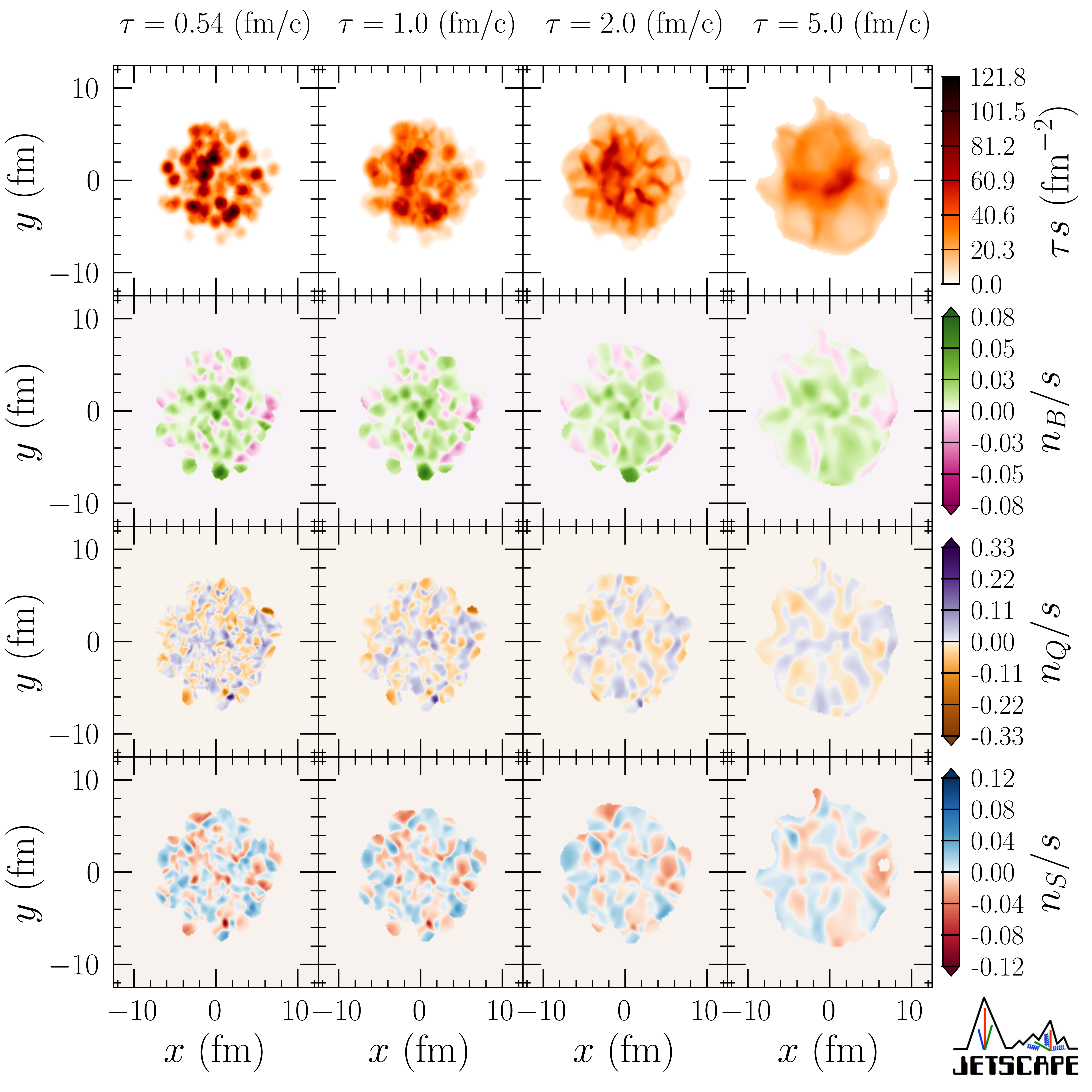}
    \caption{Evolution snapshots for scaled entropy density $\tau s$, scaled conserved charge densities ($n_B/s$, $n_Q/s$, $n_S/s$) in the $x-\eta_s$ plane at $y=0$ (left panels) and in the transverse $x-y$ plane at mid-rapidity $\eta_s = 0$ (right panels) for SMASH initial conditions at $\sqrt{s_{\rm NN}}=200\;\mathrm{GeV}$ with the covariant smearing kernel at $\tau = 0.54, 1, 2, 5$ fm/$c$ at impact parameter $b=0$.
    }
    \label{fig:snapshots_200}
\end{figure*}

Figure~\ref{fig:snapshots_200} shows the evolution snapshots for the entropy density and scaled conserved charge densities in the $x-\eta_s$ and $x-y$ planes for central Au+Au collisions at 200 GeV. We scaled the entropy density by the longitudinal proper time, which accounts for the longitudinal expansion of unit cells in the Milne coordinate. At 200 GeV, we observe that the entropy density in the $x-\eta_s$ plane shows a flux-tube-like distribution, which is reminiscent of the Lund string production~\cite{Mohs:2019iee}.
To estimate the sizes of conserved charge density profiles, we scale them with the local entropy density. The ratio of net baryon to entropy density stays mostly positive along the longitudinal direction, indicating the baryon-antibaryon pair production is subdominant compared to the contribution from initial-state baryon stopping. On the contrary, the net electric charge density distribution exhibits a lot of local fluctuations because charged $\pi^+$-$\pi^-$ pairs are relatively more abundantly produced than baryon-antibaryon pairs. At a fixed $x$ position, the net electric charge density alternates between positive and negative along $\eta_s$ as a result of the produced charge pairs from the transport model. A similar mechanism is present for the net strangeness, which is driven purely by fluctuations from local pair production within the transport model.

We note that the maximum values of $n_Q/s$ and $n_S/s$ are larger than those of $n_B/s$ from the SMASH initial condition, which is a unique feature for initial conditions based on hadronic transport models. Within hadronic degrees of freedom, the lightest particle species that carries a non-zero baryon charge is the proton. Since its mass is much heavier than that of mesons such as pions and kaons, the energy cost to produce a proton-antiproton pair is much higher than a pair of charged mesons in the string fragmentation. Therefore, this mechanism results in fluctuation-dominant density distributions for net electric charges and net strangeness, while net baryon density is dominated by the initial-stage stopping/transport. This feature differentiates the hadronic transport-based initial conditions from others~\cite{Bierlich:2018xfw, Pierog:2013ria, Shen:2017bsr, Shen:2022oyg, Werner:2023mod, Garcia-Montero:2023gex, Carzon:2019qja, Carzon:2022zpa, Carzon:2023zfp, Zhao:2022ugy, Ryu:2023bmx, Pihan:2023dsb, Pihan:2024lxw}.

The right panel of Fig.~\ref{fig:snapshots_200} shows the entropy density and scaled charged density distributions in the transverse plane at mid-rapidity $\eta_s = 0$. We can observe the Lorentz contraction from the covariant kernel in the charge density distributions at early times. As the collision system expands, the length scale of typical charge fluctuations increases with evolution time and reaches around 2 fm at $\tau = 5$ fm/$c$.

\begin{figure*}[ht!]
    \includegraphics[width=0.49\linewidth]{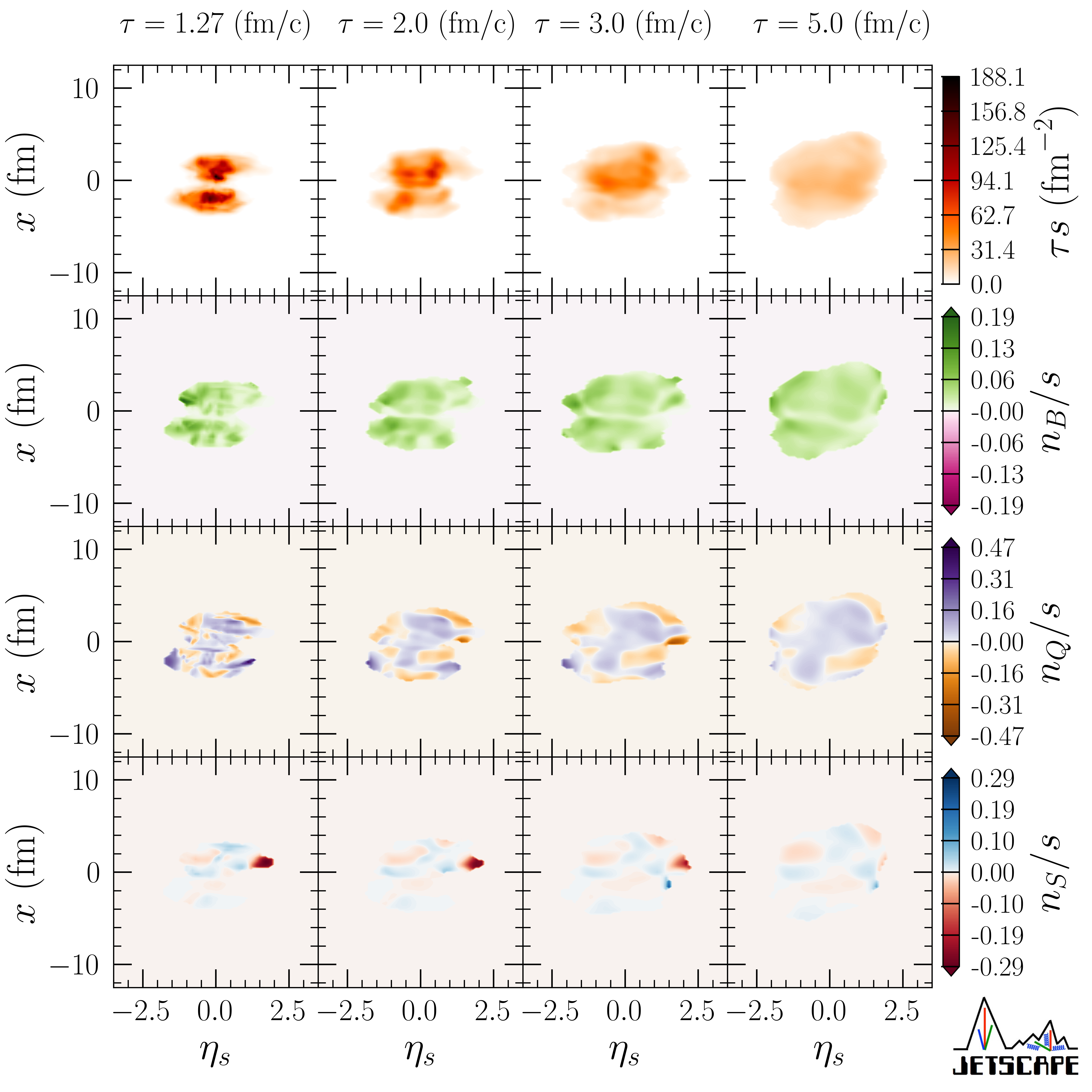}
    \includegraphics[width=0.49\linewidth]{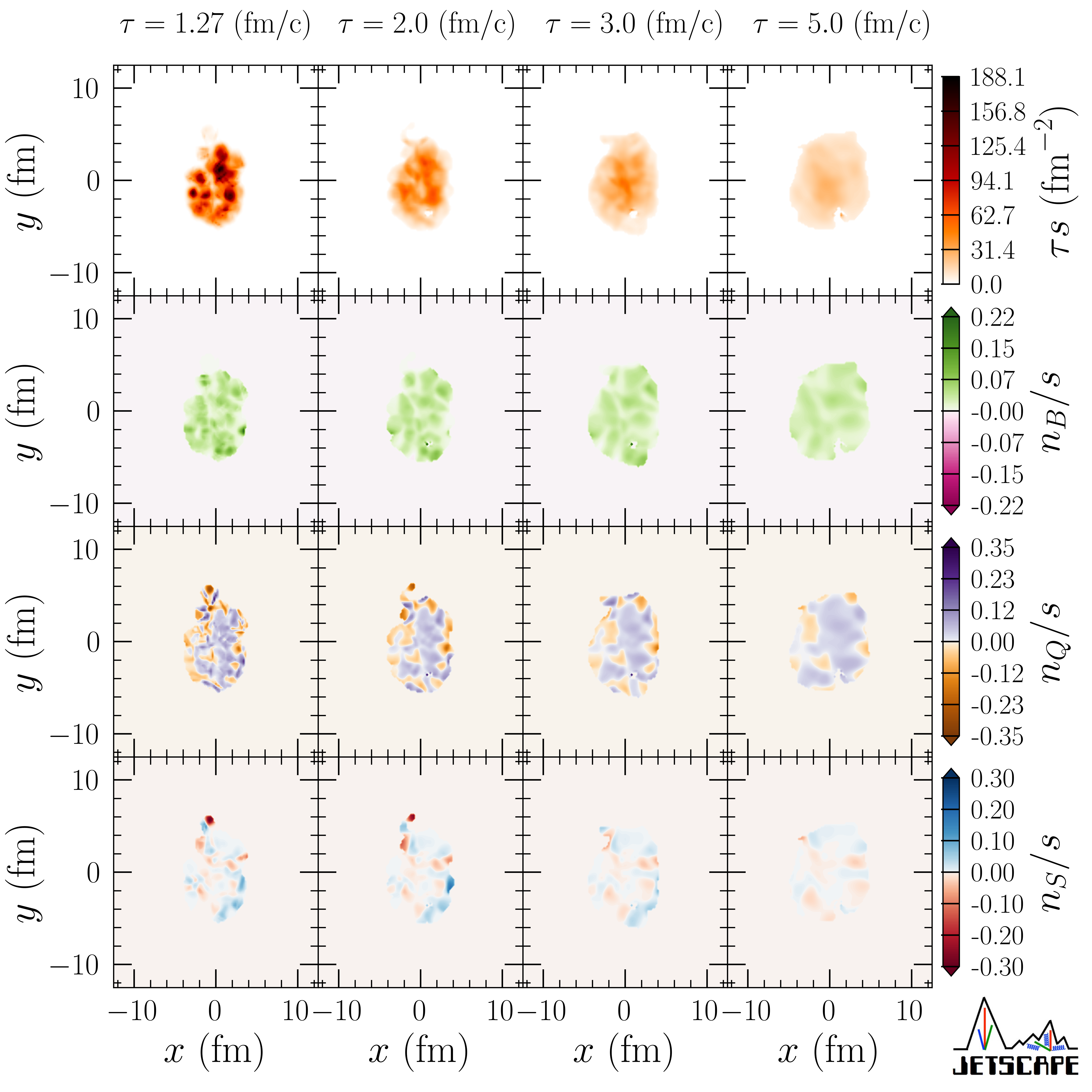}
    \caption{Evolution snapshots for scaled entropy density $\tau s$, scaled conserved charge densities ($n_B/s$, $n_Q/s$, $n_S/s$) in the $x-\eta_s$ plane at $y=0$ (left panels) and in the transverse $x-y$ plane at mid-rapidity $\eta_s = 0$ (right panels) for SMASH initial conditions at $\sqrt{s_{\rm NN}}=19.6\;\mathrm{GeV}$ with the covariant smearing kernel at $\tau = 1.27, 2, 3, 5$ fm/$c$ at a random impact parameter in the range from 6 to 8 fm.
    }
    \label{fig:snapshots_19p6}
\end{figure*}

\begin{figure*}[ht!]
    \includegraphics[width=0.49\linewidth]{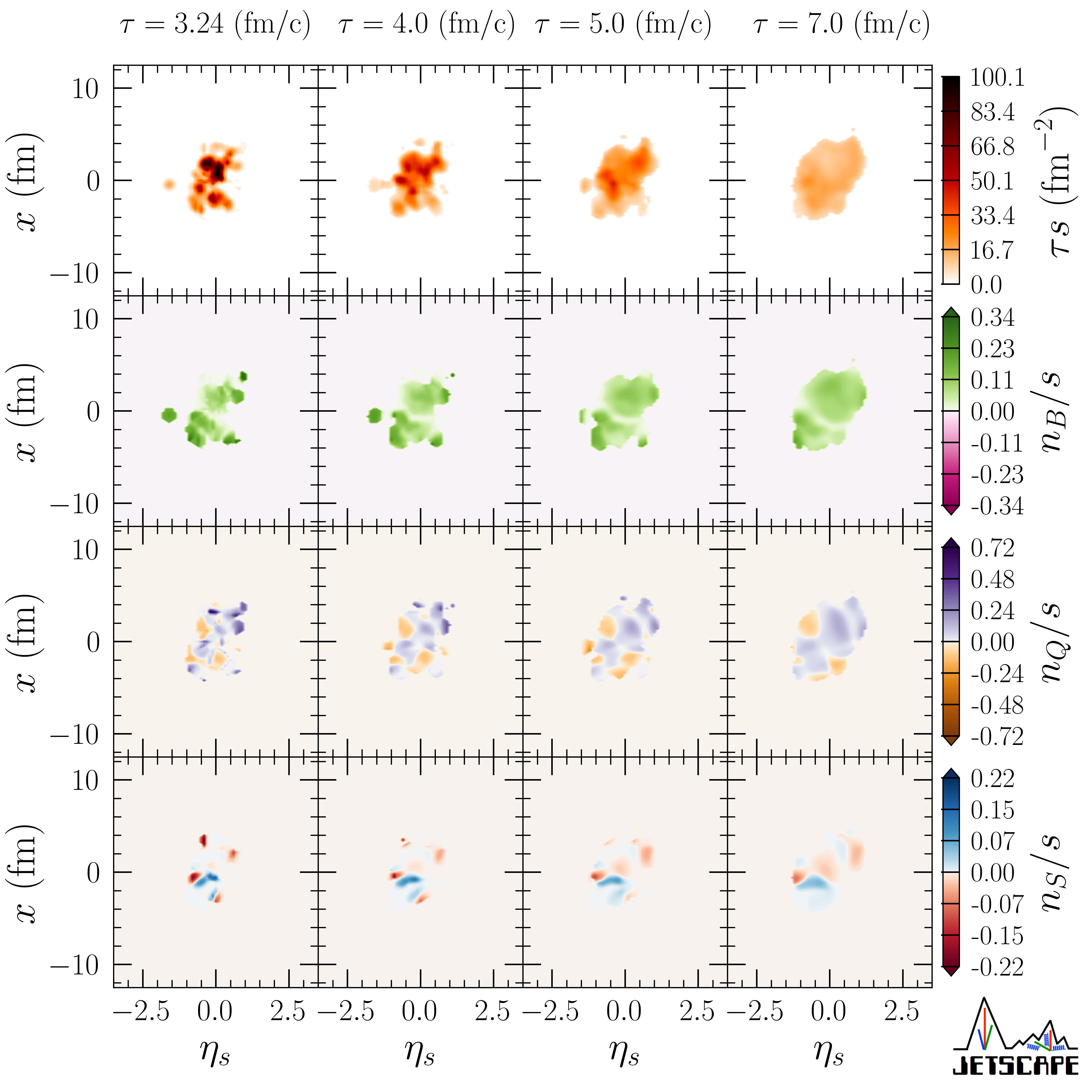}
    \includegraphics[width=0.49\linewidth]{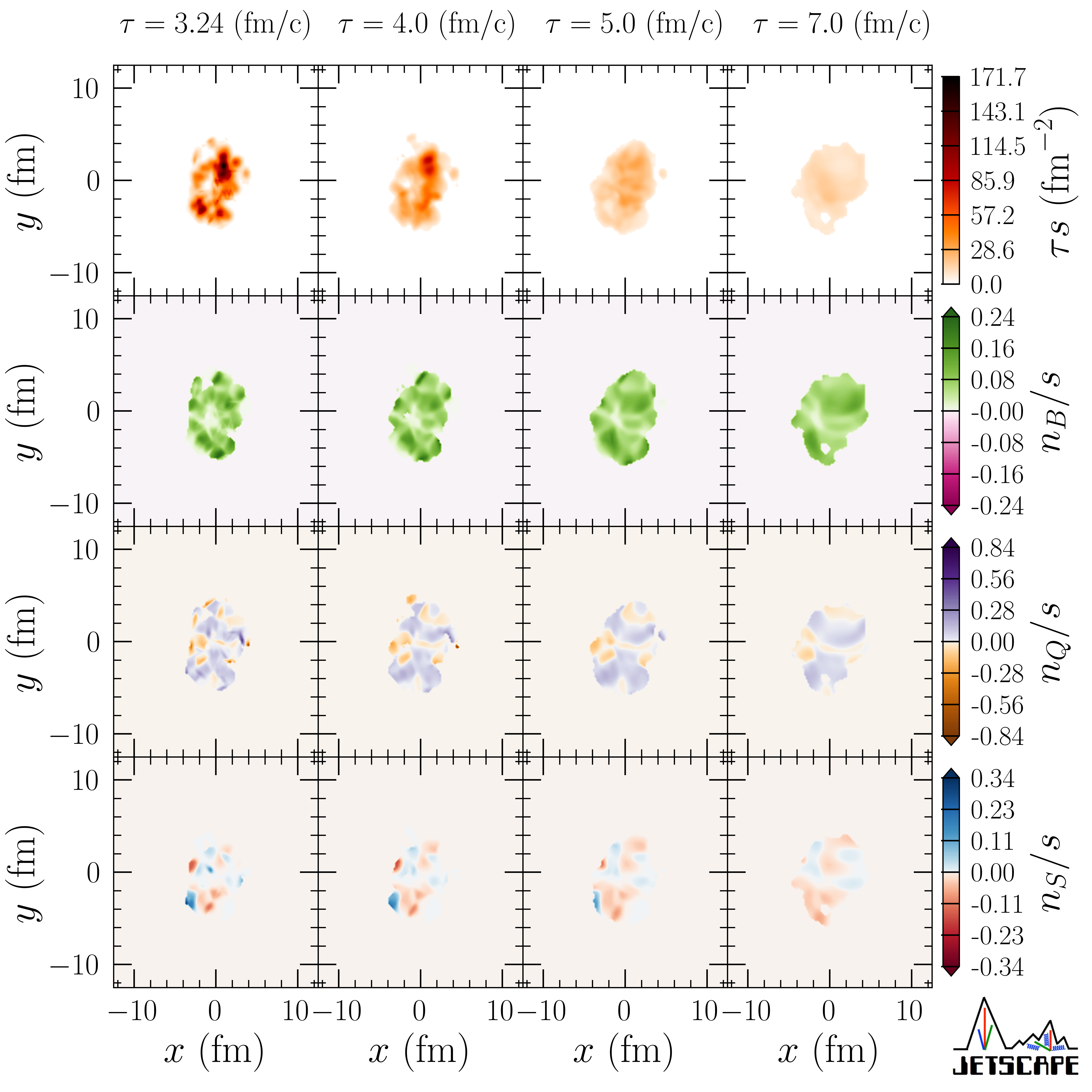}
    \caption{Evolution snapshots for scaled entropy density $\tau s$, scaled conserved charge densities ($n_B/s$, $n_Q/s$, $n_S/s$) in the $x-\eta_s$ plane at $y=0$ (left panels) and in the transverse $x-y$ plane at mid-rapidity $\eta_s = 0$ (right panels) for SMASH initial conditions at $\sqrt{s_{\rm NN}}=7.7\;\mathrm{GeV}$ with the covariant smearing kernel at $\tau = 3.24, 4, 5, 7$ fm/$c$ at a random impact parameter in the range from 6 to 8 fm.
    }
    \label{fig:snapshots_7p7}
\end{figure*}

Figures~\ref{fig:snapshots_19p6} and \ref{fig:snapshots_7p7} show similar snapshots of the hydrodynamic evolution for semi-peripheral Au+Au collisions at 19.6 and 7.7 GeV, respectively. Compared to the results at 200 GeV, we observe that the number of fluctuation sources in conserved charge distributions is significantly less at lower collision energies. However, the maximum value of net electric charges remains a factor of 2 larger than that of net baryon density. The magnitudes of $n_S/s$ values are about three times larger at low collision energies compared to those at 200 GeV. These results indicate that the collision systems probe a significant region of the QCD phase diagram along the directions of net electric charge and net strangeness density. It is crucial to employ a realistic equation of state in 4D, namely $P(e, n_B, n_Q, n_S)$, to properly take into account the variation of thermodynamic quantities at finite densities.

The right panels of Figs.~\ref{fig:snapshots_19p6} and \ref{fig:snapshots_7p7} show the transverse dynamics of the collision systems at low energies. In contrast to the rapid fireball expansion at 200 GeV shown in Fig.~\ref{fig:snapshots_200}, we observe that the development of hydrodynamic radial flow is comparable with the system's cooling rate at low energies, which results in a roughly constant size of the fireball over the entire evolution. The fireball still retains an elliptic shape at a late stage of evolution, indicating that it has not fully transformed the spatial eccentricity into flow anisotropies at low collision energies.

\subsection{Covariant smearing kernel on final-state observables}
\label{sec:observables}

In this section, we study the effects of Lorentz-contracted smearing kernels on final-state observables in heavy-ion collisions.

\begin{figure}[h!]  
    \includegraphics[width=\linewidth]{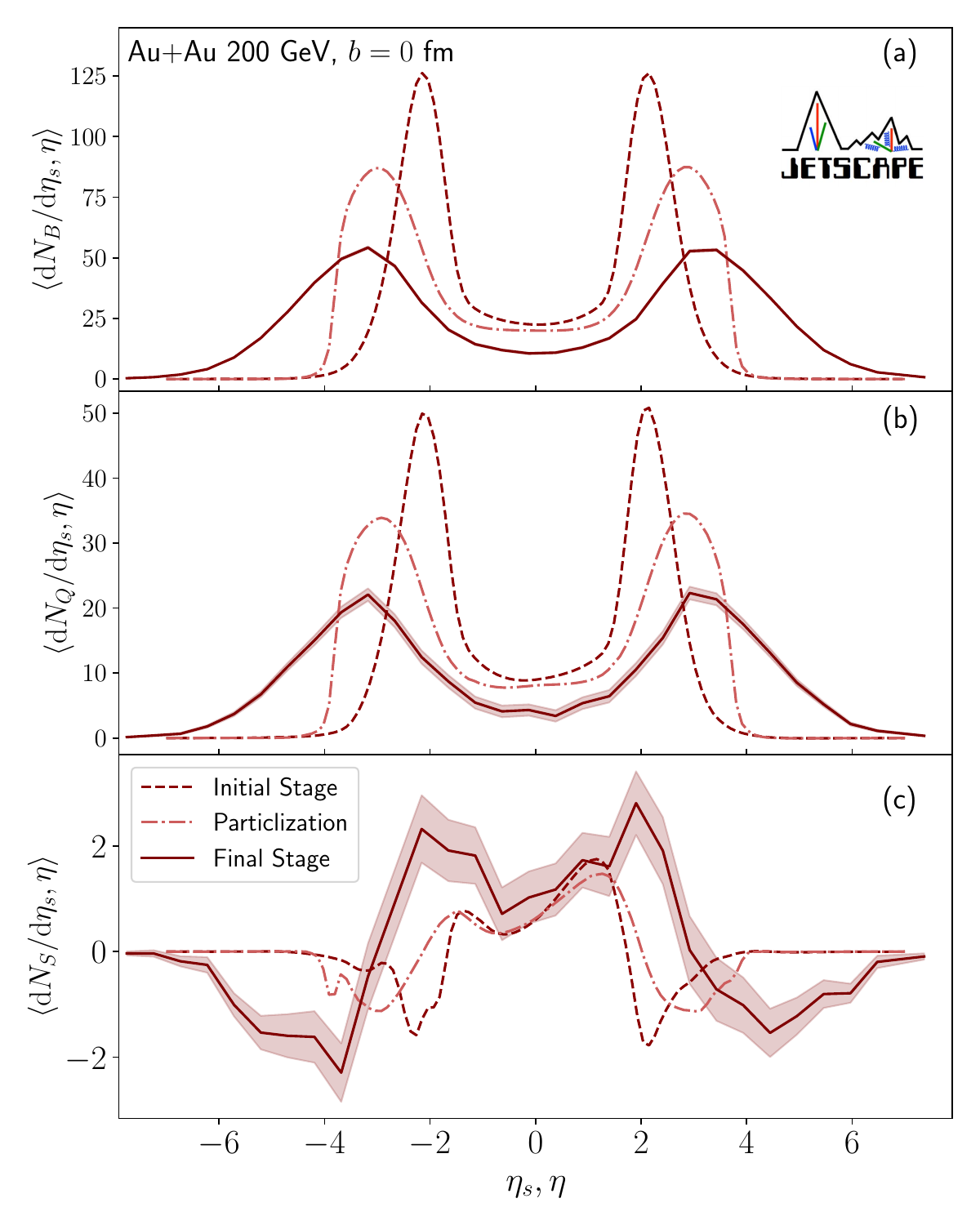}
    \caption{The space-time rapidity distributions of conserved charges at initial (dashed) and particlization (dash-dotted) from simulations with the covariant smearing kernel. They are compared with the pseudo-rapidity distributions of net baryon density (a), net electric charge density (b), and net strangeness (c) in central Au+Au collisions at 200 GeV.}
    \label{fig:dNdeta_initial_vs_final}
\end{figure}

Figure~\ref{fig:dNdeta_initial_vs_final} compares the longitudinal distributions of conserved charges at three different stages of central $(b = 0)$ Au+Au collisions at 200 GeV. The net baryon and net electric charge distributions have a similar shape at the initial stage, with most of the charges located at $|\eta_s| \sim 2$ after the colliding nuclei pass through each other, simulated by the SMASH transport model. Because of the global strangeness neutrality constraint, the net strangeness distribution presents more structures along the longitudinal direction\footnote{Because only 200 SMASH events are analyzed here, the net strangeness distribution still shows an asymmetric shape for central Au+Au collisions at the initial stage.}. The negative valleys in the forward and backward rapidities are correlated with the peak positions in the net baryon distribution, indicating the locations of strangeness baryons in the initial state. The positive strangeness near mid-rapidity represents the balanced production of strange mesons, such as kaons. 
Now, examining the conserved charge distributions on the particlization hypersurface, we observe that the hydrodynamic flow pushes more net baryons and net electric charges from central to forward and backward rapidities. The peaks of those distributions shift forward/backward by about one unit in space-time rapidity as the collision system evolves from the initial state to the particlization. The evolution of the strangeness distribution in the hydrodynamic phase is similar, with its negative valleys moving forward/backward by about one unit in space-time rapidity.
Comparing the difference between the individual conserved charge distributions at the particlization and their final pseudorapidity distributions, we can quantify the effects of thermal emission via the Cooper-Frye procedure and further hadronic rescatterings and decays in the transport phase. We observe that the tails of final-state pseudo-rapidity distributions of these conserved charges reach out to significantly higher pseudo-rapidities, about two units in $\eta$ beyond the beam rapidity at 200 GeV.
The peak positions for the final-state net baryon and net electric charges are approximately 0.5 units more forward/backward compared to their space-time rapidity immediately before particlization. For the net strangeness, we find that the maximum and minimum values increase in the final-state pseudo-rapidity distribution compared to those in the initial state.
We verified that the total numbers of net baryons, net electric charges, and net strangeness remain constant as the system evolves through these three stages.

\begin{figure}[h!]
    \includegraphics[width=\linewidth]{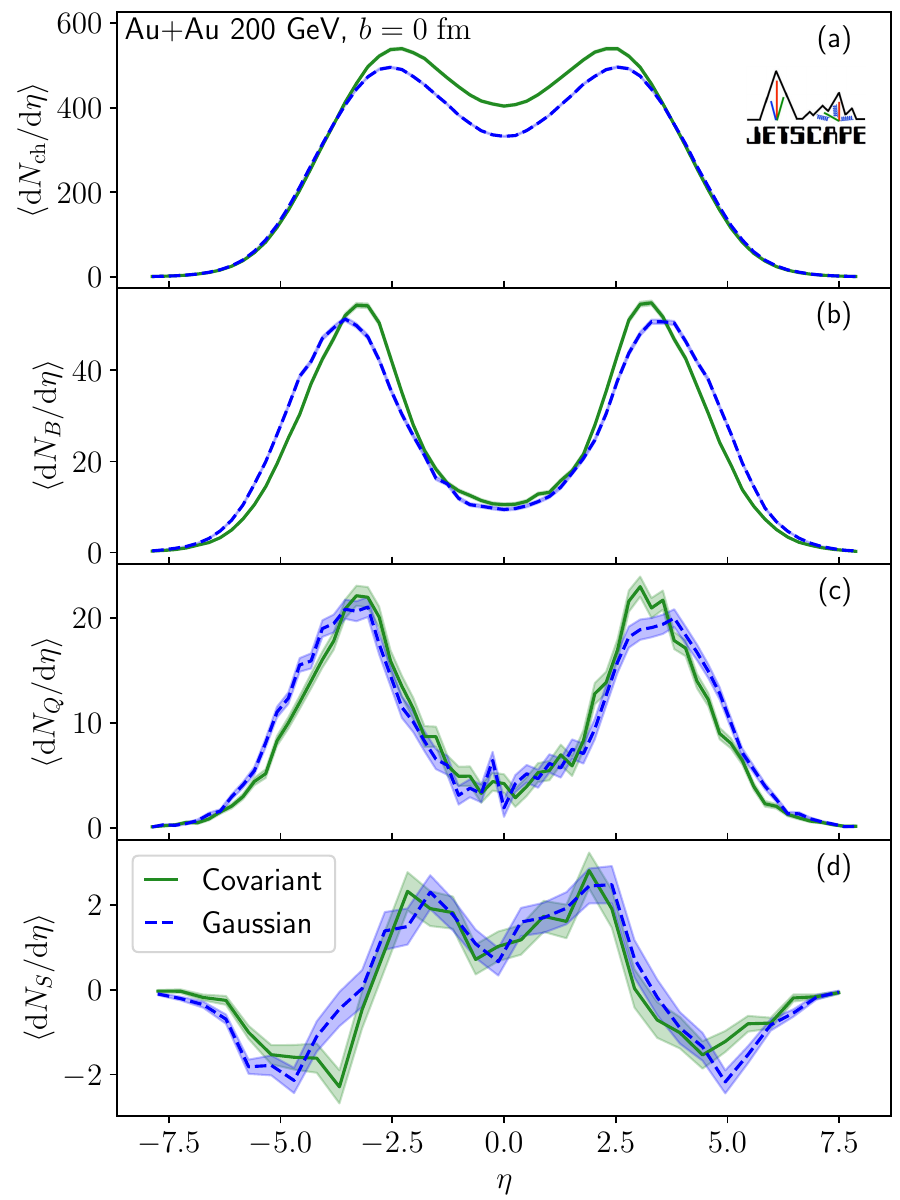}
    \caption{Event-averaged charged hadrons and net-charge distributions for $B, Q, S$ as a function of pseudorapidity in Au+Au collisions with impact parameter $b=0$ at 200 GeV using the covariant (solid) and the Gaussian (dashed) smearing kernels.}
    \label{fig:charge_distributions_b0}
\end{figure}

Figure~\ref{fig:charge_distributions_b0} compares the final-state charged hadrons and conserved charge distributions between covariant and Gaussian smearing kernels. The simulations with the covariant smearing kernel produce more charged hadrons near mid-rapidity because the sharper initial smear profiles lead to more viscous entropy production. Furthermore, the stronger shear viscous effects also decelerate longitudinal expansion, which increases particle production near mid-rapidity. On the other hand, the pseudorapidity distributions for the conserved charges are close between the results with the two smear kernels. Because both smearing kernels take into account the Lorentz contraction along the longitudinal direction, the difference in the pseudo-rapidity distributions is small. 

\begin{figure}[h!]
    \includegraphics[width=\linewidth]{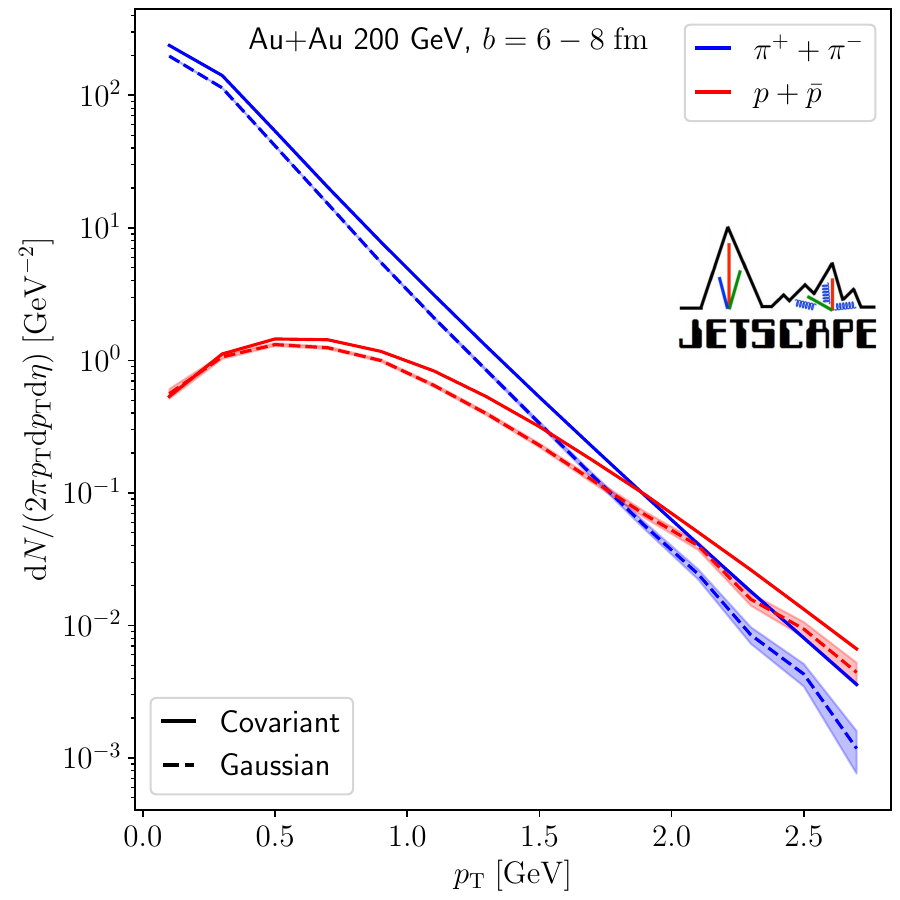}
    \caption{Event-averaged hadronic $p_{\rm T}$-differential spectra near midrapidity ($|\eta| \leq 0.5$) for final-state $\pi^+ + \pi^-$ (blue) and $p + \bar{p}$ (red) in Au+Au collisions with impact parameters $b = 6 - 8$~fm using the covariant (full) and the Gaussian (dashed) smearing kernels.}
    \label{fig:hadronic_pT_spectrum_b6p8}
\end{figure}

Figure~\ref{fig:hadronic_pT_spectrum_b6p8} compares the $p_{\rm T}$-differential spectra for pions and protons in central Au+Au collisions between the covariant and Gaussian smearing kernels. We find that the identified particle $p_{\rm T}$ spectra are flatter with the covariant kernel because the Lorentz contraction on the smearing profile in the transverse plane increases the initial pressure gradients and results in large hydrodynamic radial flow. The large pressure gradients at the early stage of the collisions also lead to a larger viscous entropy production, which increases the overall particle yields, consistent with the results shown in Fig.~\ref{fig:charge_distributions_b0}.

\begin{figure}[h!]
    \includegraphics[width=\linewidth]{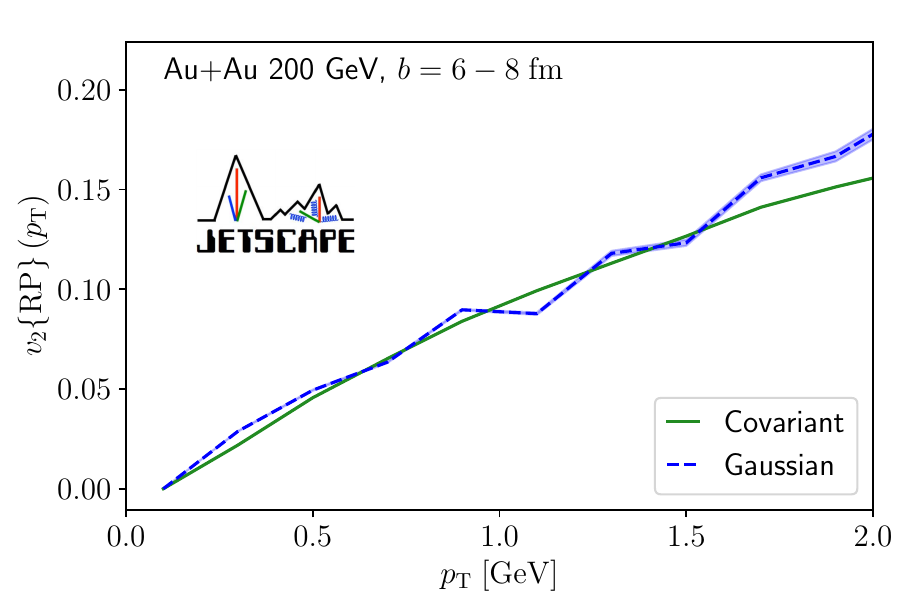}
    \caption{The reaction-plane elliptic flow for charged hadrons as a function of $p_{\rm T}$ for Au+Au collisions with impact parameters $b = 6 - 8$~fm using the covariant (full) and the Gaussian (dashed) smearing kernels.}
    \label{fig:v2_b6p8}
\end{figure}

Figure~\ref{fig:v2_b6p8} shows the reaction plane $p_{\rm T}$-differential elliptic flow in semi-central Au+Au collisions at 200 GeV. The two smearing kernels result in very similar $p_{\rm T}$-differential elliptic flow. Although the large early-stage pressure gradients with the covariant smearing kernel generate more flow anisotropy in the system, they are compensated by the stronger blue shifts from the radial flow in the $p_{\rm T}$-differential elliptic flow observable~\cite{Kestin:2008bh, Shen:2012vn, Shen:2012us}.  

\subsection{Out-of-equilibrium corrections to final-state observables at finite densities}

In this section, we study how the final-state observables depend on the form of out-of-equilibrium corrections at the particlization stage. To isolate the $\delta f$ effects from other event-by-event fluctuations, we perform particlization with different choices of $\delta f$ corrections on the same ensemble of hydrodynamic hypersurfaces. Therefore, the difference in the results stems from varying $\delta f$ corrections at particleization and their subsequent evolution in the SMASH hadronic transport phase.

\begin{figure}[h!]
    \includegraphics[width=\linewidth]{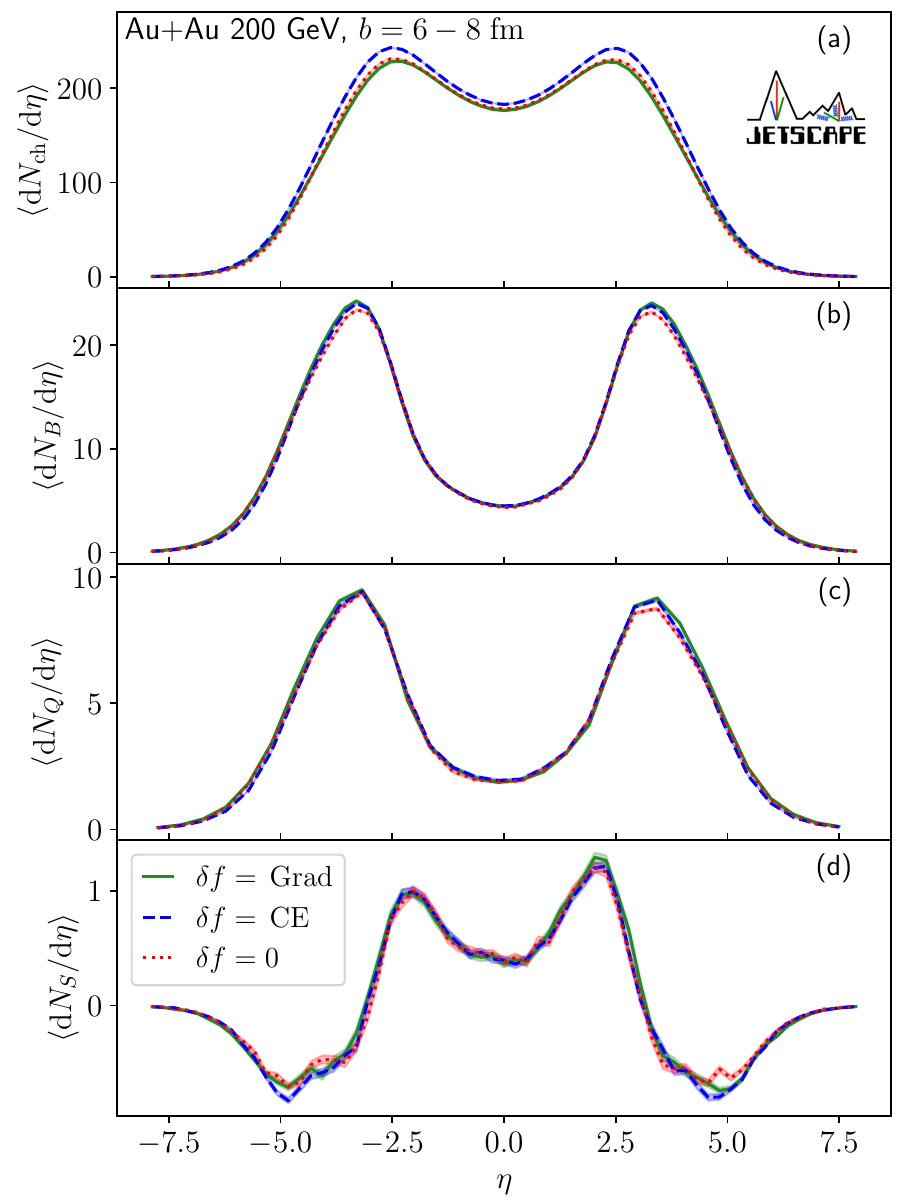}
    \caption{Event-averaged charged hadrons and net-charge distributions for $B, Q, S$ as a function of pseudorapidity for Au+Au collisions at impact parameters $b = 6 - 8$~fm without $\delta f$ corrections (dotted), with Grad's moment method (full), and with Chapman-Enskog (dashed) corrections.}
    \label{fig:charge_distributions_b6p8_df}
\end{figure}

Figure~\ref{fig:charge_distributions_b6p8_df} shows that the Chapman-Enskog $\delta f$ (dashed line) gives a $\approx 5$\% more charged hadron yields than the results using Grad's moment corrections and no corrections. We find negligible corrections from the shear and bulk $\delta f$ to the conserved charges pseudorapidity distributions because those $\delta f$ corrections do not differentiate particles vs. anti-particles.

\begin{figure}[!tb]
    \includegraphics[width=\linewidth]{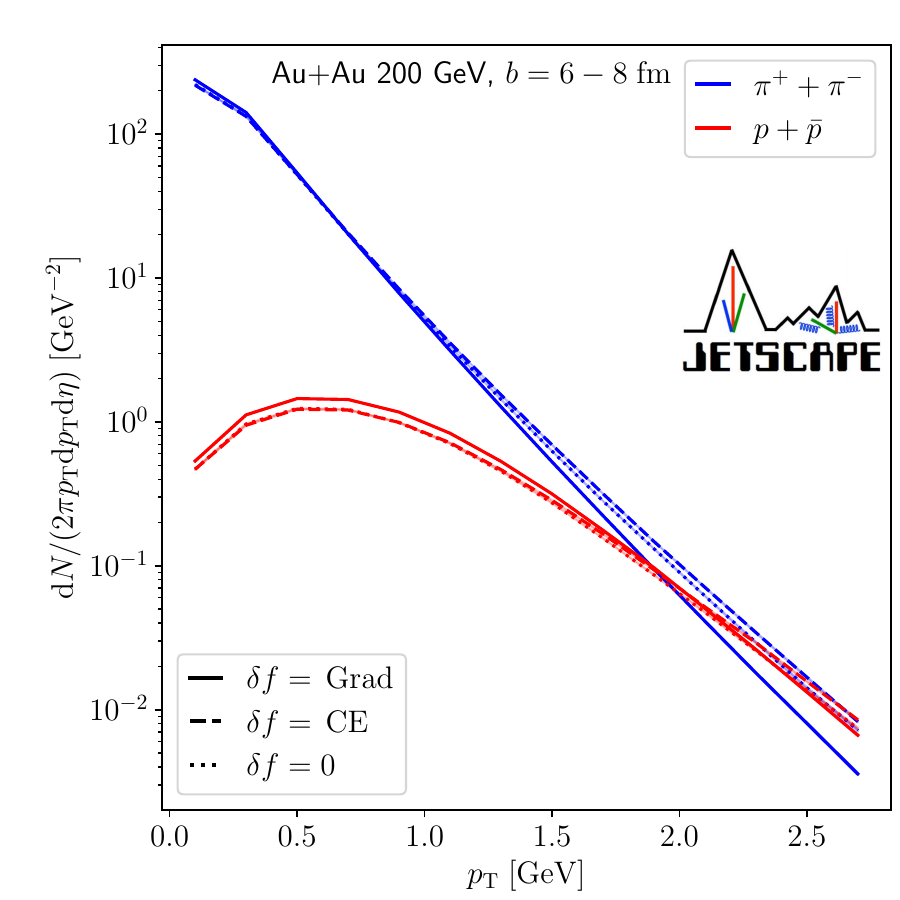}
    \caption{Event-averaged hadronic $p_{\rm T}$-differential spectra near midrapidity ($|\eta| \leq 0.5$) for $\pi^+ + \pi^-$ (blue) and $p + \bar{p}$ (red) in Au+Au collisions with impact parameter $b = 6- 8$~fm without $\delta f$ corrections (dotted), with Grad's moment (full), and with Chapman-Enskog (dashed) corrections.}
    \label{fig:hadronic_pT_spectrum_b6p8_df_comparison}
\end{figure}
\begin{figure}[h!]
    \includegraphics[width=\linewidth]{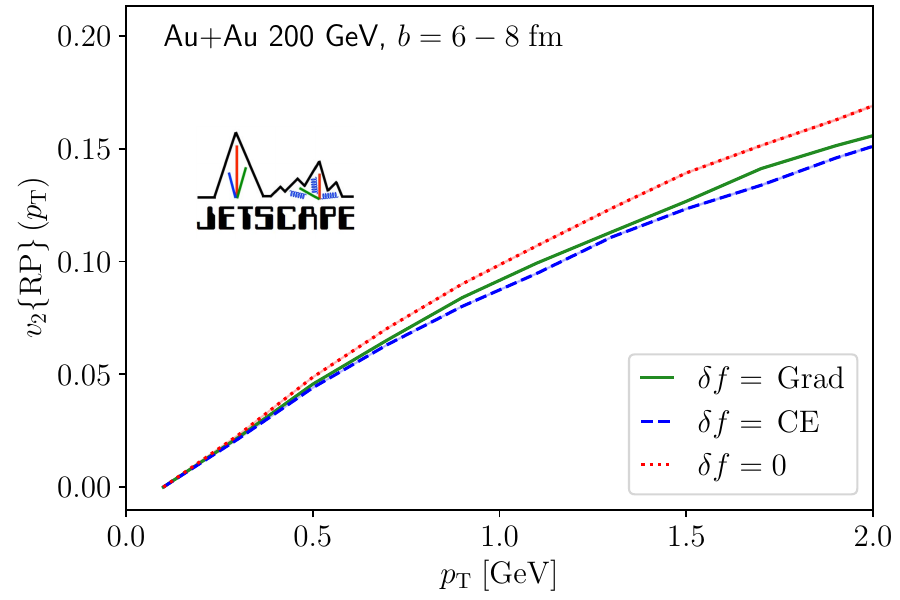}
    \caption{The reaction-plane elliptic flow for charged hadrons as a function of $p_{\rm T}$ for Au+Au collisions with impact parameters $b = 6 - 8$~fm without $\delta f$ corrections (dotted), with Grad's moment method (full), and with Chapman-Enskog (dashed) corrections.}
    \label{fig:v2_b6p8_df}
\end{figure}

Figure~\ref{fig:hadronic_pT_spectrum_b6p8_df_comparison} further compares the different $\delta f$ corrections as a function of the transverse momentum for identified particle spectra near mid-rapidity. The sizes of $\delta f$ corrections can be obtained by taking the difference with the results with $\delta f = 0$. Because the $\delta f$ from Grad's moment method grows quadratically with the transverse momentum $p_{\rm T}$, our results show larger corrections compared to those with the Chapman-Enskog $\delta f$. Both pion and proton spectra get steeper with the Grad's $\delta f$, indicating the $\delta f_\mathrm{bulk}$ overwhelms the shear ones. Interestingly, we find that the Chapman-Enskog $\delta f$ makes the particle spectra flatter, suggesting that the corrections from shear viscosity dominate over those from bulk viscous effects.

Lastly, Figure~\ref{fig:v2_b6p8_df} shows the $\delta f$ effects on the charged hadron $p_\mathrm{T}$-differential elliptic flow. Compared to the results without $\delta f$, we find that both types of $\delta f$ corrections suppress the charged hadron $v_2(p_\mathrm{T})$. For $p_\mathrm{T} < 2$ GeV, the two types of $\delta f$ give compatible results for charged hadron $v_2(p_\mathrm{T})$.

\section{Conclusions}
\label{sec:Conclusions}

In this work, we develop transport-based initial conditions using SMASH for relativistic hydrodynamic simulations. This transport-based initial-state model provides realistic, event-by-event fluctuating (3+1)D distributions for the system's energy-momentum tensor and conserved charge densities, namely the net baryon $(B)$, net electric charge $(Q)$, and net strangeness $(S)$. We find that the hadronic transport model exhibits unique features in the conserved charge density distributions, which distinguish it from other types of initial condition models. The SMASH initial conditions contain significantly more local fluctuations in net electric charges and net strangeness compared to net baryon density, which is a result of the fact that producing balancing charged meson pairs costs significantly less energy than that for baryon-antibaryon pairs. This type of initial condition indicates that the produced fireball in the hydrodynamic phase will explore a wide range of net electric charge and net strangeness densities in the QCD phase diagram. 

To capture all the finite density dependence in the full-fledged bulk dynamics simulations, we couple the SMASH initial conditions with relativistic hydrodynamics using a 4D lattice-QCD-based equation of state, in which the local pressure depends on $(e, n_B, n_Q, n_S)$. We further generalize the out-of-equilibrium corrections with all three types of conserved charges at finite densities in the Cooper-Frye particlization procedure. They are essential to ensure all the conserved macroscopic quantities are smoothly switched from fluid cells to individual hadrons. Using this new hybrid model, we conduct a pilot study of the (3+1)D dynamical evolution of all three conserved charge densities in the hydrodynamic phase and its subsequent impact on final-state observables. We investigate the dynamical evolution of the system's entropy density and the three types of conserved charges during hydrodynamic evolution at several collision energies. We observe that the local charge fluctuations remain substantial for net electric charges and net strangeness across all collision energies. The full covariant smear kernel exhibits large pressure gradients, resulting in strong hydrodynamic radial flow and flattened particle spectra. 

One numerical realization of this hybrid model has been implemented in the X-SCAPE framework version~2.0.0. It lays the foundation for future exploration of core-corona dynamics and concurrent hydrodynamics and hadronic transport descriptions for heavy-ion collisions at RHIC fixed-target, FAIR, and HADES energies.

The framework code and parameter input files used to reproduce the results of this work are available in Refs.~\cite{XSCAPErepo,XSCAPE_xml}.

\acknowledgments
\label{Ack}
This work was supported in part by the National Science Foundation (NSF) within the framework of the JETSCAPE collaboration, under grant numbers OAC-2004571 (CSSI:X-SCAPE) and OAC-2514008 (CSSI:C-SCAPE). It was also supported under  PHY-2111568 and PHY-2413003 (R.J.F., M.Ko., C.P. and A.S.); it was supported in part by the US Department of Energy, Office of Science, Office of Nuclear Physics under grant numbers \rm{DE-AC02-05CH11231} (X.-N.W. and W.Z.), \rm{DE-AC52-07NA27344} (D.A.H. and R.A.S.), \rm{DE-SC0013460} (A.M., I.S., C.Si., R.Da. and R.Do.), \rm{DE-SC0021969} (C.Sh., G.P., and W.Z.), \rm{DE-SC0024232} (C.Sh. and H.R.), \rm{DE-SC0012704} (B.S.), \rm{DE-FG02-92ER40713} (J.H.P.), \rm{DE-FG02-05ER41367} (C.Si and S.A.B.), \rm{DE-SC0024660} (R.K.E), \rm{DE-SC0024347} (J.-F.P. and M.S.). The work was also supported in part by the National Science Foundation of China (NSFC) under grant numbers 11935007, 11861131009 and 11890714 (X.-N.W.), by the Natural Sciences and Engineering Research Council of Canada (C.G., S.J., and G.V.),  by the University of Regina President's Tri-Agency Grant Support Program (G.V.), by the Canada Research Chair program (G.V. and A.K.) reference number CRC-2022-00146, by JSPS KAKENHI Grant Numbers~22K14041 and 25K07303 (Y.T.), JP19K14722 and JP24K07030 (A.Mo.), and by the S\~{a}o Paulo Research Foundation (FAPESP) under projects 2016/24029-6, 2017/05685-2 and 2018/24720-6 (M.L.). C.Sh., J.-F.P., and R.K.E. acknowledge a DOE Office of Science Early Career Award. I.~S. was funded as part of the European Research Council project ERC-2018-ADG-835105 YoctoLHC, and as a part of the Center of Excellence in Quark Matter of the Academy of Finland (project 346325).

Calculations for this work used the Wayne State Grid, generously supported by the Office of the Vice President of Research (OVPR) at Wayne State University.

\bibliography{biblio, non_inspire}

\begin{thebibliography}{107}%
\makeatletter
\providecommand \@ifxundefined [1]{%
 \@ifx{#1\undefined}
}%
\providecommand \@ifnum [1]{%
 \ifnum #1\expandafter \@firstoftwo
 \else \expandafter \@secondoftwo
 \fi
}%
\providecommand \@ifx [1]{%
 \ifx #1\expandafter \@firstoftwo
 \else \expandafter \@secondoftwo
 \fi
}%
\providecommand \natexlab [1]{#1}%
\providecommand \enquote  [1]{``#1''}%
\providecommand \bibnamefont  [1]{#1}%
\providecommand \bibfnamefont [1]{#1}%
\providecommand \citenamefont [1]{#1}%
\providecommand \href@noop [0]{\@secondoftwo}%
\providecommand \href [0]{\begingroup \@sanitize@url \@href}%
\providecommand \@href[1]{\@@startlink{#1}\@@href}%
\providecommand \@@href[1]{\endgroup#1\@@endlink}%
\providecommand \@sanitize@url [0]{\catcode `\\12\catcode `\$12\catcode `\&12\catcode `\#12\catcode `\^12\catcode `\_12\catcode `\%12\relax}%
\providecommand \@@startlink[1]{}%
\providecommand \@@endlink[0]{}%
\providecommand \url  [0]{\begingroup\@sanitize@url \@url }%
\providecommand \@url [1]{\endgroup\@href {#1}{\urlprefix }}%
\providecommand \urlprefix  [0]{URL }%
\providecommand \Eprint [0]{\href }%
\providecommand \doibase [0]{https://doi.org/}%
\providecommand \selectlanguage [0]{\@gobble}%
\providecommand \bibinfo  [0]{\@secondoftwo}%
\providecommand \bibfield  [0]{\@secondoftwo}%
\providecommand \translation [1]{[#1]}%
\providecommand \BibitemOpen [0]{}%
\providecommand \bibitemStop [0]{}%
\providecommand \bibitemNoStop [0]{.\EOS\space}%
\providecommand \EOS [0]{\spacefactor3000\relax}%
\providecommand \BibitemShut  [1]{\csname bibitem#1\endcsname}%
\let\auto@bib@innerbib\@empty
\bibitem [{\citenamefont {Busza}\ \emph {et~al.}(2018)\citenamefont {Busza}, \citenamefont {Rajagopal},\ and\ \citenamefont {van~der Schee}}]{Busza:2018rrf}%
  \BibitemOpen
  \bibfield  {author} {\bibinfo {author} {\bibfnamefont {W.}~\bibnamefont {Busza}}, \bibinfo {author} {\bibfnamefont {K.}~\bibnamefont {Rajagopal}},\ and\ \bibinfo {author} {\bibfnamefont {W.}~\bibnamefont {van~der Schee}},\ }\bibfield  {title} {\bibinfo {title} {{Heavy Ion Collisions: The Big Picture, and the Big Questions}},\ }\href {https://doi.org/10.1146/annurev-nucl-101917-020852} {\bibfield  {journal} {\bibinfo  {journal} {Ann. Rev. Nucl. Part. Sci.}\ }\textbf {\bibinfo {volume} {68}},\ \bibinfo {pages} {339} (\bibinfo {year} {2018})},\ \Eprint {https://arxiv.org/abs/1802.04801} {arXiv:1802.04801 [hep-ph]} \BibitemShut {NoStop}%
\bibitem [{\citenamefont {Shen}\ and\ \citenamefont {Yan}(2020)}]{Shen:2020mgh}%
  \BibitemOpen
  \bibfield  {author} {\bibinfo {author} {\bibfnamefont {C.}~\bibnamefont {Shen}}\ and\ \bibinfo {author} {\bibfnamefont {L.}~\bibnamefont {Yan}},\ }\bibfield  {title} {\bibinfo {title} {{Recent development of hydrodynamic modeling in heavy-ion collisions}},\ }\href {https://doi.org/10.1007/s41365-020-00829-z} {\bibfield  {journal} {\bibinfo  {journal} {Nucl. Sci. Tech.}\ }\textbf {\bibinfo {volume} {31}},\ \bibinfo {pages} {122} (\bibinfo {year} {2020})},\ \Eprint {https://arxiv.org/abs/2010.12377} {arXiv:2010.12377 [nucl-th]} \BibitemShut {NoStop}%
\bibitem [{\citenamefont {Achenbach}\ \emph {et~al.}(2024)\citenamefont {Achenbach} \emph {et~al.}}]{Achenbach:2023pba}%
  \BibitemOpen
  \bibfield  {author} {\bibinfo {author} {\bibfnamefont {P.}~\bibnamefont {Achenbach}} \emph {et~al.},\ }\bibfield  {title} {\bibinfo {title} {{The present and future of QCD}},\ }\href {https://doi.org/10.1016/j.nuclphysa.2024.122874} {\bibfield  {journal} {\bibinfo  {journal} {Nucl. Phys. A}\ }\textbf {\bibinfo {volume} {1047}},\ \bibinfo {pages} {122874} (\bibinfo {year} {2024})},\ \Eprint {https://arxiv.org/abs/2303.02579} {arXiv:2303.02579 [hep-ph]} \BibitemShut {NoStop}%
\bibitem [{\citenamefont {Arslandok}\ \emph {et~al.}(2023)\citenamefont {Arslandok} \emph {et~al.}}]{Arslandok:2023utm}%
  \BibitemOpen
  \bibfield  {author} {\bibinfo {author} {\bibfnamefont {M.}~\bibnamefont {Arslandok}} \emph {et~al.},\ }\bibfield  {title} {\bibinfo {title} {{Hot QCD White Paper}},\ }\href@noop {} {\  (\bibinfo {year} {2023})},\ \Eprint {https://arxiv.org/abs/2303.17254} {arXiv:2303.17254 [nucl-ex]} \BibitemShut {NoStop}%
\bibitem [{\citenamefont {Heinz}\ and\ \citenamefont {Snellings}(2013)}]{Heinz:2013th}%
  \BibitemOpen
  \bibfield  {author} {\bibinfo {author} {\bibfnamefont {U.}~\bibnamefont {Heinz}}\ and\ \bibinfo {author} {\bibfnamefont {R.}~\bibnamefont {Snellings}},\ }\bibfield  {title} {\bibinfo {title} {{Collective flow and viscosity in relativistic heavy-ion collisions}},\ }\href {https://doi.org/10.1146/annurev-nucl-102212-170540} {\bibfield  {journal} {\bibinfo  {journal} {Ann. Rev. Nucl. Part. Sci.}\ }\textbf {\bibinfo {volume} {63}},\ \bibinfo {pages} {123} (\bibinfo {year} {2013})},\ \Eprint {https://arxiv.org/abs/1301.2826} {arXiv:1301.2826 [nucl-th]} \BibitemShut {NoStop}%
\bibitem [{\citenamefont {Gale}\ \emph {et~al.}(2013)\citenamefont {Gale}, \citenamefont {Jeon},\ and\ \citenamefont {Schenke}}]{Gale:2013da}%
  \BibitemOpen
  \bibfield  {author} {\bibinfo {author} {\bibfnamefont {C.}~\bibnamefont {Gale}}, \bibinfo {author} {\bibfnamefont {S.}~\bibnamefont {Jeon}},\ and\ \bibinfo {author} {\bibfnamefont {B.}~\bibnamefont {Schenke}},\ }\bibfield  {title} {\bibinfo {title} {{Hydrodynamic Modeling of Heavy-Ion Collisions}},\ }\href {https://doi.org/10.1142/S0217751X13400113} {\bibfield  {journal} {\bibinfo  {journal} {Int. J. Mod. Phys. A}\ }\textbf {\bibinfo {volume} {28}},\ \bibinfo {pages} {1340011} (\bibinfo {year} {2013})},\ \Eprint {https://arxiv.org/abs/1301.5893} {arXiv:1301.5893 [nucl-th]} \BibitemShut {NoStop}%
\bibitem [{\citenamefont {Petersen}(2014)}]{Petersen:2014yqa}%
  \BibitemOpen
  \bibfield  {author} {\bibinfo {author} {\bibfnamefont {H.}~\bibnamefont {Petersen}},\ }\bibfield  {title} {\bibinfo {title} {{Anisotropic flow in transport + hydrodynamics hybrid approaches}},\ }\href {https://doi.org/10.1088/0954-3899/41/12/124005} {\bibfield  {journal} {\bibinfo  {journal} {J. Phys. G}\ }\textbf {\bibinfo {volume} {41}},\ \bibinfo {pages} {124005} (\bibinfo {year} {2014})},\ \Eprint {https://arxiv.org/abs/1404.1763} {arXiv:1404.1763 [nucl-th]} \BibitemShut {NoStop}%
\bibitem [{\citenamefont {Elfner}\ and\ \citenamefont {M{\"u}ller}(2023)}]{Elfner:2022iae}%
  \BibitemOpen
  \bibfield  {author} {\bibinfo {author} {\bibfnamefont {H.}~\bibnamefont {Elfner}}\ and\ \bibinfo {author} {\bibfnamefont {B.}~\bibnamefont {M{\"u}ller}},\ }\bibfield  {title} {\bibinfo {title} {{The exploration of hot and dense nuclear matter: introduction to relativistic heavy-ion physics}},\ }\href {https://doi.org/10.1088/1361-6471/ace824} {\bibfield  {journal} {\bibinfo  {journal} {J. Phys. G}\ }\textbf {\bibinfo {volume} {50}},\ \bibinfo {pages} {103001} (\bibinfo {year} {2023})},\ \Eprint {https://arxiv.org/abs/2210.12056} {arXiv:2210.12056 [nucl-th]} \BibitemShut {NoStop}%
\bibitem [{\citenamefont {Schenke}\ \emph {et~al.}(2020)\citenamefont {Schenke}, \citenamefont {Shen},\ and\ \citenamefont {Tribedy}}]{Schenke:2020mbo}%
  \BibitemOpen
  \bibfield  {author} {\bibinfo {author} {\bibfnamefont {B.}~\bibnamefont {Schenke}}, \bibinfo {author} {\bibfnamefont {C.}~\bibnamefont {Shen}},\ and\ \bibinfo {author} {\bibfnamefont {P.}~\bibnamefont {Tribedy}},\ }\bibfield  {title} {\bibinfo {title} {{Running the gamut of high energy nuclear collisions}},\ }\href {https://doi.org/10.1103/PhysRevC.102.044905} {\bibfield  {journal} {\bibinfo  {journal} {Phys. Rev. C}\ }\textbf {\bibinfo {volume} {102}},\ \bibinfo {pages} {044905} (\bibinfo {year} {2020})},\ \Eprint {https://arxiv.org/abs/2005.14682} {arXiv:2005.14682 [nucl-th]} \BibitemShut {NoStop}%
\bibitem [{\citenamefont {Putschke}\ \emph {et~al.}(2019)\citenamefont {Putschke} \emph {et~al.}}]{Putschke:2019yrg}%
  \BibitemOpen
  \bibfield  {author} {\bibinfo {author} {\bibfnamefont {J.~H.}\ \bibnamefont {Putschke}} \emph {et~al.},\ }\bibfield  {title} {\bibinfo {title} {{The JETSCAPE framework}},\ }\href@noop {} {\  (\bibinfo {year} {2019})},\ \Eprint {https://arxiv.org/abs/1903.07706} {arXiv:1903.07706 [nucl-th]} \BibitemShut {NoStop}%
\bibitem [{\citenamefont {Soudi}\ \emph {et~al.}(2025)\citenamefont {Soudi} \emph {et~al.}}]{JETSCAPE:2024dgu}%
  \BibitemOpen
  \bibfield  {author} {\bibinfo {author} {\bibfnamefont {I.}~\bibnamefont {Soudi}} \emph {et~al.} (\bibinfo {collaboration} {JETSCAPE}),\ }\bibfield  {title} {\bibinfo {title} {{Soft-hard framework with exact four-momentum conservation for small systems}},\ }\href {https://doi.org/10.1103/r8jt-1xpk} {\bibfield  {journal} {\bibinfo  {journal} {Phys. Rev. C}\ }\textbf {\bibinfo {volume} {112}},\ \bibinfo {pages} {014905} (\bibinfo {year} {2025})},\ \Eprint {https://arxiv.org/abs/2407.17443} {arXiv:2407.17443 [hep-ph]} \BibitemShut {NoStop}%
\bibitem [{\citenamefont {Bzdak}\ \emph {et~al.}(2020)\citenamefont {Bzdak}, \citenamefont {Esumi}, \citenamefont {Koch}, \citenamefont {Liao}, \citenamefont {Stephanov},\ and\ \citenamefont {Xu}}]{Bzdak:2019pkr}%
  \BibitemOpen
  \bibfield  {author} {\bibinfo {author} {\bibfnamefont {A.}~\bibnamefont {Bzdak}}, \bibinfo {author} {\bibfnamefont {S.}~\bibnamefont {Esumi}}, \bibinfo {author} {\bibfnamefont {V.}~\bibnamefont {Koch}}, \bibinfo {author} {\bibfnamefont {J.}~\bibnamefont {Liao}}, \bibinfo {author} {\bibfnamefont {M.}~\bibnamefont {Stephanov}},\ and\ \bibinfo {author} {\bibfnamefont {N.}~\bibnamefont {Xu}},\ }\bibfield  {title} {\bibinfo {title} {{Mapping the Phases of Quantum Chromodynamics with Beam Energy Scan}},\ }\href {https://doi.org/10.1016/j.physrep.2020.01.005} {\bibfield  {journal} {\bibinfo  {journal} {Phys. Rept.}\ }\textbf {\bibinfo {volume} {853}},\ \bibinfo {pages} {1} (\bibinfo {year} {2020})},\ \Eprint {https://arxiv.org/abs/1906.00936} {arXiv:1906.00936 [nucl-th]} \BibitemShut {NoStop}%
\bibitem [{\citenamefont {Monnai}\ \emph {et~al.}(2021)\citenamefont {Monnai}, \citenamefont {Schenke},\ and\ \citenamefont {Shen}}]{Monnai:2021kgu}%
  \BibitemOpen
  \bibfield  {author} {\bibinfo {author} {\bibfnamefont {A.}~\bibnamefont {Monnai}}, \bibinfo {author} {\bibfnamefont {B.}~\bibnamefont {Schenke}},\ and\ \bibinfo {author} {\bibfnamefont {C.}~\bibnamefont {Shen}},\ }\bibfield  {title} {\bibinfo {title} {{QCD Equation of State at Finite Chemical Potentials for Relativistic Nuclear Collisions}},\ }\href {https://doi.org/10.1142/S0217751X21300076} {\bibfield  {journal} {\bibinfo  {journal} {Int. J. Mod. Phys. A}\ }\textbf {\bibinfo {volume} {36}},\ \bibinfo {pages} {2130007} (\bibinfo {year} {2021})},\ \Eprint {https://arxiv.org/abs/2101.11591} {arXiv:2101.11591 [nucl-th]} \BibitemShut {NoStop}%
\bibitem [{\citenamefont {An}\ \emph {et~al.}(2022)\citenamefont {An} \emph {et~al.}}]{An:2021wof}%
  \BibitemOpen
  \bibfield  {author} {\bibinfo {author} {\bibfnamefont {X.}~\bibnamefont {An}} \emph {et~al.},\ }\bibfield  {title} {\bibinfo {title} {{The BEST framework for the search for the QCD critical point and the chiral magnetic effect}},\ }\href {https://doi.org/10.1016/j.nuclphysa.2021.122343} {\bibfield  {journal} {\bibinfo  {journal} {Nucl. Phys. A}\ }\textbf {\bibinfo {volume} {1017}},\ \bibinfo {pages} {122343} (\bibinfo {year} {2022})},\ \Eprint {https://arxiv.org/abs/2108.13867} {arXiv:2108.13867 [nucl-th]} \BibitemShut {NoStop}%
\bibitem [{\citenamefont {Almaalol}\ \emph {et~al.}(2022)\citenamefont {Almaalol} \emph {et~al.}}]{Almaalol:2022xwv}%
  \BibitemOpen
  \bibfield  {author} {\bibinfo {author} {\bibfnamefont {D.}~\bibnamefont {Almaalol}} \emph {et~al.},\ }\bibfield  {title} {\bibinfo {title} {{QCD Phase Structure and Interactions at High Baryon Density: Continuation of BES Physics Program with CBM at FAIR}},\ }\href@noop {} {\  (\bibinfo {year} {2022})},\ \Eprint {https://arxiv.org/abs/2209.05009} {arXiv:2209.05009 [nucl-ex]} \BibitemShut {NoStop}%
\bibitem [{\citenamefont {Sorensen}\ \emph {et~al.}(2024)\citenamefont {Sorensen} \emph {et~al.}}]{Sorensen:2023zkk}%
  \BibitemOpen
  \bibfield  {author} {\bibinfo {author} {\bibfnamefont {A.}~\bibnamefont {Sorensen}} \emph {et~al.},\ }\bibfield  {title} {\bibinfo {title} {{Dense nuclear matter equation of state from heavy-ion collisions}},\ }\href {https://doi.org/10.1016/j.ppnp.2023.104080} {\bibfield  {journal} {\bibinfo  {journal} {Prog. Part. Nucl. Phys.}\ }\textbf {\bibinfo {volume} {134}},\ \bibinfo {pages} {104080} (\bibinfo {year} {2024})},\ \Eprint {https://arxiv.org/abs/2301.13253} {arXiv:2301.13253 [nucl-th]} \BibitemShut {NoStop}%
\bibitem [{\citenamefont {Du}\ \emph {et~al.}(2024)\citenamefont {Du}, \citenamefont {Sorensen},\ and\ \citenamefont {Stephanov}}]{Du:2024wjm}%
  \BibitemOpen
  \bibfield  {author} {\bibinfo {author} {\bibfnamefont {L.}~\bibnamefont {Du}}, \bibinfo {author} {\bibfnamefont {A.}~\bibnamefont {Sorensen}},\ and\ \bibinfo {author} {\bibfnamefont {M.}~\bibnamefont {Stephanov}},\ }\bibfield  {title} {\bibinfo {title} {{The QCD phase diagram and Beam Energy Scan physics: A theory overview}},\ }\href {https://doi.org/10.1142/9789811294679_0007} {\bibfield  {journal} {\bibinfo  {journal} {Int. J. Mod. Phys. E}\ }\textbf {\bibinfo {volume} {33}},\ \bibinfo {pages} {2430008} (\bibinfo {year} {2024})},\ \Eprint {https://arxiv.org/abs/2402.10183} {arXiv:2402.10183 [nucl-th]} \BibitemShut {NoStop}%
\bibitem [{\citenamefont {Schlichting}\ and\ \citenamefont {Teaney}(2019)}]{Schlichting:2019abc}%
  \BibitemOpen
  \bibfield  {author} {\bibinfo {author} {\bibfnamefont {S.}~\bibnamefont {Schlichting}}\ and\ \bibinfo {author} {\bibfnamefont {D.}~\bibnamefont {Teaney}},\ }\bibfield  {title} {\bibinfo {title} {{The First fm/c of Heavy-Ion Collisions}},\ }\href {https://doi.org/10.1146/annurev-nucl-101918-023825} {\bibfield  {journal} {\bibinfo  {journal} {Ann. Rev. Nucl. Part. Sci.}\ }\textbf {\bibinfo {volume} {69}},\ \bibinfo {pages} {447} (\bibinfo {year} {2019})},\ \Eprint {https://arxiv.org/abs/1908.02113} {arXiv:1908.02113 [nucl-th]} \BibitemShut {NoStop}%
\bibitem [{\citenamefont {Shen}(2021)}]{Shen:2020gef}%
  \BibitemOpen
  \bibfield  {author} {\bibinfo {author} {\bibfnamefont {C.}~\bibnamefont {Shen}},\ }\bibfield  {title} {\bibinfo {title} {{Studying QGP with flow: A theory overview}},\ }\href {https://doi.org/10.1016/j.nuclphysa.2020.121788} {\bibfield  {journal} {\bibinfo  {journal} {Nucl. Phys. A}\ }\textbf {\bibinfo {volume} {1005}},\ \bibinfo {pages} {121788} (\bibinfo {year} {2021})},\ \Eprint {https://arxiv.org/abs/2001.11858} {arXiv:2001.11858 [nucl-th]} \BibitemShut {NoStop}%
\bibitem [{\citenamefont {Shen}(2022)}]{Shen:2021nbe}%
  \BibitemOpen
  \bibfield  {author} {\bibinfo {author} {\bibfnamefont {C.}~\bibnamefont {Shen}},\ }\bibfield  {title} {\bibinfo {title} {{Dynamic modeling for heavy-ion collisions}},\ }\href {https://doi.org/10.1051/epjconf/202225902001} {\bibfield  {journal} {\bibinfo  {journal} {EPJ Web Conf.}\ }\textbf {\bibinfo {volume} {259}},\ \bibinfo {pages} {02001} (\bibinfo {year} {2022})},\ \Eprint {https://arxiv.org/abs/2108.04987} {arXiv:2108.04987 [nucl-th]} \BibitemShut {NoStop}%
\bibitem [{\citenamefont {Mohanty}(2011)}]{Mohanty:2011nm}%
  \BibitemOpen
  \bibfield  {author} {\bibinfo {author} {\bibfnamefont {B.}~\bibnamefont {Mohanty}} (\bibinfo {collaboration} {STAR}),\ }\bibfield  {title} {\bibinfo {title} {{STAR experiment results from the beam energy scan program at RHIC}},\ }\href {https://doi.org/10.1088/0954-3899/38/12/124023} {\bibfield  {journal} {\bibinfo  {journal} {J. Phys. G}\ }\textbf {\bibinfo {volume} {38}},\ \bibinfo {pages} {124023} (\bibinfo {year} {2011})},\ \Eprint {https://arxiv.org/abs/1106.5902} {arXiv:1106.5902 [nucl-ex]} \BibitemShut {NoStop}%
\bibitem [{\citenamefont {Adamczyk}\ \emph {et~al.}(2017)\citenamefont {Adamczyk} \emph {et~al.}}]{STAR:2017sal}%
  \BibitemOpen
  \bibfield  {author} {\bibinfo {author} {\bibfnamefont {L.}~\bibnamefont {Adamczyk}} \emph {et~al.} (\bibinfo {collaboration} {STAR}),\ }\bibfield  {title} {\bibinfo {title} {{Bulk Properties of the Medium Produced in Relativistic Heavy-Ion Collisions from the Beam Energy Scan Program}},\ }\href {https://doi.org/10.1103/PhysRevC.96.044904} {\bibfield  {journal} {\bibinfo  {journal} {Phys. Rev. C}\ }\textbf {\bibinfo {volume} {96}},\ \bibinfo {pages} {044904} (\bibinfo {year} {2017})},\ \Eprint {https://arxiv.org/abs/1701.07065} {arXiv:1701.07065 [nucl-ex]} \BibitemShut {NoStop}%
\bibitem [{\citenamefont {Ablyazimov}\ \emph {et~al.}(2017)\citenamefont {Ablyazimov} \emph {et~al.}}]{CBM:2016kpk}%
  \BibitemOpen
  \bibfield  {author} {\bibinfo {author} {\bibfnamefont {T.}~\bibnamefont {Ablyazimov}} \emph {et~al.} (\bibinfo {collaboration} {CBM}),\ }\bibfield  {title} {\bibinfo {title} {{Challenges in QCD matter physics --The scientific programme of the Compressed Baryonic Matter experiment at FAIR}},\ }\href {https://doi.org/10.1140/epja/i2017-12248-y} {\bibfield  {journal} {\bibinfo  {journal} {Eur. Phys. J. A}\ }\textbf {\bibinfo {volume} {53}},\ \bibinfo {pages} {60} (\bibinfo {year} {2017})},\ \Eprint {https://arxiv.org/abs/1607.01487} {arXiv:1607.01487 [nucl-ex]} \BibitemShut {NoStop}%
\bibitem [{\citenamefont {Karpenko}\ \emph {et~al.}(2015)\citenamefont {Karpenko}, \citenamefont {Huovinen}, \citenamefont {Petersen},\ and\ \citenamefont {Bleicher}}]{Karpenko:2015xea}%
  \BibitemOpen
  \bibfield  {author} {\bibinfo {author} {\bibfnamefont {I.~A.}\ \bibnamefont {Karpenko}}, \bibinfo {author} {\bibfnamefont {P.}~\bibnamefont {Huovinen}}, \bibinfo {author} {\bibfnamefont {H.}~\bibnamefont {Petersen}},\ and\ \bibinfo {author} {\bibfnamefont {M.}~\bibnamefont {Bleicher}},\ }\bibfield  {title} {\bibinfo {title} {{Estimation of the shear viscosity at finite net-baryon density from $A+A$ collision data at $\sqrt{s_\mathrm{NN}} = 7.7-200$ GeV}},\ }\href {https://doi.org/10.1103/PhysRevC.91.064901} {\bibfield  {journal} {\bibinfo  {journal} {Phys. Rev. C}\ }\textbf {\bibinfo {volume} {91}},\ \bibinfo {pages} {064901} (\bibinfo {year} {2015})},\ \Eprint {https://arxiv.org/abs/1502.01978} {arXiv:1502.01978 [nucl-th]} \BibitemShut {NoStop}%
\bibitem [{\citenamefont {Shen}\ and\ \citenamefont {Schenke}(2018{\natexlab{a}})}]{Shen:2017bsr}%
  \BibitemOpen
  \bibfield  {author} {\bibinfo {author} {\bibfnamefont {C.}~\bibnamefont {Shen}}\ and\ \bibinfo {author} {\bibfnamefont {B.}~\bibnamefont {Schenke}},\ }\bibfield  {title} {\bibinfo {title} {{Dynamical initial state model for relativistic heavy-ion collisions}},\ }\href {https://doi.org/10.1103/PhysRevC.97.024907} {\bibfield  {journal} {\bibinfo  {journal} {Phys. Rev. C}\ }\textbf {\bibinfo {volume} {97}},\ \bibinfo {pages} {024907} (\bibinfo {year} {2018}{\natexlab{a}})},\ \Eprint {https://arxiv.org/abs/1710.00881} {arXiv:1710.00881 [nucl-th]} \BibitemShut {NoStop}%
\bibitem [{\citenamefont {Shen}\ and\ \citenamefont {Schenke}(2018{\natexlab{b}})}]{Shen:2017fnn}%
  \BibitemOpen
  \bibfield  {author} {\bibinfo {author} {\bibfnamefont {C.}~\bibnamefont {Shen}}\ and\ \bibinfo {author} {\bibfnamefont {B.}~\bibnamefont {Schenke}},\ }\bibfield  {title} {\bibinfo {title} {{Initial state and hydrodynamic modeling of heavy-ion collisions at RHIC BES energies}},\ }\href {https://doi.org/10.22323/1.311.0006} {\bibfield  {journal} {\bibinfo  {journal} {PoS}\ }\textbf {\bibinfo {volume} {CPOD2017}},\ \bibinfo {pages} {006} (\bibinfo {year} {2018}{\natexlab{b}})},\ \Eprint {https://arxiv.org/abs/1711.10544} {arXiv:1711.10544 [nucl-th]} \BibitemShut {NoStop}%
\bibitem [{\citenamefont {M{\"a}ntysaari}\ \emph {et~al.}(2025{\natexlab{a}})\citenamefont {M{\"a}ntysaari}, \citenamefont {Schenke}, \citenamefont {Shen},\ and\ \citenamefont {Zhao}}]{Mantysaari:2025tcg}%
  \BibitemOpen
  \bibfield  {author} {\bibinfo {author} {\bibfnamefont {H.}~\bibnamefont {M{\"a}ntysaari}}, \bibinfo {author} {\bibfnamefont {B.}~\bibnamefont {Schenke}}, \bibinfo {author} {\bibfnamefont {C.}~\bibnamefont {Shen}},\ and\ \bibinfo {author} {\bibfnamefont {W.}~\bibnamefont {Zhao}},\ }\bibfield  {title} {\bibinfo {title} {{Collision-Energy Dependence in Heavy-Ion Collisions from Nonlinear QCD Evolution}},\ }\href {https://doi.org/10.1103/gf4y-p5j7} {\bibfield  {journal} {\bibinfo  {journal} {Phys. Rev. Lett.}\ }\textbf {\bibinfo {volume} {135}},\ \bibinfo {pages} {022302} (\bibinfo {year} {2025}{\natexlab{a}})},\ \Eprint {https://arxiv.org/abs/2502.05138} {arXiv:2502.05138 [nucl-th]} \BibitemShut {NoStop}%
\bibitem [{\citenamefont {M{\"a}ntysaari}\ \emph {et~al.}(2025{\natexlab{b}})\citenamefont {M{\"a}ntysaari}, \citenamefont {Schenke}, \citenamefont {Shen},\ and\ \citenamefont {Zhao}}]{Mantysaari:2025cos}%
  \BibitemOpen
  \bibfield  {author} {\bibinfo {author} {\bibfnamefont {H.}~\bibnamefont {M{\"a}ntysaari}}, \bibinfo {author} {\bibfnamefont {B.}~\bibnamefont {Schenke}}, \bibinfo {author} {\bibfnamefont {C.}~\bibnamefont {Shen}},\ and\ \bibinfo {author} {\bibfnamefont {W.}~\bibnamefont {Zhao}},\ }\bibfield  {title} {\bibinfo {title} {{Perturbative high-energy evolution in the IP-Glasma initial state}},\ }in\ \href@noop {} {\emph {\bibinfo {booktitle} {{31st International Conference on Ultra-relativistic Nucleus-Nucleus Collisions}}}}\ (\bibinfo {year} {2025})\ \Eprint {https://arxiv.org/abs/2508.20432} {arXiv:2508.20432 [nucl-th]} \BibitemShut {NoStop}%
\bibitem [{\citenamefont {Shen}\ and\ \citenamefont {Alzhrani}(2020)}]{Shen:2020jwv}%
  \BibitemOpen
  \bibfield  {author} {\bibinfo {author} {\bibfnamefont {C.}~\bibnamefont {Shen}}\ and\ \bibinfo {author} {\bibfnamefont {S.}~\bibnamefont {Alzhrani}},\ }\bibfield  {title} {\bibinfo {title} {{Collision-geometry-based 3D initial condition for relativistic heavy-ion collisions}},\ }\href {https://doi.org/10.1103/PhysRevC.102.014909} {\bibfield  {journal} {\bibinfo  {journal} {Phys. Rev. C}\ }\textbf {\bibinfo {volume} {102}},\ \bibinfo {pages} {014909} (\bibinfo {year} {2020})},\ \Eprint {https://arxiv.org/abs/2003.05852} {arXiv:2003.05852 [nucl-th]} \BibitemShut {NoStop}%
\bibitem [{\citenamefont {Ryu}\ \emph {et~al.}(2021)\citenamefont {Ryu}, \citenamefont {Jupic},\ and\ \citenamefont {Shen}}]{Ryu:2021lnx}%
  \BibitemOpen
  \bibfield  {author} {\bibinfo {author} {\bibfnamefont {S.}~\bibnamefont {Ryu}}, \bibinfo {author} {\bibfnamefont {V.}~\bibnamefont {Jupic}},\ and\ \bibinfo {author} {\bibfnamefont {C.}~\bibnamefont {Shen}},\ }\bibfield  {title} {\bibinfo {title} {{Probing early-time longitudinal dynamics with the {\ensuremath{\Lambda}} hyperon's spin polarization in relativistic heavy-ion collisions}},\ }\href {https://doi.org/10.1103/PhysRevC.104.054908} {\bibfield  {journal} {\bibinfo  {journal} {Phys. Rev. C}\ }\textbf {\bibinfo {volume} {104}},\ \bibinfo {pages} {054908} (\bibinfo {year} {2021})},\ \Eprint {https://arxiv.org/abs/2106.08125} {arXiv:2106.08125 [nucl-th]} \BibitemShut {NoStop}%
\bibitem [{\citenamefont {Du}\ \emph {et~al.}(2023)\citenamefont {Du}, \citenamefont {Shen}, \citenamefont {Jeon},\ and\ \citenamefont {Gale}}]{Du:2022yok}%
  \BibitemOpen
  \bibfield  {author} {\bibinfo {author} {\bibfnamefont {L.}~\bibnamefont {Du}}, \bibinfo {author} {\bibfnamefont {C.}~\bibnamefont {Shen}}, \bibinfo {author} {\bibfnamefont {S.}~\bibnamefont {Jeon}},\ and\ \bibinfo {author} {\bibfnamefont {C.}~\bibnamefont {Gale}},\ }\bibfield  {title} {\bibinfo {title} {{Probing initial baryon stopping and equation~of state with rapidity-dependent directed flow of identified particles}},\ }\href {https://doi.org/10.1103/PhysRevC.108.L041901} {\bibfield  {journal} {\bibinfo  {journal} {Phys. Rev. C}\ }\textbf {\bibinfo {volume} {108}},\ \bibinfo {pages} {L041901} (\bibinfo {year} {2023})},\ \Eprint {https://arxiv.org/abs/2211.16408} {arXiv:2211.16408 [nucl-th]} \BibitemShut {NoStop}%
\bibitem [{\citenamefont {Bass}\ \emph {et~al.}(1998)\citenamefont {Bass} \emph {et~al.}}]{Bass:1998ca}%
  \BibitemOpen
  \bibfield  {author} {\bibinfo {author} {\bibfnamefont {S.~A.}\ \bibnamefont {Bass}} \emph {et~al.},\ }\bibfield  {title} {\bibinfo {title} {{Microscopic models for ultrarelativistic heavy ion collisions}},\ }\href {https://doi.org/10.1016/S0146-6410(98)00058-1} {\bibfield  {journal} {\bibinfo  {journal} {Prog. Part. Nucl. Phys.}\ }\textbf {\bibinfo {volume} {41}},\ \bibinfo {pages} {255} (\bibinfo {year} {1998})},\ \Eprint {https://arxiv.org/abs/nucl-th/9803035} {arXiv:nucl-th/9803035} \BibitemShut {NoStop}%
\bibitem [{\citenamefont {Bleicher}\ \emph {et~al.}(1999)\citenamefont {Bleicher} \emph {et~al.}}]{Bleicher:1999xi}%
  \BibitemOpen
  \bibfield  {author} {\bibinfo {author} {\bibfnamefont {M.}~\bibnamefont {Bleicher}} \emph {et~al.},\ }\bibfield  {title} {\bibinfo {title} {{Relativistic hadron hadron collisions in the ultrarelativistic quantum molecular dynamics model}},\ }\href {https://doi.org/10.1088/0954-3899/25/9/308} {\bibfield  {journal} {\bibinfo  {journal} {J. Phys. G}\ }\textbf {\bibinfo {volume} {25}},\ \bibinfo {pages} {1859} (\bibinfo {year} {1999})},\ \Eprint {https://arxiv.org/abs/hep-ph/9909407} {arXiv:hep-ph/9909407} \BibitemShut {NoStop}%
\bibitem [{\citenamefont {Nara}\ \emph {et~al.}(2000)\citenamefont {Nara}, \citenamefont {Otuka}, \citenamefont {Ohnishi}, \citenamefont {Niita},\ and\ \citenamefont {Chiba}}]{Nara:1999dz}%
  \BibitemOpen
  \bibfield  {author} {\bibinfo {author} {\bibfnamefont {Y.}~\bibnamefont {Nara}}, \bibinfo {author} {\bibfnamefont {N.}~\bibnamefont {Otuka}}, \bibinfo {author} {\bibfnamefont {A.}~\bibnamefont {Ohnishi}}, \bibinfo {author} {\bibfnamefont {K.}~\bibnamefont {Niita}},\ and\ \bibinfo {author} {\bibfnamefont {S.}~\bibnamefont {Chiba}},\ }\bibfield  {title} {\bibinfo {title} {{Study of relativistic nuclear collisions at AGS energies from p + Be to Au + Au with hadronic cascade model}},\ }\href {https://doi.org/10.1103/PhysRevC.61.024901} {\bibfield  {journal} {\bibinfo  {journal} {Phys. Rev. C}\ }\textbf {\bibinfo {volume} {61}},\ \bibinfo {pages} {024901} (\bibinfo {year} {2000})},\ \Eprint {https://arxiv.org/abs/nucl-th/9904059} {arXiv:nucl-th/9904059} \BibitemShut {NoStop}%
\bibitem [{\citenamefont {Weil}\ \emph {et~al.}(2016)\citenamefont {Weil} \emph {et~al.}}]{SMASH:2016zqf}%
  \BibitemOpen
  \bibfield  {author} {\bibinfo {author} {\bibfnamefont {J.}~\bibnamefont {Weil}} \emph {et~al.} (\bibinfo {collaboration} {SMASH}),\ }\bibfield  {title} {\bibinfo {title} {{Particle production and equilibrium properties within a new hadron transport approach for heavy-ion collisions}},\ }\href {https://doi.org/10.1103/PhysRevC.94.054905} {\bibfield  {journal} {\bibinfo  {journal} {Phys. Rev. C}\ }\textbf {\bibinfo {volume} {94}},\ \bibinfo {pages} {054905} (\bibinfo {year} {2016})},\ \Eprint {https://arxiv.org/abs/1606.06642} {arXiv:1606.06642 [nucl-th]} \BibitemShut {NoStop}%
\bibitem [{\citenamefont {Lin}\ \emph {et~al.}(2005)\citenamefont {Lin}, \citenamefont {Ko}, \citenamefont {Li}, \citenamefont {Zhang},\ and\ \citenamefont {Pal}}]{Lin:2004en}%
  \BibitemOpen
  \bibfield  {author} {\bibinfo {author} {\bibfnamefont {Z.-W.}\ \bibnamefont {Lin}}, \bibinfo {author} {\bibfnamefont {C.~M.}\ \bibnamefont {Ko}}, \bibinfo {author} {\bibfnamefont {B.-A.}\ \bibnamefont {Li}}, \bibinfo {author} {\bibfnamefont {B.}~\bibnamefont {Zhang}},\ and\ \bibinfo {author} {\bibfnamefont {S.}~\bibnamefont {Pal}},\ }\bibfield  {title} {\bibinfo {title} {{A Multi-phase transport model for relativistic heavy ion collisions}},\ }\href {https://doi.org/10.1103/PhysRevC.72.064901} {\bibfield  {journal} {\bibinfo  {journal} {Phys. Rev. C}\ }\textbf {\bibinfo {volume} {72}},\ \bibinfo {pages} {064901} (\bibinfo {year} {2005})},\ \Eprint {https://arxiv.org/abs/nucl-th/0411110} {arXiv:nucl-th/0411110} \BibitemShut {NoStop}%
\bibitem [{\citenamefont {Petersen}\ \emph {et~al.}(2008)\citenamefont {Petersen}, \citenamefont {Steinheimer}, \citenamefont {Burau}, \citenamefont {Bleicher},\ and\ \citenamefont {St{\"o}cker}}]{Petersen:2008dd}%
  \BibitemOpen
  \bibfield  {author} {\bibinfo {author} {\bibfnamefont {H.}~\bibnamefont {Petersen}}, \bibinfo {author} {\bibfnamefont {J.}~\bibnamefont {Steinheimer}}, \bibinfo {author} {\bibfnamefont {G.}~\bibnamefont {Burau}}, \bibinfo {author} {\bibfnamefont {M.}~\bibnamefont {Bleicher}},\ and\ \bibinfo {author} {\bibfnamefont {H.}~\bibnamefont {St{\"o}cker}},\ }\bibfield  {title} {\bibinfo {title} {{A Fully Integrated Transport Approach to Heavy Ion Reactions with an Intermediate Hydrodynamic Stage}},\ }\href {https://doi.org/10.1103/PhysRevC.78.044901} {\bibfield  {journal} {\bibinfo  {journal} {Phys. Rev. C}\ }\textbf {\bibinfo {volume} {78}},\ \bibinfo {pages} {044901} (\bibinfo {year} {2008})},\ \Eprint {https://arxiv.org/abs/0806.1695} {arXiv:0806.1695 [nucl-th]} \BibitemShut {NoStop}%
\bibitem [{\citenamefont {Pang}\ \emph {et~al.}(2012)\citenamefont {Pang}, \citenamefont {Wang},\ and\ \citenamefont {Wang}}]{Pang:2012he}%
  \BibitemOpen
  \bibfield  {author} {\bibinfo {author} {\bibfnamefont {L.}~\bibnamefont {Pang}}, \bibinfo {author} {\bibfnamefont {Q.}~\bibnamefont {Wang}},\ and\ \bibinfo {author} {\bibfnamefont {X.-N.}\ \bibnamefont {Wang}},\ }\bibfield  {title} {\bibinfo {title} {{Effects of initial flow velocity fluctuation in event-by-event (3+1)D hydrodynamics}},\ }\href {https://doi.org/10.1103/PhysRevC.86.024911} {\bibfield  {journal} {\bibinfo  {journal} {Phys. Rev. C}\ }\textbf {\bibinfo {volume} {86}},\ \bibinfo {pages} {024911} (\bibinfo {year} {2012})},\ \Eprint {https://arxiv.org/abs/1205.5019} {arXiv:1205.5019 [nucl-th]} \BibitemShut {NoStop}%
\bibitem [{\citenamefont {Pang}\ \emph {et~al.}(2016)\citenamefont {Pang}, \citenamefont {Petersen}, \citenamefont {Qin}, \citenamefont {Roy},\ and\ \citenamefont {Wang}}]{Pang:2015zrq}%
  \BibitemOpen
  \bibfield  {author} {\bibinfo {author} {\bibfnamefont {L.-G.}\ \bibnamefont {Pang}}, \bibinfo {author} {\bibfnamefont {H.}~\bibnamefont {Petersen}}, \bibinfo {author} {\bibfnamefont {G.-Y.}\ \bibnamefont {Qin}}, \bibinfo {author} {\bibfnamefont {V.}~\bibnamefont {Roy}},\ and\ \bibinfo {author} {\bibfnamefont {X.-N.}\ \bibnamefont {Wang}},\ }\bibfield  {title} {\bibinfo {title} {{Decorrelation of anisotropic flow along the longitudinal direction}},\ }\href {https://doi.org/10.1140/epja/i2016-16097-x} {\bibfield  {journal} {\bibinfo  {journal} {Eur. Phys. J. A}\ }\textbf {\bibinfo {volume} {52}},\ \bibinfo {pages} {97} (\bibinfo {year} {2016})},\ \Eprint {https://arxiv.org/abs/1511.04131} {arXiv:1511.04131 [nucl-th]} \BibitemShut {NoStop}%
\bibitem [{\citenamefont {Du}\ \emph {et~al.}(2019)\citenamefont {Du}, \citenamefont {Heinz},\ and\ \citenamefont {Vujanovic}}]{Du:2018mpf}%
  \BibitemOpen
  \bibfield  {author} {\bibinfo {author} {\bibfnamefont {L.}~\bibnamefont {Du}}, \bibinfo {author} {\bibfnamefont {U.}~\bibnamefont {Heinz}},\ and\ \bibinfo {author} {\bibfnamefont {G.}~\bibnamefont {Vujanovic}},\ }\bibfield  {title} {\bibinfo {title} {{Hybrid model with dynamical sources for heavy-ion collisions at BES energies}},\ }\href {https://doi.org/10.1016/j.nuclphysa.2018.09.015} {\bibfield  {journal} {\bibinfo  {journal} {Nucl. Phys. A}\ }\textbf {\bibinfo {volume} {982}},\ \bibinfo {pages} {407} (\bibinfo {year} {2019})},\ \Eprint {https://arxiv.org/abs/1807.04721} {arXiv:1807.04721 [nucl-th]} \BibitemShut {NoStop}%
\bibitem [{\citenamefont {Akamatsu}\ \emph {et~al.}(2018)\citenamefont {Akamatsu}, \citenamefont {Asakawa}, \citenamefont {Hirano}, \citenamefont {Kitazawa}, \citenamefont {Morita}, \citenamefont {Murase}, \citenamefont {Nara}, \citenamefont {Nonaka},\ and\ \citenamefont {Ohnishi}}]{Akamatsu:2018olk}%
  \BibitemOpen
  \bibfield  {author} {\bibinfo {author} {\bibfnamefont {Y.}~\bibnamefont {Akamatsu}}, \bibinfo {author} {\bibfnamefont {M.}~\bibnamefont {Asakawa}}, \bibinfo {author} {\bibfnamefont {T.}~\bibnamefont {Hirano}}, \bibinfo {author} {\bibfnamefont {M.}~\bibnamefont {Kitazawa}}, \bibinfo {author} {\bibfnamefont {K.}~\bibnamefont {Morita}}, \bibinfo {author} {\bibfnamefont {K.}~\bibnamefont {Murase}}, \bibinfo {author} {\bibfnamefont {Y.}~\bibnamefont {Nara}}, \bibinfo {author} {\bibfnamefont {C.}~\bibnamefont {Nonaka}},\ and\ \bibinfo {author} {\bibfnamefont {A.}~\bibnamefont {Ohnishi}},\ }\bibfield  {title} {\bibinfo {title} {{Dynamically integrated transport approach for heavy-ion collisions at high baryon density}},\ }\href {https://doi.org/10.1103/PhysRevC.98.024909} {\bibfield  {journal} {\bibinfo  {journal} {Phys. Rev. C}\ }\textbf {\bibinfo {volume} {98}},\ \bibinfo {pages} {024909} (\bibinfo {year} {2018})},\ \Eprint {https://arxiv.org/abs/1805.09024} {arXiv:1805.09024 [nucl-th]} \BibitemShut {NoStop}%
\bibitem [{\citenamefont {Sch{\"a}fer}\ \emph {et~al.}(2022)\citenamefont {Sch{\"a}fer}, \citenamefont {Karpenko}, \citenamefont {Wu}, \citenamefont {Hammelmann},\ and\ \citenamefont {Elfner}}]{Schafer:2021csj}%
  \BibitemOpen
  \bibfield  {author} {\bibinfo {author} {\bibfnamefont {A.}~\bibnamefont {Sch{\"a}fer}}, \bibinfo {author} {\bibfnamefont {I.}~\bibnamefont {Karpenko}}, \bibinfo {author} {\bibfnamefont {X.-Y.}\ \bibnamefont {Wu}}, \bibinfo {author} {\bibfnamefont {J.}~\bibnamefont {Hammelmann}},\ and\ \bibinfo {author} {\bibfnamefont {H.}~\bibnamefont {Elfner}} (\bibinfo {collaboration} {SMASH}),\ }\bibfield  {title} {\bibinfo {title} {{Particle production in a hybrid approach for a beam energy scan of Au+Au/Pb+Pb collisions between $\sqrt{s_\textrm{NN}}$ = 4.3 GeV and $\sqrt{s_\textrm{NN}}$ = 200.0 GeV}},\ }\href {https://doi.org/10.1140/epja/s10050-022-00872-x} {\bibfield  {journal} {\bibinfo  {journal} {Eur. Phys. J. A}\ }\textbf {\bibinfo {volume} {58}},\ \bibinfo {pages} {230} (\bibinfo {year} {2022})},\ \Eprint {https://arxiv.org/abs/2112.08724} {arXiv:2112.08724 [hep-ph]} \BibitemShut {NoStop}%
\bibitem [{\citenamefont {Cassing}\ and\ \citenamefont {Bratkovskaya}(2009)}]{Cassing:2009vt}%
  \BibitemOpen
  \bibfield  {author} {\bibinfo {author} {\bibfnamefont {W.}~\bibnamefont {Cassing}}\ and\ \bibinfo {author} {\bibfnamefont {E.~L.}\ \bibnamefont {Bratkovskaya}},\ }\bibfield  {title} {\bibinfo {title} {{Parton-Hadron-String Dynamics: an off-shell transport approach for relativistic energies}},\ }\href {https://doi.org/10.1016/j.nuclphysa.2009.09.007} {\bibfield  {journal} {\bibinfo  {journal} {Nucl. Phys. A}\ }\textbf {\bibinfo {volume} {831}},\ \bibinfo {pages} {215} (\bibinfo {year} {2009})},\ \Eprint {https://arxiv.org/abs/0907.5331} {arXiv:0907.5331 [nucl-th]} \BibitemShut {NoStop}%
\bibitem [{\citenamefont {Kanakubo}\ \emph {et~al.}(2020)\citenamefont {Kanakubo}, \citenamefont {Tachibana},\ and\ \citenamefont {Hirano}}]{Kanakubo:2019ogh}%
  \BibitemOpen
  \bibfield  {author} {\bibinfo {author} {\bibfnamefont {Y.}~\bibnamefont {Kanakubo}}, \bibinfo {author} {\bibfnamefont {Y.}~\bibnamefont {Tachibana}},\ and\ \bibinfo {author} {\bibfnamefont {T.}~\bibnamefont {Hirano}},\ }\bibfield  {title} {\bibinfo {title} {{Unified description of hadron yield ratios from dynamical core-corona initialization}},\ }\href {https://doi.org/10.1103/PhysRevC.101.024912} {\bibfield  {journal} {\bibinfo  {journal} {Phys. Rev. C}\ }\textbf {\bibinfo {volume} {101}},\ \bibinfo {pages} {024912} (\bibinfo {year} {2020})},\ \Eprint {https://arxiv.org/abs/1910.10556} {arXiv:1910.10556 [nucl-th]} \BibitemShut {NoStop}%
\bibitem [{\citenamefont {Galatyuk}\ \emph {et~al.}(2016)\citenamefont {Galatyuk}, \citenamefont {Hohler}, \citenamefont {Rapp}, \citenamefont {Seck},\ and\ \citenamefont {Stroth}}]{Galatyuk:2015pkq}%
  \BibitemOpen
  \bibfield  {author} {\bibinfo {author} {\bibfnamefont {T.}~\bibnamefont {Galatyuk}}, \bibinfo {author} {\bibfnamefont {P.~M.}\ \bibnamefont {Hohler}}, \bibinfo {author} {\bibfnamefont {R.}~\bibnamefont {Rapp}}, \bibinfo {author} {\bibfnamefont {F.}~\bibnamefont {Seck}},\ and\ \bibinfo {author} {\bibfnamefont {J.}~\bibnamefont {Stroth}},\ }\bibfield  {title} {\bibinfo {title} {{Thermal Dileptons from Coarse-Grained Transport as Fireball Probes at SIS Energies}},\ }\href {https://doi.org/10.1140/epja/i2016-16131-1} {\bibfield  {journal} {\bibinfo  {journal} {Eur. Phys. J. A}\ }\textbf {\bibinfo {volume} {52}},\ \bibinfo {pages} {131} (\bibinfo {year} {2016})},\ \Eprint {https://arxiv.org/abs/1512.08688} {arXiv:1512.08688 [nucl-th]} \BibitemShut {NoStop}%
\bibitem [{\citenamefont {Inghirami}\ and\ \citenamefont {Elfner}(2022)}]{Inghirami:2022afu}%
  \BibitemOpen
  \bibfield  {author} {\bibinfo {author} {\bibfnamefont {G.}~\bibnamefont {Inghirami}}\ and\ \bibinfo {author} {\bibfnamefont {H.}~\bibnamefont {Elfner}},\ }\bibfield  {title} {\bibinfo {title} {{The applicability of hydrodynamics in heavy ion collisions at $\sqrt{s_\mathrm{NN}}$~=~2.4{\textendash}7.7~GeV}},\ }\href {https://doi.org/10.1140/epjc/s10052-022-10718-x} {\bibfield  {journal} {\bibinfo  {journal} {Eur. Phys. J. C}\ }\textbf {\bibinfo {volume} {82}},\ \bibinfo {pages} {796} (\bibinfo {year} {2022})},\ \Eprint {https://arxiv.org/abs/2201.05934} {arXiv:2201.05934 [hep-ph]} \BibitemShut {NoStop}%
\bibitem [{\citenamefont {Greif}\ \emph {et~al.}(2018)\citenamefont {Greif}, \citenamefont {Fotakis}, \citenamefont {Denicol},\ and\ \citenamefont {Greiner}}]{Greif:2017byw}%
  \BibitemOpen
  \bibfield  {author} {\bibinfo {author} {\bibfnamefont {M.}~\bibnamefont {Greif}}, \bibinfo {author} {\bibfnamefont {J.~A.}\ \bibnamefont {Fotakis}}, \bibinfo {author} {\bibfnamefont {G.~S.}\ \bibnamefont {Denicol}},\ and\ \bibinfo {author} {\bibfnamefont {C.}~\bibnamefont {Greiner}},\ }\bibfield  {title} {\bibinfo {title} {{Diffusion of conserved charges in relativistic heavy ion collisions}},\ }\href {https://doi.org/10.1103/PhysRevLett.120.242301} {\bibfield  {journal} {\bibinfo  {journal} {Phys. Rev. Lett.}\ }\textbf {\bibinfo {volume} {120}},\ \bibinfo {pages} {242301} (\bibinfo {year} {2018})},\ \Eprint {https://arxiv.org/abs/1711.08680} {arXiv:1711.08680 [hep-ph]} \BibitemShut {NoStop}%
\bibitem [{\citenamefont {Fotakis}\ \emph {et~al.}(2020)\citenamefont {Fotakis}, \citenamefont {Greif}, \citenamefont {Greiner}, \citenamefont {Denicol},\ and\ \citenamefont {Niemi}}]{Fotakis:2019nbq}%
  \BibitemOpen
  \bibfield  {author} {\bibinfo {author} {\bibfnamefont {J.~A.}\ \bibnamefont {Fotakis}}, \bibinfo {author} {\bibfnamefont {M.}~\bibnamefont {Greif}}, \bibinfo {author} {\bibfnamefont {C.}~\bibnamefont {Greiner}}, \bibinfo {author} {\bibfnamefont {G.~S.}\ \bibnamefont {Denicol}},\ and\ \bibinfo {author} {\bibfnamefont {H.}~\bibnamefont {Niemi}},\ }\bibfield  {title} {\bibinfo {title} {{Diffusion processes involving multiple conserved charges: A study from kinetic theory and implications to the fluid-dynamical modeling of heavy ion collisions}},\ }\href {https://doi.org/10.1103/PhysRevD.101.076007} {\bibfield  {journal} {\bibinfo  {journal} {Phys. Rev. D}\ }\textbf {\bibinfo {volume} {101}},\ \bibinfo {pages} {076007} (\bibinfo {year} {2020})},\ \Eprint {https://arxiv.org/abs/1912.09103} {arXiv:1912.09103 [hep-ph]} \BibitemShut {NoStop}%
\bibitem [{\citenamefont {Danhoni}\ and\ \citenamefont {Moore}(2023)}]{Danhoni:2022xmt}%
  \BibitemOpen
  \bibfield  {author} {\bibinfo {author} {\bibfnamefont {I.}~\bibnamefont {Danhoni}}\ and\ \bibinfo {author} {\bibfnamefont {G.~D.}\ \bibnamefont {Moore}},\ }\bibfield  {title} {\bibinfo {title} {{Hot and dense QCD shear viscosity at leading log}},\ }\href {https://doi.org/10.1007/JHEP02(2023)124} {\bibfield  {journal} {\bibinfo  {journal} {JHEP}\ }\textbf {\bibinfo {volume} {02}},\ \bibinfo {pages} {124}},\ \Eprint {https://arxiv.org/abs/2212.02325} {arXiv:2212.02325 [hep-ph]} \BibitemShut {NoStop}%
\bibitem [{\citenamefont {Danhoni}\ \emph {et~al.}(2025)\citenamefont {Danhoni}, \citenamefont {Martin},\ and\ \citenamefont {Noronha-Hostler}}]{Danhoni:2024kgi}%
  \BibitemOpen
  \bibfield  {author} {\bibinfo {author} {\bibfnamefont {I.}~\bibnamefont {Danhoni}}, \bibinfo {author} {\bibfnamefont {J.~S.~S.}\ \bibnamefont {Martin}},\ and\ \bibinfo {author} {\bibfnamefont {J.}~\bibnamefont {Noronha-Hostler}},\ }\bibfield  {title} {\bibinfo {title} {{Shear viscosity from perturbative quantum chromodynamics to the hadron resonance gas at finite baryon, strangeness, and electric charge densities}},\ }\href {https://doi.org/10.1103/gckr-c2yg} {\bibfield  {journal} {\bibinfo  {journal} {Phys. Rev. D}\ }\textbf {\bibinfo {volume} {111}},\ \bibinfo {pages} {116006} (\bibinfo {year} {2025})},\ \Eprint {https://arxiv.org/abs/2406.04968} {arXiv:2406.04968 [hep-ph]} \BibitemShut {NoStop}%
\bibitem [{\citenamefont {Plumberg}\ \emph {et~al.}(2025)\citenamefont {Plumberg} \emph {et~al.}}]{Plumberg:2024leb}%
  \BibitemOpen
  \bibfield  {author} {\bibinfo {author} {\bibfnamefont {C.}~\bibnamefont {Plumberg}} \emph {et~al.},\ }\bibfield  {title} {\bibinfo {title} {{Conservation of B, S, and Q charges in relativistic viscous hydrodynamics solved with smoothed particle hydrodynamics}},\ }\href {https://doi.org/10.1103/PhysRevC.111.044905} {\bibfield  {journal} {\bibinfo  {journal} {Phys. Rev. C}\ }\textbf {\bibinfo {volume} {111}},\ \bibinfo {pages} {044905} (\bibinfo {year} {2025})},\ \Eprint {https://arxiv.org/abs/2405.09648} {arXiv:2405.09648 [nucl-th]} \BibitemShut {NoStop}%
\bibitem [{\citenamefont {Monnai}\ \emph {et~al.}(2024)\citenamefont {Monnai}, \citenamefont {Pihan}, \citenamefont {Schenke},\ and\ \citenamefont {Shen}}]{Monnai:2024pvy}%
  \BibitemOpen
  \bibfield  {author} {\bibinfo {author} {\bibfnamefont {A.}~\bibnamefont {Monnai}}, \bibinfo {author} {\bibfnamefont {G.}~\bibnamefont {Pihan}}, \bibinfo {author} {\bibfnamefont {B.}~\bibnamefont {Schenke}},\ and\ \bibinfo {author} {\bibfnamefont {C.}~\bibnamefont {Shen}},\ }\bibfield  {title} {\bibinfo {title} {{Four-dimensional QCD equation~of state with multiple chemical potentials}},\ }\href {https://doi.org/10.1103/PhysRevC.110.044905} {\bibfield  {journal} {\bibinfo  {journal} {Phys. Rev. C}\ }\textbf {\bibinfo {volume} {110}},\ \bibinfo {pages} {044905} (\bibinfo {year} {2024})},\ \Eprint {https://arxiv.org/abs/2406.11610} {arXiv:2406.11610 [nucl-th]} \BibitemShut {NoStop}%
\bibitem [{\citenamefont {Abuali}\ \emph {et~al.}(2025)\citenamefont {Abuali}, \citenamefont {Bors{\'a}nyi}, \citenamefont {Fodor}, \citenamefont {Jahan}, \citenamefont {Kahangirwe}, \citenamefont {Parotto}, \citenamefont {P{\'a}sztor}, \citenamefont {Ratti}, \citenamefont {Shah},\ and\ \citenamefont {Trabulsi}}]{Abuali:2025tbd}%
  \BibitemOpen
  \bibfield  {author} {\bibinfo {author} {\bibfnamefont {A.}~\bibnamefont {Abuali}}, \bibinfo {author} {\bibfnamefont {S.}~\bibnamefont {Bors{\'a}nyi}}, \bibinfo {author} {\bibfnamefont {Z.}~\bibnamefont {Fodor}}, \bibinfo {author} {\bibfnamefont {J.}~\bibnamefont {Jahan}}, \bibinfo {author} {\bibfnamefont {M.}~\bibnamefont {Kahangirwe}}, \bibinfo {author} {\bibfnamefont {P.}~\bibnamefont {Parotto}}, \bibinfo {author} {\bibfnamefont {A.}~\bibnamefont {P{\'a}sztor}}, \bibinfo {author} {\bibfnamefont {C.}~\bibnamefont {Ratti}}, \bibinfo {author} {\bibfnamefont {H.}~\bibnamefont {Shah}},\ and\ \bibinfo {author} {\bibfnamefont {S.~A.}\ \bibnamefont {Trabulsi}},\ }\bibfield  {title} {\bibinfo {title} {{New 4D lattice QCD equation of state: Extended density coverage from a generalized T' expansion}},\ }\href {https://doi.org/10.1103/2dmh-26yh} {\bibfield  {journal} {\bibinfo  {journal} {Phys. Rev. D}\ }\textbf {\bibinfo {volume} {112}},\ \bibinfo {pages} {054502} (\bibinfo {year} {2025})},\ \Eprint
  {https://arxiv.org/abs/2504.01881} {arXiv:2504.01881 [hep-lat]} \BibitemShut {NoStop}%
\bibitem [{\citenamefont {Shen}\ \emph {et~al.}(2024)\citenamefont {Shen}, \citenamefont {Noble}, \citenamefont {Paquet}, \citenamefont {Schenke},\ and\ \citenamefont {Gale}}]{Shen:2023aeg}%
  \BibitemOpen
  \bibfield  {author} {\bibinfo {author} {\bibfnamefont {C.}~\bibnamefont {Shen}}, \bibinfo {author} {\bibfnamefont {A.}~\bibnamefont {Noble}}, \bibinfo {author} {\bibfnamefont {J.-F.}\ \bibnamefont {Paquet}}, \bibinfo {author} {\bibfnamefont {B.}~\bibnamefont {Schenke}},\ and\ \bibinfo {author} {\bibfnamefont {C.}~\bibnamefont {Gale}},\ }\bibfield  {title} {\bibinfo {title} {{Illuminating early-stage dynamics of heavy-ion collisions through photons at RHIC BES energies}},\ }\href {https://doi.org/10.22323/1.438.0042} {\bibfield  {journal} {\bibinfo  {journal} {PoS}\ }\textbf {\bibinfo {volume} {HardProbes2023}},\ \bibinfo {pages} {042} (\bibinfo {year} {2024})},\ \Eprint {https://arxiv.org/abs/2307.08498} {arXiv:2307.08498 [nucl-th]} \BibitemShut {NoStop}%
\bibitem [{\citenamefont {Wu}\ \emph {et~al.}(2025)\citenamefont {Wu}, \citenamefont {Gale}, \citenamefont {Jeon}, \citenamefont {Paquet}, \citenamefont {Schenke},\ and\ \citenamefont {Shen}}]{Wu:2025psu}%
  \BibitemOpen
  \bibfield  {author} {\bibinfo {author} {\bibfnamefont {X.-Y.}\ \bibnamefont {Wu}}, \bibinfo {author} {\bibfnamefont {C.}~\bibnamefont {Gale}}, \bibinfo {author} {\bibfnamefont {S.}~\bibnamefont {Jeon}}, \bibinfo {author} {\bibfnamefont {J.-F.}\ \bibnamefont {Paquet}}, \bibinfo {author} {\bibfnamefont {B.}~\bibnamefont {Schenke}},\ and\ \bibinfo {author} {\bibfnamefont {C.}~\bibnamefont {Shen}},\ }\bibfield  {title} {\bibinfo {title} {{Electromagnetic radiation from Quark-Gluon Plasma at finite baryon density}},\ }in\ \href@noop {} {\emph {\bibinfo {booktitle} {{31st International Conference on Ultra-relativistic Nucleus-Nucleus Collisions}}}}\ (\bibinfo {year} {2025})\ \Eprint {https://arxiv.org/abs/2509.03289} {arXiv:2509.03289 [nucl-th]} \BibitemShut {NoStop}%
\bibitem [{\citenamefont {Monnai}\ \emph {et~al.}(2025)\citenamefont {Monnai}, \citenamefont {Pihan}, \citenamefont {Schenke},\ and\ \citenamefont {Shen}}]{Monnai:2025nyg}%
  \BibitemOpen
  \bibfield  {author} {\bibinfo {author} {\bibfnamefont {A.}~\bibnamefont {Monnai}}, \bibinfo {author} {\bibfnamefont {G.}~\bibnamefont {Pihan}}, \bibinfo {author} {\bibfnamefont {B.}~\bibnamefont {Schenke}},\ and\ \bibinfo {author} {\bibfnamefont {C.}~\bibnamefont {Shen}},\ }\bibfield  {title} {\bibinfo {title} {{Four-dimensional QCD equation of state at finite chemical potentials}},\ }in\ \href@noop {} {\emph {\bibinfo {booktitle} {{16th Conference on Quark Confinement and the Hadron Spectrum}}}}\ (\bibinfo {year} {2025})\ \Eprint {https://arxiv.org/abs/2503.03566} {arXiv:2503.03566 [nucl-th]} \BibitemShut {NoStop}%
\bibitem [{\citenamefont {Cooper}\ and\ \citenamefont {Frye}(1974)}]{Cooper:1974mv}%
  \BibitemOpen
  \bibfield  {author} {\bibinfo {author} {\bibfnamefont {F.}~\bibnamefont {Cooper}}\ and\ \bibinfo {author} {\bibfnamefont {G.}~\bibnamefont {Frye}},\ }\bibfield  {title} {\bibinfo {title} {{Comment on the Single Particle Distribution in the Hydrodynamic and Statistical Thermodynamic Models of Multiparticle Production}},\ }\href {https://doi.org/10.1103/PhysRevD.10.186} {\bibfield  {journal} {\bibinfo  {journal} {Phys. Rev. D}\ }\textbf {\bibinfo {volume} {10}},\ \bibinfo {pages} {186} (\bibinfo {year} {1974})}\BibitemShut {NoStop}%
\bibitem [{\citenamefont {Elfner}\ and\ \citenamefont {G{\'o}es-Hirayama}(2025)}]{Elfner:2025ojd}%
  \BibitemOpen
  \bibfield  {author} {\bibinfo {author} {\bibfnamefont {H.}~\bibnamefont {Elfner}}\ and\ \bibinfo {author} {\bibfnamefont {R.}~\bibnamefont {G{\'o}es-Hirayama}},\ }\bibfield  {title} {\bibinfo {title} {{SMASH: Results from hadronic transport for heavy-ion collisions at high densities}},\ }\href@noop {} {\  (\bibinfo {year} {2025})},\ \Eprint {https://arxiv.org/abs/2508.21477} {arXiv:2508.21477 [nucl-th]} \BibitemShut {NoStop}%
\bibitem [{\citenamefont {Rathod}\ \emph {et~al.}(2025)\citenamefont {Rathod}, \citenamefont {Sinyukov}, \citenamefont {Adzhymambetov},\ and\ \citenamefont {Zbroszczyk}}]{Rathod:2025gvj}%
  \BibitemOpen
  \bibfield  {author} {\bibinfo {author} {\bibfnamefont {N.}~\bibnamefont {Rathod}}, \bibinfo {author} {\bibfnamefont {Y.}~\bibnamefont {Sinyukov}}, \bibinfo {author} {\bibfnamefont {M.}~\bibnamefont {Adzhymambetov}},\ and\ \bibinfo {author} {\bibfnamefont {H.}~\bibnamefont {Zbroszczyk}},\ }\bibfield  {title} {\bibinfo {title} {{Particle Spectra in the Integrated HydroKinetic Model at RHIC BES Energies}},\ }\href@noop {} {\  (\bibinfo {year} {2025})},\ \Eprint {https://arxiv.org/abs/2506.13944} {arXiv:2506.13944 [nucl-th]} \BibitemShut {NoStop}%
\bibitem [{\citenamefont {Weil}\ \emph {et~al.}(2025)\citenamefont {Weil}, \citenamefont {Tindall}, \citenamefont {Steinberg}, \citenamefont {Staudenmaier}, \citenamefont {Sorensen}, \citenamefont {Sciarra}, \citenamefont {Schäfer}, \citenamefont {Sattler}, \citenamefont {Sass}, \citenamefont {Ryu}, \citenamefont {Rothermel}, \citenamefont {Rosenkvist}, \citenamefont {Rose}, \citenamefont {Roch}, \citenamefont {Prinz}, \citenamefont {Paulinyova}, \citenamefont {Pang}, \citenamefont {Oliinychenko}, \citenamefont {Mohs}, \citenamefont {Mitrovic}, \citenamefont {Mayer}, \citenamefont {Li}, \citenamefont {Kübler}, \citenamefont {Kretz}, \citenamefont {Kehrenberg}, \citenamefont {Inghirami}, \citenamefont {Hammelmann}, \citenamefont {Götz}, \citenamefont {Góes-Hirayama}, \citenamefont {Groebel}, \citenamefont {Goldschmidt}, \citenamefont {Geiger}, \citenamefont {Garcia-Montero}, \citenamefont {Elfner}, \citenamefont {Ehlert}, \citenamefont {Constantin}, \citenamefont {Christensen}, \citenamefont {Bäuchle},
  \citenamefont {Auvinen},\ and\ \citenamefont {Attems}}]{weil_2025_15837933}%
  \BibitemOpen
  \bibfield  {author} {\bibinfo {author} {\bibfnamefont {J.}~\bibnamefont {Weil}}, \bibinfo {author} {\bibfnamefont {J.}~\bibnamefont {Tindall}}, \bibinfo {author} {\bibfnamefont {V.}~\bibnamefont {Steinberg}}, \bibinfo {author} {\bibfnamefont {J.}~\bibnamefont {Staudenmaier}}, \bibinfo {author} {\bibfnamefont {A.}~\bibnamefont {Sorensen}}, \bibinfo {author} {\bibfnamefont {A.}~\bibnamefont {Sciarra}}, \bibinfo {author} {\bibfnamefont {A.}~\bibnamefont {Schäfer}}, \bibinfo {author} {\bibfnamefont {R.}~\bibnamefont {Sattler}}, \bibinfo {author} {\bibfnamefont {N.}~\bibnamefont {Sass}}, \bibinfo {author} {\bibfnamefont {S.}~\bibnamefont {Ryu}}, \bibinfo {author} {\bibfnamefont {J.}~\bibnamefont {Rothermel}}, \bibinfo {author} {\bibfnamefont {C.~B.}\ \bibnamefont {Rosenkvist}}, \bibinfo {author} {\bibfnamefont {J.-B.}\ \bibnamefont {Rose}}, \bibinfo {author} {\bibfnamefont {H.}~\bibnamefont {Roch}}, \bibinfo {author} {\bibfnamefont {L.}~\bibnamefont {Prinz}}, \bibinfo {author} {\bibfnamefont {Z.}~\bibnamefont
  {Paulinyova}}, \bibinfo {author} {\bibfnamefont {L.-G.}\ \bibnamefont {Pang}}, \bibinfo {author} {\bibfnamefont {D.}~\bibnamefont {Oliinychenko}}, \bibinfo {author} {\bibfnamefont {J.}~\bibnamefont {Mohs}}, \bibinfo {author} {\bibfnamefont {D.}~\bibnamefont {Mitrovic}}, \bibinfo {author} {\bibfnamefont {M.}~\bibnamefont {Mayer}}, \bibinfo {author} {\bibfnamefont {F.}~\bibnamefont {Li}}, \bibinfo {author} {\bibfnamefont {N.}~\bibnamefont {Kübler}}, \bibinfo {author} {\bibfnamefont {M.}~\bibnamefont {Kretz}}, \bibinfo {author} {\bibfnamefont {T.}~\bibnamefont {Kehrenberg}}, \bibinfo {author} {\bibfnamefont {G.}~\bibnamefont {Inghirami}}, \bibinfo {author} {\bibfnamefont {J.}~\bibnamefont {Hammelmann}}, \bibinfo {author} {\bibfnamefont {N.}~\bibnamefont {Götz}}, \bibinfo {author} {\bibfnamefont {R.}~\bibnamefont {Góes-Hirayama}}, \bibinfo {author} {\bibfnamefont {J.}~\bibnamefont {Groebel}}, \bibinfo {author} {\bibfnamefont {A.}~\bibnamefont {Goldschmidt}}, \bibinfo {author} {\bibfnamefont {L.}~\bibnamefont
  {Geiger}}, \bibinfo {author} {\bibfnamefont {O.}~\bibnamefont {Garcia-Montero}}, \bibinfo {author} {\bibfnamefont {H.}~\bibnamefont {Elfner}}, \bibinfo {author} {\bibfnamefont {N.}~\bibnamefont {Ehlert}}, \bibinfo {author} {\bibfnamefont {L.}~\bibnamefont {Constantin}}, \bibinfo {author} {\bibfnamefont {C.~H.}\ \bibnamefont {Christensen}}, \bibinfo {author} {\bibfnamefont {B.}~\bibnamefont {Bäuchle}}, \bibinfo {author} {\bibfnamefont {J.}~\bibnamefont {Auvinen}},\ and\ \bibinfo {author} {\bibfnamefont {M.}~\bibnamefont {Attems}},\ }\href {https://doi.org/10.5281/zenodo.15837933} {\bibinfo {title} {smash-transport/smash: Smash-3.2.1}} (\bibinfo {year} {2025})\BibitemShut {NoStop}%
\bibitem [{\citenamefont {Tanabashi}\ \emph {et~al.}(2018)\citenamefont {Tanabashi} \emph {et~al.}}]{ParticleDataGroup:2018ovx}%
  \BibitemOpen
  \bibfield  {author} {\bibinfo {author} {\bibfnamefont {M.}~\bibnamefont {Tanabashi}} \emph {et~al.} (\bibinfo {collaboration} {Particle Data Group}),\ }\bibfield  {title} {\bibinfo {title} {{Review of Particle Physics}},\ }\href {https://doi.org/10.1103/PhysRevD.98.030001} {\bibfield  {journal} {\bibinfo  {journal} {Phys. Rev. D}\ }\textbf {\bibinfo {volume} {98}},\ \bibinfo {pages} {030001} (\bibinfo {year} {2018})}\BibitemShut {NoStop}%
\bibitem [{\citenamefont {Sjostrand}\ \emph {et~al.}(2006)\citenamefont {Sjostrand}, \citenamefont {Mrenna},\ and\ \citenamefont {Skands}}]{Sjostrand:2006za}%
  \BibitemOpen
  \bibfield  {author} {\bibinfo {author} {\bibfnamefont {T.}~\bibnamefont {Sjostrand}}, \bibinfo {author} {\bibfnamefont {S.}~\bibnamefont {Mrenna}},\ and\ \bibinfo {author} {\bibfnamefont {P.~Z.}\ \bibnamefont {Skands}},\ }\bibfield  {title} {\bibinfo {title} {{PYTHIA 6.4 Physics and Manual}},\ }\href {https://doi.org/10.1088/1126-6708/2006/05/026} {\bibfield  {journal} {\bibinfo  {journal} {JHEP}\ }\textbf {\bibinfo {volume} {05}},\ \bibinfo {pages} {026}},\ \Eprint {https://arxiv.org/abs/hep-ph/0603175} {arXiv:hep-ph/0603175} \BibitemShut {NoStop}%
\bibitem [{\citenamefont {Sjostrand}\ \emph {et~al.}(2008)\citenamefont {Sjostrand}, \citenamefont {Mrenna},\ and\ \citenamefont {Skands}}]{Sjostrand:2007gs}%
  \BibitemOpen
  \bibfield  {author} {\bibinfo {author} {\bibfnamefont {T.}~\bibnamefont {Sjostrand}}, \bibinfo {author} {\bibfnamefont {S.}~\bibnamefont {Mrenna}},\ and\ \bibinfo {author} {\bibfnamefont {P.~Z.}\ \bibnamefont {Skands}},\ }\bibfield  {title} {\bibinfo {title} {{A Brief Introduction to PYTHIA 8.1}},\ }\href {https://doi.org/10.1016/j.cpc.2008.01.036} {\bibfield  {journal} {\bibinfo  {journal} {Comput. Phys. Commun.}\ }\textbf {\bibinfo {volume} {178}},\ \bibinfo {pages} {852} (\bibinfo {year} {2008})},\ \Eprint {https://arxiv.org/abs/0710.3820} {arXiv:0710.3820 [hep-ph]} \BibitemShut {NoStop}%
\bibitem [{\citenamefont {Manley}\ and\ \citenamefont {Saleski}(1992)}]{Manley:1992yb}%
  \BibitemOpen
  \bibfield  {author} {\bibinfo {author} {\bibfnamefont {D.~M.}\ \bibnamefont {Manley}}\ and\ \bibinfo {author} {\bibfnamefont {E.~M.}\ \bibnamefont {Saleski}},\ }\bibfield  {title} {\bibinfo {title} {{Multichannel resonance parametrization of pi N scattering amplitudes}},\ }\href {https://doi.org/10.1103/PhysRevD.45.4002} {\bibfield  {journal} {\bibinfo  {journal} {Phys. Rev. D}\ }\textbf {\bibinfo {volume} {45}},\ \bibinfo {pages} {4002} (\bibinfo {year} {1992})}\BibitemShut {NoStop}%
\bibitem [{\citenamefont {Hirano}\ and\ \citenamefont {Nara}(2012)}]{Hirano:2012yy}%
  \BibitemOpen
  \bibfield  {author} {\bibinfo {author} {\bibfnamefont {T.}~\bibnamefont {Hirano}}\ and\ \bibinfo {author} {\bibfnamefont {Y.}~\bibnamefont {Nara}},\ }\bibfield  {title} {\bibinfo {title} {{Dynamical modeling of high energy heavy ion collisions}},\ }\href {https://doi.org/10.1093/ptep/pts007} {\bibfield  {journal} {\bibinfo  {journal} {PTEP}\ }\textbf {\bibinfo {volume} {2012}},\ \bibinfo {pages} {01A203} (\bibinfo {year} {2012})},\ \Eprint {https://arxiv.org/abs/1203.4418} {arXiv:1203.4418 [nucl-th]} \BibitemShut {NoStop}%
\bibitem [{\citenamefont {Mohs}\ \emph {et~al.}(2020)\citenamefont {Mohs}, \citenamefont {Ryu},\ and\ \citenamefont {Elfner}}]{Mohs:2019iee}%
  \BibitemOpen
  \bibfield  {author} {\bibinfo {author} {\bibfnamefont {J.}~\bibnamefont {Mohs}}, \bibinfo {author} {\bibfnamefont {S.}~\bibnamefont {Ryu}},\ and\ \bibinfo {author} {\bibfnamefont {H.}~\bibnamefont {Elfner}} (\bibinfo {collaboration} {SMASH}),\ }\bibfield  {title} {\bibinfo {title} {{Particle Production via Strings and Baryon Stopping within a Hadronic Transport Approach}},\ }\href {https://doi.org/10.1088/1361-6471/ab7bd1} {\bibfield  {journal} {\bibinfo  {journal} {J. Phys. G}\ }\textbf {\bibinfo {volume} {47}},\ \bibinfo {pages} {065101} (\bibinfo {year} {2020})},\ \Eprint {https://arxiv.org/abs/1909.05586} {arXiv:1909.05586 [nucl-th]} \BibitemShut {NoStop}%
\bibitem [{\citenamefont {Bierlich}(2024)}]{Bierlich:2024odg}%
  \BibitemOpen
  \bibfield  {author} {\bibinfo {author} {\bibfnamefont {C.}~\bibnamefont {Bierlich}},\ }\bibfield  {title} {\bibinfo {title} {{String Interactions as a Source of Collective Behaviour}},\ }\href {https://doi.org/10.3390/universe10010046} {\bibfield  {journal} {\bibinfo  {journal} {Universe}\ }\textbf {\bibinfo {volume} {10}},\ \bibinfo {pages} {46} (\bibinfo {year} {2024})},\ \Eprint {https://arxiv.org/abs/2401.07585} {arXiv:2401.07585 [hep-ph]} \BibitemShut {NoStop}%
\bibitem [{\citenamefont {Schenke}\ \emph {et~al.}(2010)\citenamefont {Schenke}, \citenamefont {Jeon},\ and\ \citenamefont {Gale}}]{Schenke:2010nt}%
  \BibitemOpen
  \bibfield  {author} {\bibinfo {author} {\bibfnamefont {B.}~\bibnamefont {Schenke}}, \bibinfo {author} {\bibfnamefont {S.}~\bibnamefont {Jeon}},\ and\ \bibinfo {author} {\bibfnamefont {C.}~\bibnamefont {Gale}},\ }\bibfield  {title} {\bibinfo {title} {{(3+1)D hydrodynamic simulation of relativistic heavy-ion collisions}},\ }\href {https://doi.org/10.1103/PhysRevC.82.014903} {\bibfield  {journal} {\bibinfo  {journal} {Phys. Rev. C}\ }\textbf {\bibinfo {volume} {82}},\ \bibinfo {pages} {014903} (\bibinfo {year} {2010})},\ \Eprint {https://arxiv.org/abs/1004.1408} {arXiv:1004.1408 [hep-ph]} \BibitemShut {NoStop}%
\bibitem [{\citenamefont {Schenke}\ \emph {et~al.}(2011)\citenamefont {Schenke}, \citenamefont {Jeon},\ and\ \citenamefont {Gale}}]{Schenke:2010rr}%
  \BibitemOpen
  \bibfield  {author} {\bibinfo {author} {\bibfnamefont {B.}~\bibnamefont {Schenke}}, \bibinfo {author} {\bibfnamefont {S.}~\bibnamefont {Jeon}},\ and\ \bibinfo {author} {\bibfnamefont {C.}~\bibnamefont {Gale}},\ }\bibfield  {title} {\bibinfo {title} {{Elliptic and triangular flow in event-by-event (3+1)D viscous hydrodynamics}},\ }\href {https://doi.org/10.1103/PhysRevLett.106.042301} {\bibfield  {journal} {\bibinfo  {journal} {Phys. Rev. Lett.}\ }\textbf {\bibinfo {volume} {106}},\ \bibinfo {pages} {042301} (\bibinfo {year} {2011})},\ \Eprint {https://arxiv.org/abs/1009.3244} {arXiv:1009.3244 [hep-ph]} \BibitemShut {NoStop}%
\bibitem [{\citenamefont {Paquet}\ \emph {et~al.}(2016)\citenamefont {Paquet}, \citenamefont {Shen}, \citenamefont {Denicol}, \citenamefont {Luzum}, \citenamefont {Schenke}, \citenamefont {Jeon},\ and\ \citenamefont {Gale}}]{Paquet:2015lta}%
  \BibitemOpen
  \bibfield  {author} {\bibinfo {author} {\bibfnamefont {J.-F.}\ \bibnamefont {Paquet}}, \bibinfo {author} {\bibfnamefont {C.}~\bibnamefont {Shen}}, \bibinfo {author} {\bibfnamefont {G.~S.}\ \bibnamefont {Denicol}}, \bibinfo {author} {\bibfnamefont {M.}~\bibnamefont {Luzum}}, \bibinfo {author} {\bibfnamefont {B.}~\bibnamefont {Schenke}}, \bibinfo {author} {\bibfnamefont {S.}~\bibnamefont {Jeon}},\ and\ \bibinfo {author} {\bibfnamefont {C.}~\bibnamefont {Gale}},\ }\bibfield  {title} {\bibinfo {title} {{Production of photons in relativistic heavy-ion collisions}},\ }\href {https://doi.org/10.1103/PhysRevC.93.044906} {\bibfield  {journal} {\bibinfo  {journal} {Phys. Rev. C}\ }\textbf {\bibinfo {volume} {93}},\ \bibinfo {pages} {044906} (\bibinfo {year} {2016})},\ \Eprint {https://arxiv.org/abs/1509.06738} {arXiv:1509.06738 [hep-ph]} \BibitemShut {NoStop}%
\bibitem [{\citenamefont {Denicol}\ \emph {et~al.}(2018)\citenamefont {Denicol}, \citenamefont {Gale}, \citenamefont {Jeon}, \citenamefont {Monnai}, \citenamefont {Schenke},\ and\ \citenamefont {Shen}}]{Denicol:2018wdp}%
  \BibitemOpen
  \bibfield  {author} {\bibinfo {author} {\bibfnamefont {G.~S.}\ \bibnamefont {Denicol}}, \bibinfo {author} {\bibfnamefont {C.}~\bibnamefont {Gale}}, \bibinfo {author} {\bibfnamefont {S.}~\bibnamefont {Jeon}}, \bibinfo {author} {\bibfnamefont {A.}~\bibnamefont {Monnai}}, \bibinfo {author} {\bibfnamefont {B.}~\bibnamefont {Schenke}},\ and\ \bibinfo {author} {\bibfnamefont {C.}~\bibnamefont {Shen}},\ }\bibfield  {title} {\bibinfo {title} {{Net baryon diffusion in fluid dynamic simulations of relativistic heavy-ion collisions}},\ }\href {https://doi.org/10.1103/PhysRevC.98.034916} {\bibfield  {journal} {\bibinfo  {journal} {Phys. Rev. C}\ }\textbf {\bibinfo {volume} {98}},\ \bibinfo {pages} {034916} (\bibinfo {year} {2018})},\ \Eprint {https://arxiv.org/abs/1804.10557} {arXiv:1804.10557 [nucl-th]} \BibitemShut {NoStop}%
\bibitem [{\citenamefont {Oliinychenko}\ and\ \citenamefont {Petersen}(2016)}]{Oliinychenko:2015lva}%
  \BibitemOpen
  \bibfield  {author} {\bibinfo {author} {\bibfnamefont {D.}~\bibnamefont {Oliinychenko}}\ and\ \bibinfo {author} {\bibfnamefont {H.}~\bibnamefont {Petersen}},\ }\bibfield  {title} {\bibinfo {title} {{Deviations of the Energy-Momentum Tensor from Equilibrium in the Initial State for Hydrodynamics from Transport Approaches}},\ }\href {https://doi.org/10.1103/PhysRevC.93.034905} {\bibfield  {journal} {\bibinfo  {journal} {Phys. Rev. C}\ }\textbf {\bibinfo {volume} {93}},\ \bibinfo {pages} {034905} (\bibinfo {year} {2016})},\ \Eprint {https://arxiv.org/abs/1508.04378} {arXiv:1508.04378 [nucl-th]} \BibitemShut {NoStop}%
\bibitem [{\citenamefont {Du}(2025)}]{Du:2025fdp}%
  \BibitemOpen
  \bibfield  {author} {\bibinfo {author} {\bibfnamefont {L.}~\bibnamefont {Du}},\ }\bibfield  {title} {\bibinfo {title} {{Efficient calculation of thermodynamic properties of baryon-rich QCD matter from heavy-ion transport models}},\ }\href {https://doi.org/10.1016/j.cpc.2025.109578} {\bibfield  {journal} {\bibinfo  {journal} {Comput. Phys. Commun.}\ }\textbf {\bibinfo {volume} {312}},\ \bibinfo {pages} {109578} (\bibinfo {year} {2025})},\ \Eprint {https://arxiv.org/abs/2506.19766} {arXiv:2506.19766 [physics.comp-ph]} \BibitemShut {NoStop}%
\bibitem [{\citenamefont {Denicol}\ \emph {et~al.}(2012)\citenamefont {Denicol}, \citenamefont {Niemi}, \citenamefont {Molnar},\ and\ \citenamefont {Rischke}}]{Denicol:2012cn}%
  \BibitemOpen
  \bibfield  {author} {\bibinfo {author} {\bibfnamefont {G.~S.}\ \bibnamefont {Denicol}}, \bibinfo {author} {\bibfnamefont {H.}~\bibnamefont {Niemi}}, \bibinfo {author} {\bibfnamefont {E.}~\bibnamefont {Molnar}},\ and\ \bibinfo {author} {\bibfnamefont {D.~H.}\ \bibnamefont {Rischke}},\ }\bibfield  {title} {\bibinfo {title} {{Derivation of transient relativistic fluid dynamics from the Boltzmann equation}},\ }\href {https://doi.org/10.1103/PhysRevD.85.114047} {\bibfield  {journal} {\bibinfo  {journal} {Phys. Rev. D}\ }\textbf {\bibinfo {volume} {85}},\ \bibinfo {pages} {114047} (\bibinfo {year} {2012})},\ \bibinfo {note} {[Erratum: Phys.Rev.D 91, 039902 (2015)]},\ \Eprint {https://arxiv.org/abs/1202.4551} {arXiv:1202.4551 [nucl-th]} \BibitemShut {NoStop}%
\bibitem [{\citenamefont {Denicol}\ \emph {et~al.}(2014)\citenamefont {Denicol}, \citenamefont {Jeon},\ and\ \citenamefont {Gale}}]{Denicol:2014vaa}%
  \BibitemOpen
  \bibfield  {author} {\bibinfo {author} {\bibfnamefont {G.~S.}\ \bibnamefont {Denicol}}, \bibinfo {author} {\bibfnamefont {S.}~\bibnamefont {Jeon}},\ and\ \bibinfo {author} {\bibfnamefont {C.}~\bibnamefont {Gale}},\ }\bibfield  {title} {\bibinfo {title} {{Transport Coefficients of Bulk Viscous Pressure in the 14-moment approximation}},\ }\href {https://doi.org/10.1103/PhysRevC.90.024912} {\bibfield  {journal} {\bibinfo  {journal} {Phys. Rev. C}\ }\textbf {\bibinfo {volume} {90}},\ \bibinfo {pages} {024912} (\bibinfo {year} {2014})},\ \Eprint {https://arxiv.org/abs/1403.0962} {arXiv:1403.0962 [nucl-th]} \BibitemShut {NoStop}%
\bibitem [{\citenamefont {Pihan}\ \emph {et~al.}(2024{\natexlab{a}})\citenamefont {Pihan}, \citenamefont {Monnai}, \citenamefont {Schenke},\ and\ \citenamefont {Shen}}]{Pihan:2023dsb}%
  \BibitemOpen
  \bibfield  {author} {\bibinfo {author} {\bibfnamefont {G.}~\bibnamefont {Pihan}}, \bibinfo {author} {\bibfnamefont {A.}~\bibnamefont {Monnai}}, \bibinfo {author} {\bibfnamefont {B.}~\bibnamefont {Schenke}},\ and\ \bibinfo {author} {\bibfnamefont {C.}~\bibnamefont {Shen}},\ }\bibfield  {title} {\bibinfo {title} {{Tracing baryon and electric charge transport in isobar collisions}},\ }\href {https://doi.org/10.1051/epjconf/202429605005} {\bibfield  {journal} {\bibinfo  {journal} {EPJ Web Conf.}\ }\textbf {\bibinfo {volume} {296}},\ \bibinfo {pages} {05005} (\bibinfo {year} {2024}{\natexlab{a}})},\ \Eprint {https://arxiv.org/abs/2312.12376} {arXiv:2312.12376 [nucl-th]} \BibitemShut {NoStop}%
\bibitem [{\citenamefont {Pihan}\ \emph {et~al.}(2024{\natexlab{b}})\citenamefont {Pihan}, \citenamefont {Monnai}, \citenamefont {Schenke},\ and\ \citenamefont {Shen}}]{Pihan:2024lxw}%
  \BibitemOpen
  \bibfield  {author} {\bibinfo {author} {\bibfnamefont {G.}~\bibnamefont {Pihan}}, \bibinfo {author} {\bibfnamefont {A.}~\bibnamefont {Monnai}}, \bibinfo {author} {\bibfnamefont {B.}~\bibnamefont {Schenke}},\ and\ \bibinfo {author} {\bibfnamefont {C.}~\bibnamefont {Shen}},\ }\bibfield  {title} {\bibinfo {title} {{Unveiling Baryon Charge Carriers through Charge Stopping in Isobar Collisions}},\ }\href {https://doi.org/10.1103/PhysRevLett.133.182301} {\bibfield  {journal} {\bibinfo  {journal} {Phys. Rev. Lett.}\ }\textbf {\bibinfo {volume} {133}},\ \bibinfo {pages} {182301} (\bibinfo {year} {2024}{\natexlab{b}})},\ \Eprint {https://arxiv.org/abs/2405.19439} {arXiv:2405.19439 [nucl-th]} \BibitemShut {NoStop}%
\bibitem [{\citenamefont {Jahan}\ \emph {et~al.}(2024)\citenamefont {Jahan}, \citenamefont {Roch},\ and\ \citenamefont {Shen}}]{Jahan:2024wpj}%
  \BibitemOpen
  \bibfield  {author} {\bibinfo {author} {\bibfnamefont {S.~A.}\ \bibnamefont {Jahan}}, \bibinfo {author} {\bibfnamefont {H.}~\bibnamefont {Roch}},\ and\ \bibinfo {author} {\bibfnamefont {C.}~\bibnamefont {Shen}},\ }\bibfield  {title} {\bibinfo {title} {{Bayesian analysis of (3+1)D relativistic nuclear dynamics with the RHIC beam energy scan data}},\ }\href {https://doi.org/10.1103/PhysRevC.110.054905} {\bibfield  {journal} {\bibinfo  {journal} {Phys. Rev. C}\ }\textbf {\bibinfo {volume} {110}},\ \bibinfo {pages} {054905} (\bibinfo {year} {2024})},\ \Eprint {https://arxiv.org/abs/2408.00537} {arXiv:2408.00537 [nucl-th]} \BibitemShut {NoStop}%
\bibitem [{\citenamefont {Huovinen}\ and\ \citenamefont {Petersen}(2012)}]{Huovinen:2012is}%
  \BibitemOpen
  \bibfield  {author} {\bibinfo {author} {\bibfnamefont {P.}~\bibnamefont {Huovinen}}\ and\ \bibinfo {author} {\bibfnamefont {H.}~\bibnamefont {Petersen}},\ }\bibfield  {title} {\bibinfo {title} {{Particlization in hybrid models}},\ }\href {https://doi.org/10.1140/epja/i2012-12171-9} {\bibfield  {journal} {\bibinfo  {journal} {Eur. Phys. J. A}\ }\textbf {\bibinfo {volume} {48}},\ \bibinfo {pages} {171} (\bibinfo {year} {2012})},\ \Eprint {https://arxiv.org/abs/1206.3371} {arXiv:1206.3371 [nucl-th]} \BibitemShut {NoStop}%
\bibitem [{\citenamefont {Venugopalan}\ and\ \citenamefont {Prakash}(1992)}]{Venugopalan:1992hy}%
  \BibitemOpen
  \bibfield  {author} {\bibinfo {author} {\bibfnamefont {R.}~\bibnamefont {Venugopalan}}\ and\ \bibinfo {author} {\bibfnamefont {M.}~\bibnamefont {Prakash}},\ }\bibfield  {title} {\bibinfo {title} {{Thermal properties of interacting hadrons}},\ }\href {https://doi.org/10.1016/0375-9474(92)90005-5} {\bibfield  {journal} {\bibinfo  {journal} {Nucl. Phys. A}\ }\textbf {\bibinfo {volume} {546}},\ \bibinfo {pages} {718} (\bibinfo {year} {1992})}\BibitemShut {NoStop}%
\bibitem [{\citenamefont {Everett}\ \emph {et~al.}(2021)\citenamefont {Everett} \emph {et~al.}}]{JETSCAPE:2020mzn}%
  \BibitemOpen
  \bibfield  {author} {\bibinfo {author} {\bibfnamefont {D.}~\bibnamefont {Everett}} \emph {et~al.} (\bibinfo {collaboration} {JETSCAPE}),\ }\bibfield  {title} {\bibinfo {title} {{Multisystem Bayesian constraints on the transport coefficients of QCD matter}},\ }\href {https://doi.org/10.1103/PhysRevC.103.054904} {\bibfield  {journal} {\bibinfo  {journal} {Phys. Rev. C}\ }\textbf {\bibinfo {volume} {103}},\ \bibinfo {pages} {054904} (\bibinfo {year} {2021})},\ \Eprint {https://arxiv.org/abs/2011.01430} {arXiv:2011.01430 [hep-ph]} \BibitemShut {NoStop}%
\bibitem [{\citenamefont {Fotakis}\ \emph {et~al.}(2022)\citenamefont {Fotakis}, \citenamefont {Moln{\'a}r}, \citenamefont {Niemi}, \citenamefont {Greiner},\ and\ \citenamefont {Rischke}}]{Fotakis:2022usk}%
  \BibitemOpen
  \bibfield  {author} {\bibinfo {author} {\bibfnamefont {J.~A.}\ \bibnamefont {Fotakis}}, \bibinfo {author} {\bibfnamefont {E.}~\bibnamefont {Moln{\'a}r}}, \bibinfo {author} {\bibfnamefont {H.}~\bibnamefont {Niemi}}, \bibinfo {author} {\bibfnamefont {C.}~\bibnamefont {Greiner}},\ and\ \bibinfo {author} {\bibfnamefont {D.~H.}\ \bibnamefont {Rischke}},\ }\bibfield  {title} {\bibinfo {title} {{Multicomponent relativistic dissipative fluid dynamics from the Boltzmann equation}},\ }\href {https://doi.org/10.1103/PhysRevD.106.036009} {\bibfield  {journal} {\bibinfo  {journal} {Phys. Rev. D}\ }\textbf {\bibinfo {volume} {106}},\ \bibinfo {pages} {036009} (\bibinfo {year} {2022})},\ \Eprint {https://arxiv.org/abs/2203.11549} {arXiv:2203.11549 [nucl-th]} \BibitemShut {NoStop}%
\bibitem [{\citenamefont {Monnai}\ and\ \citenamefont {Hirano}(2009)}]{Monnai:2009ad}%
  \BibitemOpen
  \bibfield  {author} {\bibinfo {author} {\bibfnamefont {A.}~\bibnamefont {Monnai}}\ and\ \bibinfo {author} {\bibfnamefont {T.}~\bibnamefont {Hirano}},\ }\bibfield  {title} {\bibinfo {title} {{Effects of Bulk Viscosity at Freezeout}},\ }\href {https://doi.org/10.1103/PhysRevC.80.054906} {\bibfield  {journal} {\bibinfo  {journal} {Phys. Rev. C}\ }\textbf {\bibinfo {volume} {80}},\ \bibinfo {pages} {054906} (\bibinfo {year} {2009})},\ \Eprint {https://arxiv.org/abs/0903.4436} {arXiv:0903.4436 [nucl-th]} \BibitemShut {NoStop}%
\bibitem [{\citenamefont {Monnai}\ and\ \citenamefont {Hirano}(2010)}]{Monnai:2010qp}%
  \BibitemOpen
  \bibfield  {author} {\bibinfo {author} {\bibfnamefont {A.}~\bibnamefont {Monnai}}\ and\ \bibinfo {author} {\bibfnamefont {T.}~\bibnamefont {Hirano}},\ }\bibfield  {title} {\bibinfo {title} {{Relativistic Dissipative Hydrodynamic Equations at the Second Order for Multi-Component Systems with Multiple Conserved Currents}},\ }\href {https://doi.org/10.1016/j.nuclphysa.2010.08.002} {\bibfield  {journal} {\bibinfo  {journal} {Nucl. Phys. A}\ }\textbf {\bibinfo {volume} {847}},\ \bibinfo {pages} {283} (\bibinfo {year} {2010})},\ \Eprint {https://arxiv.org/abs/1003.3087} {arXiv:1003.3087 [nucl-th]} \BibitemShut {NoStop}%
\bibitem [{\citenamefont {McNelis}\ \emph {et~al.}(2021)\citenamefont {McNelis}, \citenamefont {Everett},\ and\ \citenamefont {Heinz}}]{McNelis:2019auj}%
  \BibitemOpen
  \bibfield  {author} {\bibinfo {author} {\bibfnamefont {M.}~\bibnamefont {McNelis}}, \bibinfo {author} {\bibfnamefont {D.}~\bibnamefont {Everett}},\ and\ \bibinfo {author} {\bibfnamefont {U.}~\bibnamefont {Heinz}},\ }\bibfield  {title} {\bibinfo {title} {{Particlization in fluid dynamical simulations of heavy-ion collisions: The i S3D module}},\ }\href {https://doi.org/10.1016/j.cpc.2020.107604} {\bibfield  {journal} {\bibinfo  {journal} {Comput. Phys. Commun.}\ }\textbf {\bibinfo {volume} {258}},\ \bibinfo {pages} {107604} (\bibinfo {year} {2021})},\ \Eprint {https://arxiv.org/abs/1912.08271} {arXiv:1912.08271 [nucl-th]} \BibitemShut {NoStop}%
\bibitem [{\citenamefont {Czajka}\ \emph {et~al.}(2018)\citenamefont {Czajka}, \citenamefont {Hauksson}, \citenamefont {Shen}, \citenamefont {Jeon},\ and\ \citenamefont {Gale}}]{Czajka:2017wdo}%
  \BibitemOpen
  \bibfield  {author} {\bibinfo {author} {\bibfnamefont {A.}~\bibnamefont {Czajka}}, \bibinfo {author} {\bibfnamefont {S.}~\bibnamefont {Hauksson}}, \bibinfo {author} {\bibfnamefont {C.}~\bibnamefont {Shen}}, \bibinfo {author} {\bibfnamefont {S.}~\bibnamefont {Jeon}},\ and\ \bibinfo {author} {\bibfnamefont {C.}~\bibnamefont {Gale}},\ }\bibfield  {title} {\bibinfo {title} {{Bulk viscosity of strongly interacting matter in the relaxation time approximation}},\ }\href {https://doi.org/10.1103/PhysRevC.97.044914} {\bibfield  {journal} {\bibinfo  {journal} {Phys. Rev. C}\ }\textbf {\bibinfo {volume} {97}},\ \bibinfo {pages} {044914} (\bibinfo {year} {2018})},\ \Eprint {https://arxiv.org/abs/1712.05905} {arXiv:1712.05905 [nucl-th]} \BibitemShut {NoStop}%
\bibitem [{\citenamefont {Czajka}\ \emph {et~al.}(2021)\citenamefont {Czajka}, \citenamefont {Shen}, \citenamefont {Hauksson}, \citenamefont {Jeon},\ and\ \citenamefont {Gale}}]{Czajka:2020mho}%
  \BibitemOpen
  \bibfield  {author} {\bibinfo {author} {\bibfnamefont {A.}~\bibnamefont {Czajka}}, \bibinfo {author} {\bibfnamefont {C.}~\bibnamefont {Shen}}, \bibinfo {author} {\bibfnamefont {S.}~\bibnamefont {Hauksson}}, \bibinfo {author} {\bibfnamefont {S.}~\bibnamefont {Jeon}},\ and\ \bibinfo {author} {\bibfnamefont {C.}~\bibnamefont {Gale}},\ }\bibfield  {title} {\bibinfo {title} {{Effects of the Mean Field on Fluid Dynamics in the Relaxation Time Approximation}},\ }\href {https://doi.org/10.5506/APhysPolBSupp.14.169} {\bibfield  {journal} {\bibinfo  {journal} {Acta Phys. Polon. Supp.}\ }\textbf {\bibinfo {volume} {14}},\ \bibinfo {pages} {169} (\bibinfo {year} {2021})},\ \Eprint {https://arxiv.org/abs/2004.11920} {arXiv:2004.11920 [nucl-th]} \BibitemShut {NoStop}%
\bibitem [{\citenamefont {Ryu}\ and\ \citenamefont {Shen}(2025)}]{ryu_2025_17088248}%
  \BibitemOpen
  \bibfield  {author} {\bibinfo {author} {\bibfnamefont {S.}~\bibnamefont {Ryu}}\ and\ \bibinfo {author} {\bibfnamefont {C.}~\bibnamefont {Shen}},\ }\href {https://doi.org/10.5281/zenodo.17088248} {\bibinfo {title} {Thermodynamic integrals for heavy-ion collisions}} (\bibinfo {year} {2025})\BibitemShut {NoStop}%
\bibitem [{\citenamefont {Shen}\ \emph {et~al.}(2016)\citenamefont {Shen}, \citenamefont {Qiu}, \citenamefont {Song}, \citenamefont {Bernhard}, \citenamefont {Bass},\ and\ \citenamefont {Heinz}}]{Shen:2014vra}%
  \BibitemOpen
  \bibfield  {author} {\bibinfo {author} {\bibfnamefont {C.}~\bibnamefont {Shen}}, \bibinfo {author} {\bibfnamefont {Z.}~\bibnamefont {Qiu}}, \bibinfo {author} {\bibfnamefont {H.}~\bibnamefont {Song}}, \bibinfo {author} {\bibfnamefont {J.}~\bibnamefont {Bernhard}}, \bibinfo {author} {\bibfnamefont {S.}~\bibnamefont {Bass}},\ and\ \bibinfo {author} {\bibfnamefont {U.}~\bibnamefont {Heinz}},\ }\bibfield  {title} {\bibinfo {title} {{The iEBE-VISHNU code package for relativistic heavy-ion collisions}},\ }\href {https://doi.org/10.1016/j.cpc.2015.08.039} {\bibfield  {journal} {\bibinfo  {journal} {Comput. Phys. Commun.}\ }\textbf {\bibinfo {volume} {199}},\ \bibinfo {pages} {61} (\bibinfo {year} {2016})},\ \Eprint {https://arxiv.org/abs/1409.8164} {arXiv:1409.8164 [nucl-th]} \BibitemShut {NoStop}%
\bibitem [{\citenamefont {{JETSCAPE Collaboration}}(2025{\natexlab{a}})}]{XSCAPErepo}%
  \BibitemOpen
  \bibfield  {author} {\bibinfo {author} {\bibnamefont {{JETSCAPE Collaboration}}},\ }\href@noop {} {\bibinfo {title} {{X-SCAPE GitHub Repository}}} (\bibinfo {year} {2025}{\natexlab{a}}),\ \bibinfo {note} {\url{https://github.com/JETSCAPE/X-SCAPE/tree/v2.0}.}\BibitemShut {Stop}%
\bibitem [{\citenamefont {Courant}\ \emph {et~al.}(1928)\citenamefont {Courant}, \citenamefont {Friedrichs},\ and\ \citenamefont {Lewy}}]{Courant1928}%
  \BibitemOpen
  \bibfield  {author} {\bibinfo {author} {\bibfnamefont {R.}~\bibnamefont {Courant}}, \bibinfo {author} {\bibfnamefont {K.}~\bibnamefont {Friedrichs}},\ and\ \bibinfo {author} {\bibfnamefont {H.}~\bibnamefont {Lewy}},\ }\bibfield  {title} {\bibinfo {title} {{\"U}ber die partiellen differenzengleichungen der mathematischen physik},\ }\href {https://doi.org/10.1007/BF01448839} {\bibfield  {journal} {\bibinfo  {journal} {Mathematische Annalen}\ }\textbf {\bibinfo {volume} {100}},\ \bibinfo {pages} {32} (\bibinfo {year} {1928})}\BibitemShut {NoStop}%
\bibitem [{\citenamefont {Sass}\ \emph {et~al.}(2026)\citenamefont {Sass}, \citenamefont {Roch}, \citenamefont {G{\"o}tz}, \citenamefont {Krupczak},\ and\ \citenamefont {Rosenkvist}}]{Sass:2025opk}%
  \BibitemOpen
  \bibfield  {author} {\bibinfo {author} {\bibfnamefont {N.}~\bibnamefont {Sass}}, \bibinfo {author} {\bibfnamefont {H.}~\bibnamefont {Roch}}, \bibinfo {author} {\bibfnamefont {N.}~\bibnamefont {G{\"o}tz}}, \bibinfo {author} {\bibfnamefont {R.}~\bibnamefont {Krupczak}},\ and\ \bibinfo {author} {\bibfnamefont {C.~B.}\ \bibnamefont {Rosenkvist}},\ }\bibfield  {title} {\bibinfo {title} {{SPARKX: a software package for analyzing relativistic kinematics in collision experiments}},\ }\href {https://doi.org/10.1140/epjc/s10052-025-15258-8} {\bibfield  {journal} {\bibinfo  {journal} {Eur. Phys. J. C}\ }\textbf {\bibinfo {volume} {86}},\ \bibinfo {pages} {27} (\bibinfo {year} {2026})},\ \Eprint {https://arxiv.org/abs/2503.09415} {arXiv:2503.09415 [physics.data-an]} \BibitemShut {NoStop}%
\bibitem [{\citenamefont {Roch}\ \emph {et~al.}(2025)\citenamefont {Roch}, \citenamefont {Götz}, \citenamefont {Saß}, \citenamefont {Krupczak},\ and\ \citenamefont {Carl-Rosenkvist}}]{hendrik_roch_2025_15838371}%
  \BibitemOpen
  \bibfield  {author} {\bibinfo {author} {\bibfnamefont {H.}~\bibnamefont {Roch}}, \bibinfo {author} {\bibfnamefont {N.}~\bibnamefont {Götz}}, \bibinfo {author} {\bibfnamefont {N.}~\bibnamefont {Saß}}, \bibinfo {author} {\bibfnamefont {R.}~\bibnamefont {Krupczak}},\ and\ \bibinfo {author} {\bibnamefont {Carl-Rosenkvist}},\ }\href {https://doi.org/10.5281/zenodo.15838371} {\bibinfo {title} {smash-transport/sparkx: v2.1.1-chatelet}} (\bibinfo {year} {2025})\BibitemShut {NoStop}%
\bibitem [{\citenamefont {Bierlich}\ \emph {et~al.}(2018)\citenamefont {Bierlich}, \citenamefont {Gustafson}, \citenamefont {L{\"o}nnblad},\ and\ \citenamefont {Shah}}]{Bierlich:2018xfw}%
  \BibitemOpen
  \bibfield  {author} {\bibinfo {author} {\bibfnamefont {C.}~\bibnamefont {Bierlich}}, \bibinfo {author} {\bibfnamefont {G.}~\bibnamefont {Gustafson}}, \bibinfo {author} {\bibfnamefont {L.}~\bibnamefont {L{\"o}nnblad}},\ and\ \bibinfo {author} {\bibfnamefont {H.}~\bibnamefont {Shah}},\ }\bibfield  {title} {\bibinfo {title} {{The Angantyr model for Heavy-Ion Collisions in PYTHIA8}},\ }\href {https://doi.org/10.1007/JHEP10(2018)134} {\bibfield  {journal} {\bibinfo  {journal} {JHEP}\ }\textbf {\bibinfo {volume} {10}},\ \bibinfo {pages} {134}},\ \Eprint {https://arxiv.org/abs/1806.10820} {arXiv:1806.10820 [hep-ph]} \BibitemShut {NoStop}%
\bibitem [{\citenamefont {Pierog}\ \emph {et~al.}(2015)\citenamefont {Pierog}, \citenamefont {Karpenko}, \citenamefont {Katzy}, \citenamefont {Yatsenko},\ and\ \citenamefont {Werner}}]{Pierog:2013ria}%
  \BibitemOpen
  \bibfield  {author} {\bibinfo {author} {\bibfnamefont {T.}~\bibnamefont {Pierog}}, \bibinfo {author} {\bibfnamefont {I.}~\bibnamefont {Karpenko}}, \bibinfo {author} {\bibfnamefont {J.~M.}\ \bibnamefont {Katzy}}, \bibinfo {author} {\bibfnamefont {E.}~\bibnamefont {Yatsenko}},\ and\ \bibinfo {author} {\bibfnamefont {K.}~\bibnamefont {Werner}},\ }\bibfield  {title} {\bibinfo {title} {{EPOS LHC: Test of collective hadronization with data measured at the CERN Large Hadron Collider}},\ }\href {https://doi.org/10.1103/PhysRevC.92.034906} {\bibfield  {journal} {\bibinfo  {journal} {Phys. Rev. C}\ }\textbf {\bibinfo {volume} {92}},\ \bibinfo {pages} {034906} (\bibinfo {year} {2015})},\ \Eprint {https://arxiv.org/abs/1306.0121} {arXiv:1306.0121 [hep-ph]} \BibitemShut {NoStop}%
\bibitem [{\citenamefont {Shen}\ and\ \citenamefont {Schenke}(2022)}]{Shen:2022oyg}%
  \BibitemOpen
  \bibfield  {author} {\bibinfo {author} {\bibfnamefont {C.}~\bibnamefont {Shen}}\ and\ \bibinfo {author} {\bibfnamefont {B.}~\bibnamefont {Schenke}},\ }\bibfield  {title} {\bibinfo {title} {{Longitudinal dynamics and particle production in relativistic nuclear collisions}},\ }\href {https://doi.org/10.1103/PhysRevC.105.064905} {\bibfield  {journal} {\bibinfo  {journal} {Phys. Rev. C}\ }\textbf {\bibinfo {volume} {105}},\ \bibinfo {pages} {064905} (\bibinfo {year} {2022})},\ \Eprint {https://arxiv.org/abs/2203.04685} {arXiv:2203.04685 [nucl-th]} \BibitemShut {NoStop}%
\bibitem [{\citenamefont {Werner}(2024)}]{Werner:2023mod}%
  \BibitemOpen
  \bibfield  {author} {\bibinfo {author} {\bibfnamefont {K.}~\bibnamefont {Werner}},\ }\bibfield  {title} {\bibinfo {title} {{Parallel scattering, saturation, and generalized Abramovskii-Gribov-Kancheli (AGK) theorem in the EPOS4 framework, with applications for heavy-ion collisions at sNN of 5.02 TeV and 200 GeV}},\ }\href {https://doi.org/10.1103/PhysRevC.109.034918} {\bibfield  {journal} {\bibinfo  {journal} {Phys. Rev. C}\ }\textbf {\bibinfo {volume} {109}},\ \bibinfo {pages} {034918} (\bibinfo {year} {2024})},\ \Eprint {https://arxiv.org/abs/2310.09380} {arXiv:2310.09380 [hep-ph]} \BibitemShut {NoStop}%
\bibitem [{\citenamefont {Garcia-Montero}\ \emph {et~al.}(2024)\citenamefont {Garcia-Montero}, \citenamefont {Elfner},\ and\ \citenamefont {Schlichting}}]{Garcia-Montero:2023gex}%
  \BibitemOpen
  \bibfield  {author} {\bibinfo {author} {\bibfnamefont {O.}~\bibnamefont {Garcia-Montero}}, \bibinfo {author} {\bibfnamefont {H.}~\bibnamefont {Elfner}},\ and\ \bibinfo {author} {\bibfnamefont {S.}~\bibnamefont {Schlichting}},\ }\bibfield  {title} {\bibinfo {title} {{McDIPPER: A novel saturation-based 3+1D initial-state model for heavy ion collisions}},\ }\href {https://doi.org/10.1103/PhysRevC.109.044916} {\bibfield  {journal} {\bibinfo  {journal} {Phys. Rev. C}\ }\textbf {\bibinfo {volume} {109}},\ \bibinfo {pages} {044916} (\bibinfo {year} {2024})},\ \Eprint {https://arxiv.org/abs/2308.11713} {arXiv:2308.11713 [hep-ph]} \BibitemShut {NoStop}%
\bibitem [{\citenamefont {Carzon}\ \emph {et~al.}(2022{\natexlab{a}})\citenamefont {Carzon}, \citenamefont {Martinez}, \citenamefont {Sievert}, \citenamefont {Wertepny},\ and\ \citenamefont {Noronha-Hostler}}]{Carzon:2019qja}%
  \BibitemOpen
  \bibfield  {author} {\bibinfo {author} {\bibfnamefont {P.}~\bibnamefont {Carzon}}, \bibinfo {author} {\bibfnamefont {M.}~\bibnamefont {Martinez}}, \bibinfo {author} {\bibfnamefont {M.~D.}\ \bibnamefont {Sievert}}, \bibinfo {author} {\bibfnamefont {D.~E.}\ \bibnamefont {Wertepny}},\ and\ \bibinfo {author} {\bibfnamefont {J.}~\bibnamefont {Noronha-Hostler}},\ }\bibfield  {title} {\bibinfo {title} {{Monte Carlo event generator for initial conditions of conserved charges in nuclear geometry}},\ }\href {https://doi.org/10.1103/PhysRevC.105.034908} {\bibfield  {journal} {\bibinfo  {journal} {Phys. Rev. C}\ }\textbf {\bibinfo {volume} {105}},\ \bibinfo {pages} {034908} (\bibinfo {year} {2022}{\natexlab{a}})},\ \Eprint {https://arxiv.org/abs/1911.12454} {arXiv:1911.12454 [nucl-th]} \BibitemShut {NoStop}%
\bibitem [{\citenamefont {Carzon}\ \emph {et~al.}(2022{\natexlab{b}})\citenamefont {Carzon}, \citenamefont {Martinez}, \citenamefont {Sievert}, \citenamefont {Wertepny},\ and\ \citenamefont {Noronha-Hostler}}]{Carzon:2022zpa}%
  \BibitemOpen
  \bibfield  {author} {\bibinfo {author} {\bibfnamefont {P.}~\bibnamefont {Carzon}}, \bibinfo {author} {\bibfnamefont {M.}~\bibnamefont {Martinez}}, \bibinfo {author} {\bibfnamefont {M.~D.}\ \bibnamefont {Sievert}}, \bibinfo {author} {\bibfnamefont {D.~E.}\ \bibnamefont {Wertepny}},\ and\ \bibinfo {author} {\bibfnamefont {J.}~\bibnamefont {Noronha-Hostler}},\ }\bibfield  {title} {\bibinfo {title} {{Initializing BSQ with Open-Source ICCING}},\ }\href {https://doi.org/10.31349/SuplRevMexFis.3.040912} {\bibfield  {journal} {\bibinfo  {journal} {Rev. Mex. Fis. Suppl.}\ }\textbf {\bibinfo {volume} {3}},\ \bibinfo {pages} {040912} (\bibinfo {year} {2022}{\natexlab{b}})},\ \Eprint {https://arxiv.org/abs/2207.09604} {arXiv:2207.09604 [nucl-th]} \BibitemShut {NoStop}%
\bibitem [{\citenamefont {Carzon}\ \emph {et~al.}(2023)\citenamefont {Carzon}, \citenamefont {Martinez}, \citenamefont {Noronha-Hostler}, \citenamefont {Plaschke}, \citenamefont {Schlichting},\ and\ \citenamefont {Sievert}}]{Carzon:2023zfp}%
  \BibitemOpen
  \bibfield  {author} {\bibinfo {author} {\bibfnamefont {P.}~\bibnamefont {Carzon}}, \bibinfo {author} {\bibfnamefont {M.}~\bibnamefont {Martinez}}, \bibinfo {author} {\bibfnamefont {J.}~\bibnamefont {Noronha-Hostler}}, \bibinfo {author} {\bibfnamefont {P.}~\bibnamefont {Plaschke}}, \bibinfo {author} {\bibfnamefont {S.}~\bibnamefont {Schlichting}},\ and\ \bibinfo {author} {\bibfnamefont {M.}~\bibnamefont {Sievert}},\ }\bibfield  {title} {\bibinfo {title} {{Pre-equilibrium evolution of conserved charges with initial conditions in the ICCING Monte Carlo event generator}},\ }\href {https://doi.org/10.1103/PhysRevC.108.064905} {\bibfield  {journal} {\bibinfo  {journal} {Phys. Rev. C}\ }\textbf {\bibinfo {volume} {108}},\ \bibinfo {pages} {064905} (\bibinfo {year} {2023})},\ \Eprint {https://arxiv.org/abs/2301.04572} {arXiv:2301.04572 [nucl-th]} \BibitemShut {NoStop}%
\bibitem [{\citenamefont {Zhao}\ \emph {et~al.}(2023)\citenamefont {Zhao}, \citenamefont {Ryu}, \citenamefont {Shen},\ and\ \citenamefont {Schenke}}]{Zhao:2022ugy}%
  \BibitemOpen
  \bibfield  {author} {\bibinfo {author} {\bibfnamefont {W.}~\bibnamefont {Zhao}}, \bibinfo {author} {\bibfnamefont {S.}~\bibnamefont {Ryu}}, \bibinfo {author} {\bibfnamefont {C.}~\bibnamefont {Shen}},\ and\ \bibinfo {author} {\bibfnamefont {B.}~\bibnamefont {Schenke}},\ }\bibfield  {title} {\bibinfo {title} {{3D structure of anisotropic flow in small collision systems at energies available at the BNL Relativistic Heavy Ion Collider}},\ }\href {https://doi.org/10.1103/PhysRevC.107.014904} {\bibfield  {journal} {\bibinfo  {journal} {Phys. Rev. C}\ }\textbf {\bibinfo {volume} {107}},\ \bibinfo {pages} {014904} (\bibinfo {year} {2023})},\ \Eprint {https://arxiv.org/abs/2211.16376} {arXiv:2211.16376 [nucl-th]} \BibitemShut {NoStop}%
\bibitem [{\citenamefont {Ryu}\ \emph {et~al.}(2024)\citenamefont {Ryu}, \citenamefont {Schenke}, \citenamefont {Shen},\ and\ \citenamefont {Zhao}}]{Ryu:2023bmx}%
  \BibitemOpen
  \bibfield  {author} {\bibinfo {author} {\bibfnamefont {S.}~\bibnamefont {Ryu}}, \bibinfo {author} {\bibfnamefont {B.}~\bibnamefont {Schenke}}, \bibinfo {author} {\bibfnamefont {C.}~\bibnamefont {Shen}},\ and\ \bibinfo {author} {\bibfnamefont {W.}~\bibnamefont {Zhao}},\ }\bibfield  {title} {\bibinfo {title} {{The role of longitudinal decorrelations for measurements of anisotropic flow in small collision systems}},\ }\href {https://doi.org/10.1051/epjconf/202429615001} {\bibfield  {journal} {\bibinfo  {journal} {EPJ Web Conf.}\ }\textbf {\bibinfo {volume} {296}},\ \bibinfo {pages} {15001} (\bibinfo {year} {2024})},\ \Eprint {https://arxiv.org/abs/2312.12595} {arXiv:2312.12595 [nucl-th]} \BibitemShut {NoStop}%
\bibitem [{\citenamefont {Kestin}\ and\ \citenamefont {Heinz}(2009)}]{Kestin:2008bh}%
  \BibitemOpen
  \bibfield  {author} {\bibinfo {author} {\bibfnamefont {G.}~\bibnamefont {Kestin}}\ and\ \bibinfo {author} {\bibfnamefont {U.~W.}\ \bibnamefont {Heinz}},\ }\bibfield  {title} {\bibinfo {title} {{Hydrodynamic radial and elliptic flow in heavy-ion collisions from AGS to LHC energies}},\ }\href {https://doi.org/10.1140/epjc/s10052-008-0832-y} {\bibfield  {journal} {\bibinfo  {journal} {Eur. Phys. J. C}\ }\textbf {\bibinfo {volume} {61}},\ \bibinfo {pages} {545} (\bibinfo {year} {2009})},\ \Eprint {https://arxiv.org/abs/0806.4539} {arXiv:0806.4539 [nucl-th]} \BibitemShut {NoStop}%
\bibitem [{\citenamefont {Shen}\ and\ \citenamefont {Heinz}(2012)}]{Shen:2012vn}%
  \BibitemOpen
  \bibfield  {author} {\bibinfo {author} {\bibfnamefont {C.}~\bibnamefont {Shen}}\ and\ \bibinfo {author} {\bibfnamefont {U.}~\bibnamefont {Heinz}},\ }\bibfield  {title} {\bibinfo {title} {{Collision Energy Dependence of Viscous Hydrodynamic Flow in Relativistic Heavy-Ion Collisions}},\ }\href {https://doi.org/10.1103/PhysRevC.85.054902} {\bibfield  {journal} {\bibinfo  {journal} {Phys. Rev. C}\ }\textbf {\bibinfo {volume} {85}},\ \bibinfo {pages} {054902} (\bibinfo {year} {2012})},\ \bibinfo {note} {[Erratum: Phys.Rev.C 86, 049903 (2012)]},\ \Eprint {https://arxiv.org/abs/1202.6620} {arXiv:1202.6620 [nucl-th]} \BibitemShut {NoStop}%
\bibitem [{\citenamefont {Shen}\ and\ \citenamefont {Heinz}(2013)}]{Shen:2012us}%
  \BibitemOpen
  \bibfield  {author} {\bibinfo {author} {\bibfnamefont {C.}~\bibnamefont {Shen}}\ and\ \bibinfo {author} {\bibfnamefont {U.}~\bibnamefont {Heinz}},\ }\bibfield  {title} {\bibinfo {title} {{Viscous Flow in Heavy-Ion Collisions from RHIC to LHC}},\ }\href {https://doi.org/10.1016/j.nuclphysa.2013.02.024} {\bibfield  {journal} {\bibinfo  {journal} {Nucl. Phys. A}\ }\textbf {\bibinfo {volume} {904-905}},\ \bibinfo {pages} {361c} (\bibinfo {year} {2013})},\ \Eprint {https://arxiv.org/abs/1210.2074} {arXiv:1210.2074 [nucl-th]} \BibitemShut {NoStop}%
\bibitem [{\citenamefont {{JETSCAPE Collaboration}}(2025{\natexlab{b}})}]{XSCAPE_xml}%
  \BibitemOpen
  \bibfield  {author} {\bibinfo {author} {\bibnamefont {{JETSCAPE Collaboration}}},\ }\href@noop {} {\bibinfo {title} {{X-SCAPE XML Input Files}}} (\bibinfo {year} {2025}{\natexlab{b}}),\ \bibinfo {note} {\url{https://github.com/JETSCAPE/Default-tunes}.}\BibitemShut {Stop}%
\end{thebibliography}%

\end{document}